\documentclass[structabstract]{aa}

\usepackage{graphicx}
\usepackage{txfonts}
\usepackage{natbib}
\usepackage{hyperref}
\usepackage{xspace}
\usepackage{lineno}
\usepackage{rotating}
\bibpunct{(}{)}{;}{a}{}{,} % to follow the A&A style
\hypersetup{
  colorlinks=true,
  citecolor=blue,
  linkcolor=blue
}
\newcommand{\bwb}{bluer-when-brighter\xspace}
\newcommand{\rwb}{redder-when-brighter\xspace}
\newcommand{\Bwb}{Bluer-when-brighter\xspace}

\begin{document}

\title{Longterm Optical Monitoring of Bright BL Lacertae Objects with ATOM: Spectral Variability and Multiwavelength Correlations}

\titlerunning{Optical Monitoring of BL Lac objects with ATOM}

\author{
  Alicja Wierzcholska\inst{1,2,3}, Micha{\l} Ostrowski\inst1, {\L}ukasz Stawarz\inst{4,1}, Stefan Wagner\inst3, Marcus Hauser\inst3.
}

\institute{
  \inst1Astronomical Observatory, Jagiellonian University, ul. Orla 171, 30-244 Krak\'{o}w, Poland\\
  \inst2Insitute of Nuclear Physics, Polish Academy of Science, ul. Radzikowskiego 152, 31-342 Krak\'{o}w, Poland\\
  \inst3Landessternwarte, Universit\"at Heidelberg, K\"onigstuhl, D 69117 Heidelberg, Germany \\
  \inst4Institute of Space and Astronautical Science JAXA, 3-1-1 Yoshinodai, Chuo-ku, Sagamihara, Kanagawa 252-5210, Japan \\
  \email{alicja.wierzcholska@ifj.edu.pl}
}

\authorrunning{A. Wierzcholska et al.}

\abstract
% context
{Blazars are the established sources of an intense and variable non-thermal radiation extending from radio wavelengths up to high and very high energy $\gamma$-rays. Understanding the spectral evolution of blazars in selected frequency ranges, as well as multi-frequency correlations in various types of blazar sources, is of a primary importance for constraining the blazar physics.}
% aims
{Here we present the results of a long-term optical monitoring of a sample of 30 blazars of the BL Lac type, most of which are the confirmed TeV emitters. We study the optical color--magnitude correlation patterns emerging in the analyzed sample, and compare the optical properties of the targets with the high-energy $\gamma$-ray and high-frequency radio data.}
% methods
{The optical observations were carried out in $R$ and $B$ filters using the Automatic Telescope for Optical Monitoring (ATOM) located at the site of the H.E.S.S. Array. Each object in the sample was observed during at least 20 nights in the period 2007--2012. }
% results
{We find significant global color--magnitude correlations (meaning \bwb spectral evolution) in 40\% of the sample. The sources which do not display any clear chromatism in the gathered entire datasets often do exhibit \bwb behavior but only in isolated shorter time intervals. We also discovered spectral state transitions at optical wavelengths in several analyzed sources.  Finally, we find that the radio, optical, and $\gamma$-ray luminosities of the sources in the sample obey almost linear correlations, which seem however induced, at least partly, by the redshift dependance, and may be also affected by non-simultaneousness of the analyzed multifrequency dataset.}
% conclusions
{We argue that the observed \bwb behavior is intrinsic to the jet emission regions, at least for some of the analyzed blazars, rather than resulting from the contamination of the measured flux by the starlight of host galaxies. We also conclude that the significance of color--magnitude scalings does not correlate with the optical color, but instead seems to depend on the source luminosity, in a sense that these are the lowest-luminosity BL Lac objects which display the strongest correlations.}


\maketitle

\section{Introduction}
\label{intro}

Blazars  are radio-loud Active Galactic Nuclei (AGN) with relativistic jets pointing at small angles to the line of sight \citep[e.g.][]{begelman84}. The Doppler-boosted non-thermal emission of these sources is observed in a wide range of frequencies of the electromagnetic spectrum, from radio to X-rays and, in the case of the brightest objects, up to high and very high energy $\gamma$-rays \citep[e.g.,][]{2LAC}\footnote{see also \texttt{http://tevcat.uchicago.edu}}. Rapid variability of blazars is routinely detected at different wavelengths on timescales down to hours or even minutes, albeit with often drastically different amplitudes \citep[e.g.,][]{wagner,2155flare,sasada08,saito}. Based on the optical classification, blazars can be divided into two subclasses: Flat Spectrum Radio Quasars (FSRQs) and BL Lacertae-type objects (BL Lac objects). Optical/UV spectra of FSRQs are characterized by the presence of prominent broad and narrow emission lines; BL Lac objects are instead dominated by a continuum 
emission in the 
optical band \citep{urry}. BL Lac objects can be subdivided further into high-, intermediate-, and low-energy peaked sources (HBLs, IBLs, and LBLs, respectively), depending on the position of their synchrotron peak frequencies \citep[see, e.g.,][]{padovani95,fossati98,abdo10}.

The broad-band spectral energy distribution (SED, meaning the $\nu - \nu F_{\nu}$ spectral representation) of blazars has a characteristic double-hump structure. The first hump extends from radio up to the optical or X-ray range, and is due to the synchrotron emission of \emph{in-situ} accelerated jet electrons; the second hump is located in the $\gamma$-ray part of the spectrum, and is most widely believed to be due to the inverse-Compton emission of the same electron population \citep[e.g.,][]{konigl,marscher,dermer,sikora}. The most recent multiwavelength campaigns targeting the brightest or dramatically flaring blazars have challenged to some extent the simplest (homogeneous one-zone) versions of this standard leptonic model, but truly simultaneous, broad-band, and long-term monitoring data for a representative sample of the blazar population, which are of a primary importance for constraining the blazar physics, are still relatively sparse.

Previous observational studies of blazars (BL Lac objects, in particular) at optical frequencies revealed complex flux--color correlation patterns. For example, the 10-year-long photometric monitoring of optically bright OJ~287 and BL~Lacertae presented by \cite{Carini1992} did not indicate any universal and persistent relation between spectral and flux changes, although some hints for a general trend that the sources appear \bwb were noted. Interestingly, \citeauthor{Carini1992} detected a microvariability in both blazars, at the level of 0.08~mag$/$hr for OJ~287 and 0.01~mag$/$hr for BL~Lacertae. In a different dataset, a strong correlation between $V-R$ color and $R$ magnitude has been found in 11 nights of the optical outburst of BL~Lacertae in 1997, with the correlation coefficient $\simeq 0.7$ \citep{Clements}. The analysis of the longterm monitoring of the same object by \cite{Villata2002} and \cite{Villata2004} confirmed that \bwb relation is hardly present in long-timescale variations but in short 
isolated 
outbursts. An analogous color evolution of OJ~287 during the flaring state in years  2005$-$2006 was reported by \cite{Dai}, who found the \bwb chromatism for the source with the correlation coefficient $\simeq 0.67$.

Similar results regarding color-magnitude (hereafter CM) correlations --- namely, clear \bwb behavior during the rapid flares and basically achromatic longterm variability --- were also reported for the another exceptionally bright and active (at optical frequencies) blazar S5~0716+714 \citep{dai13,ghisellini97,Gu}. We note that this source is known for its persistent variability down to the timescales of hours and minutes, both in total and also polarized optical fluxes \citep{sasada08,dai13,bhatta13}. 

Besides the three aforementioned well-known BL Lac objects, the positive CM correlations were also detected in, e.g., 3C~66A \citep{Ghosh}, AO~0235+164 \citep{Raiteri2001}, and PKS~0735+178 \citep{Gu}. More recently, \cite{Ikejiri2011} presented a rich dataset gathered with the KANAT telescope from 2008 till 2010 for a sample of 42 blazars including both FSRQs and BL Lac objects. The authors established that majority of the studied sources ($88\%$ of the sample) do exhibit a universal \bwb trend in the optical band, with a few exceptions only, mostly FSRQs, which reveal the opposite \rwb behavior \citep[see in this context also][for the results of the SMARTS optical and infrared blazar monitoring program]{bonning12}. 

In addition to the CM correlation patterns, the optical polarization variability of bright blazars is now being widely studied \citep[e.g.,][]{sasada08,Ikejiri2011,gaur14}. The successful operation of the Large Area Telescope \citep[LAT;][]{LAT} onboard the \emph{Fermi} satellite enabled, on the other hand, the first in-depth investigations of the optical/$\gamma$-ray correlations for a large number of blazar sources detected on a daily basis at GeV energies. The extensive optical monitoring of the selected BL Lac objects revealed in particular significant flux-flux correlations during the flaring states (e.g., \citealt{raiteri13} for the GASP-WEBT campaign on BL~Lacertae; see also \citealt{chatterjee12}), as well as the overall optical/$\gamma$-ray luminosity correlations for the LAT-detected objects \citep{hovatta14}.

In this paper, we present the data gathered with the Automatic Telescope for Optical Monitoring \citep[ATOM;][]{hauser} during the last six years (2007--2012) for a sample of 30 blazars of the BL Lac type. The targets were selected in the southern hemisphere using various criteria as promising TeV candidates or already established TeV emitters. We study the $B-R$ color versus $R$ magnitude correlations, along with the radio, optical, and GeV luminosity-luminosity or luminosity-spectral scalings and dependencies. We emphasize that the analyzed multi-wavelength dataset is not simultaneous, the discussed blazar sample is not complete, and the source light-curves are sampled with ATOM non-uniformly. Still, several general trends and correlation patterns are revealed, allowing for a novel insight into the variability properties of bright BL Lac objects.

\section{Data Analysis and Results} 

\subsection{ATOM Observations and Target Selection}
\label{data}

ATOM is the $75$-cm optical telescope \citep[see][]{hauser}, which is operating in Namibia since 2006 at the site of the High Energy Stereoscopic System \citep[H.E.S.S.; see][]{crab}. The telescope works in a fully robotic way in four filters in the Johnson-Cousins $UBVRI$ broad-band photometric system defined by \cite{Bessell90}: $B$ (440\,nm), $V$ (550\,nm), $R$ (640\,nm), and $I$ (790\,nm). The main scientific goal of ATOM observations is an automatic monitoring of $\gamma$-ray sources in the southern hemisphere, and in particular of the potential H.E.S.S. targets. The durations of single pointings analyzed in this paper are between 100\,s and 1000\,s; this enables us to limit the uncertainty of each single observation to below 0.1\,mag. The photometric flux scale was calibrated using the reference stars. In addition to the standard automatic analysis of the collected data, some of the raw images were also checked manually, and a limited number of bad-quality pointings (e.g., due to bad weather conditions 
or a moon light contamination) were rejected from the final dataset. This process of discarding some of the observations does not affect the final results presented below. The observed magnitudes have been corrected against the Galactic extinction based on the model by \cite{Schlegel98} with the most recent recalibration by \cite{Schlafly}.\footnote{See the corresponding $A_B$ and  $A_R$  extinction coefficients collected in Table~\ref{table_1}.}

The analyzed blazar sample includes 30 sources of the BL Lac type (mostly HBLs), all monitored frequently with the ATOM telescope, majority of which (21) are the established TeV emitters. In particular, we require each target in the sample to be observed for at least 20 --- but not necessarily consecutive --- nights, \emph{simultaneously} in $B$ and $R$ filters, in a time span from 2007 till 2012. In the case when a given source was observed more than once during the same night, the average value of the pointings was used. A list of thus selected objects is given in Table \ref{table_1}, and the particular dates of the ATOM observations are given in Table~\ref{table_2}.

\begin{figure*}[]
\centering{\includegraphics[width=0.43\textwidth]{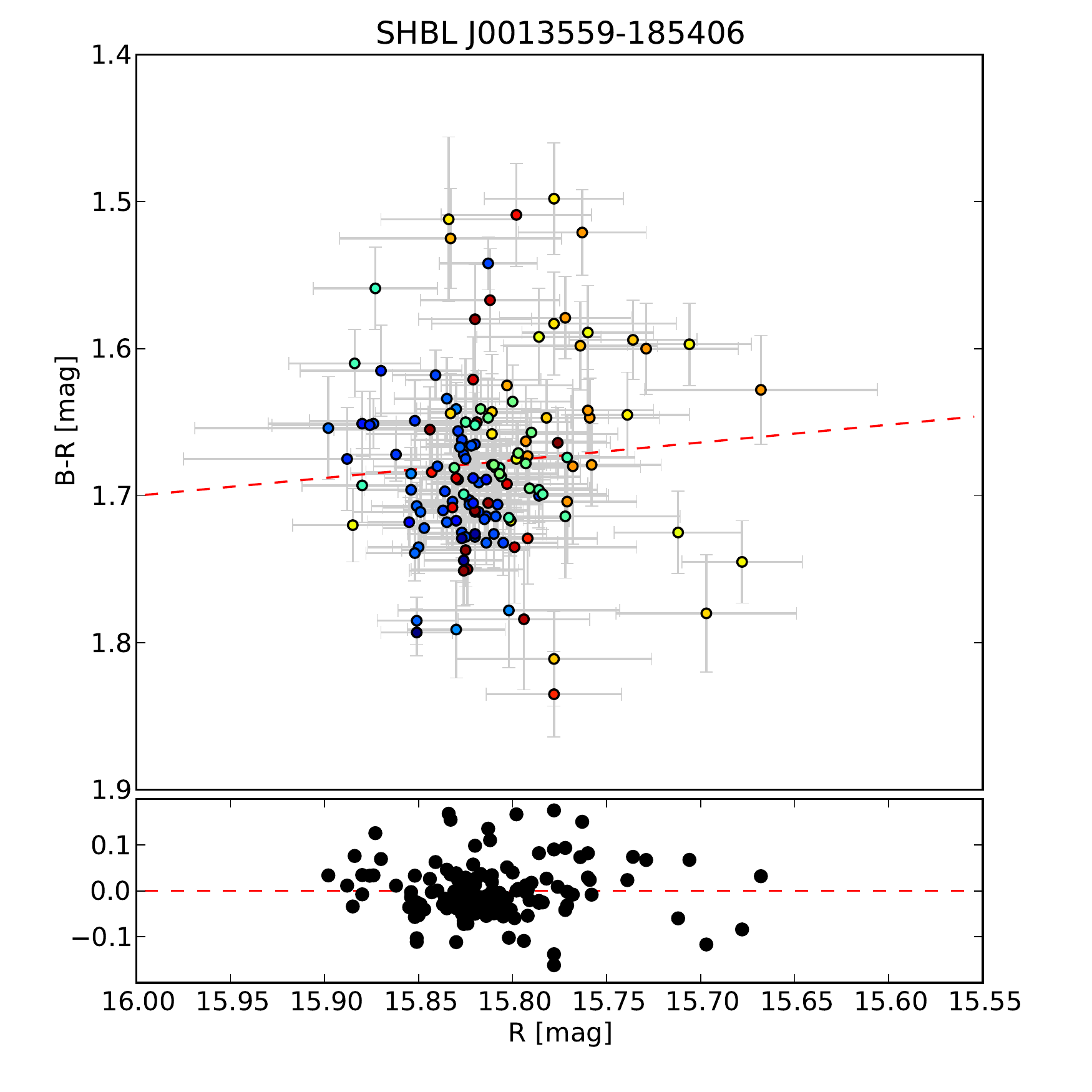}}
\centering{\includegraphics[width=0.43\textwidth]{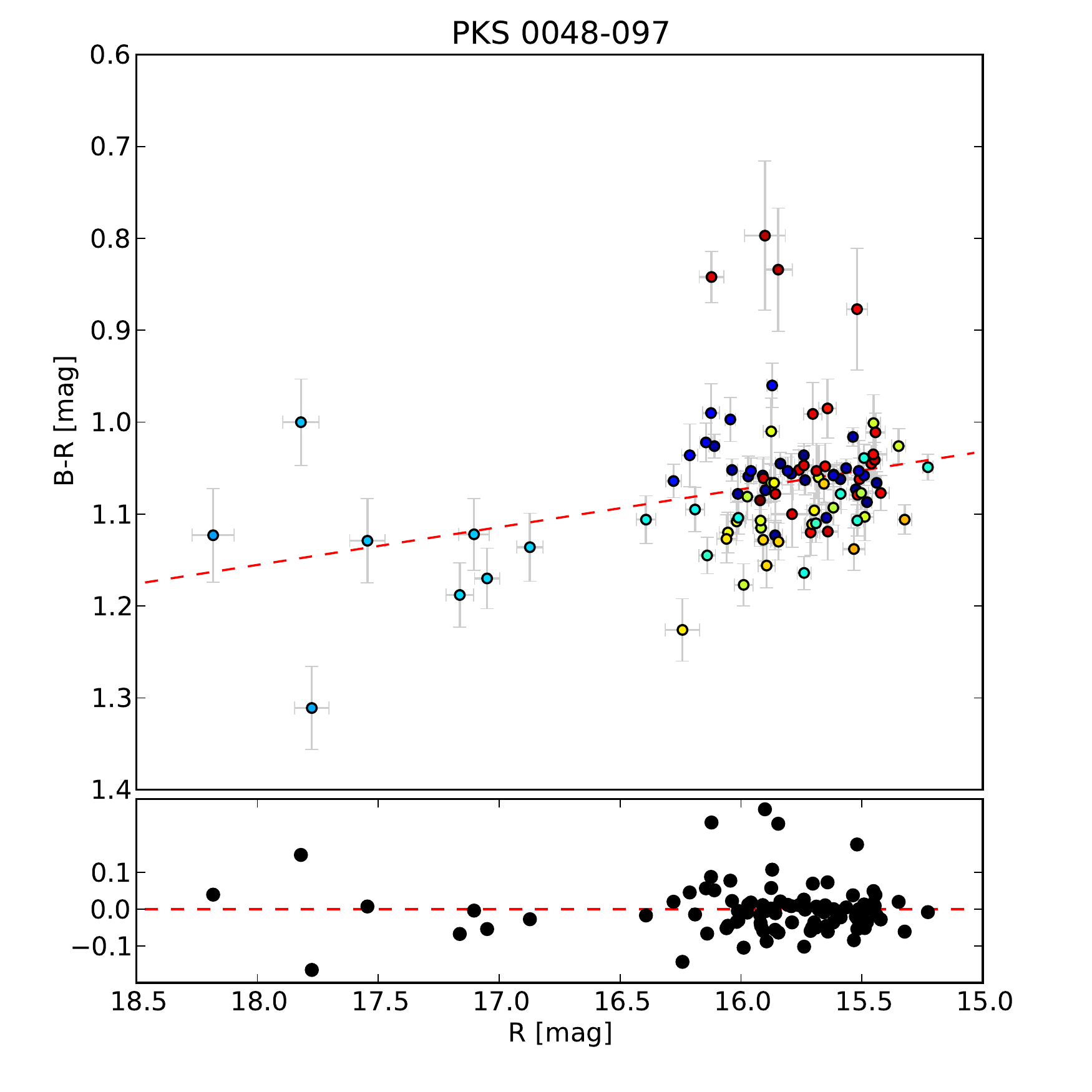}}\\
\centering{\includegraphics[width=0.43\textwidth]{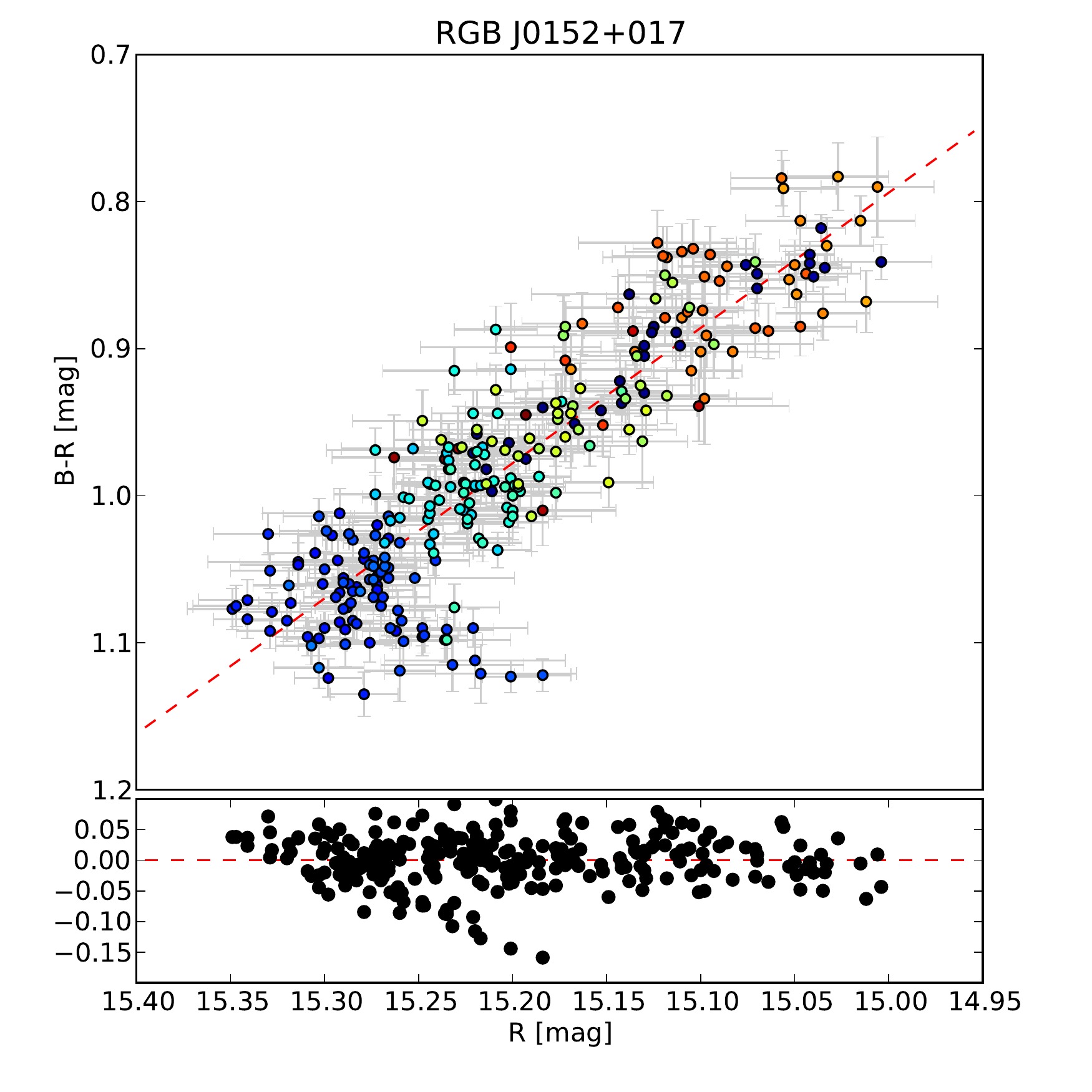}}
\centering{\includegraphics[width=0.43\textwidth]{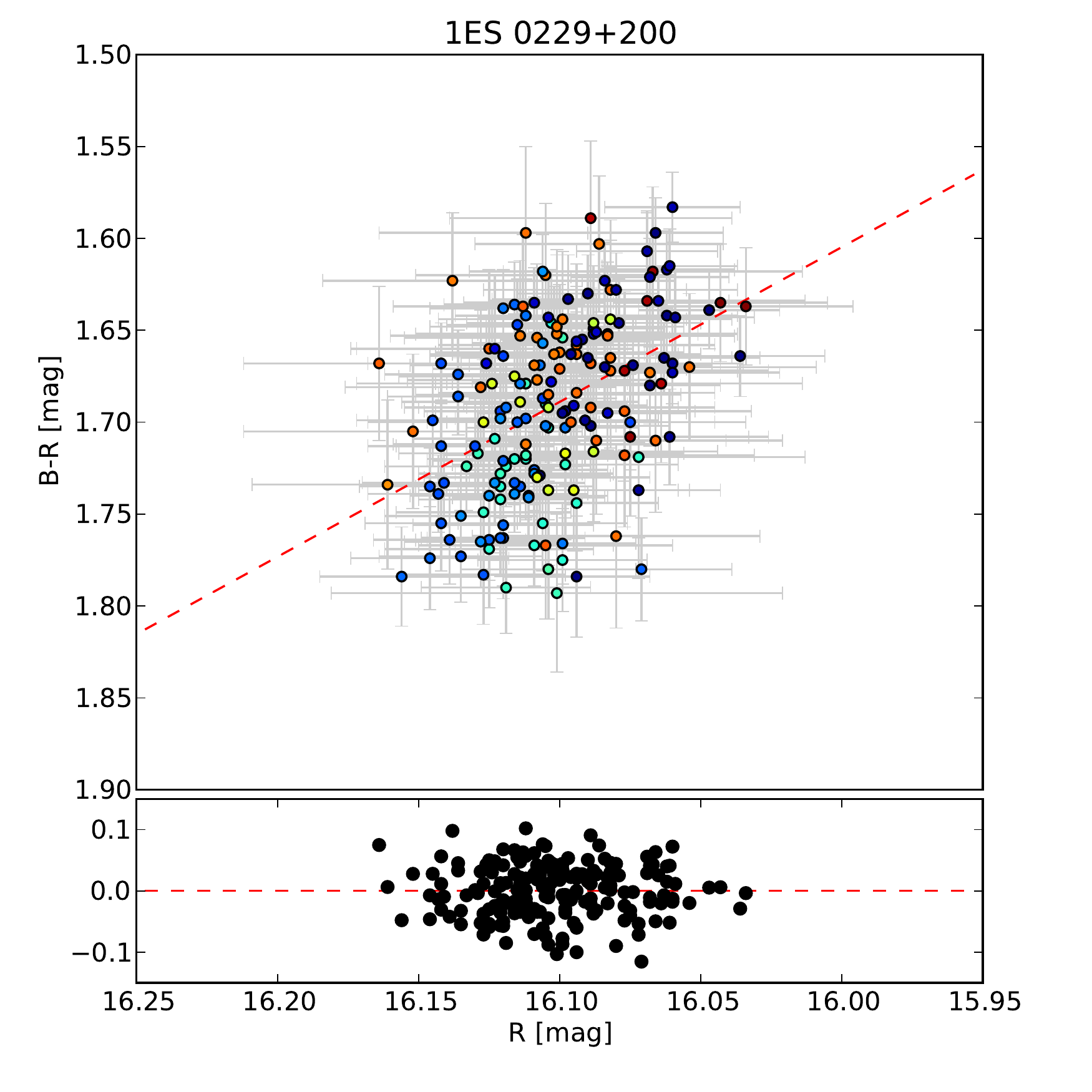}}\\
\centering{\includegraphics[width=0.43\textwidth]{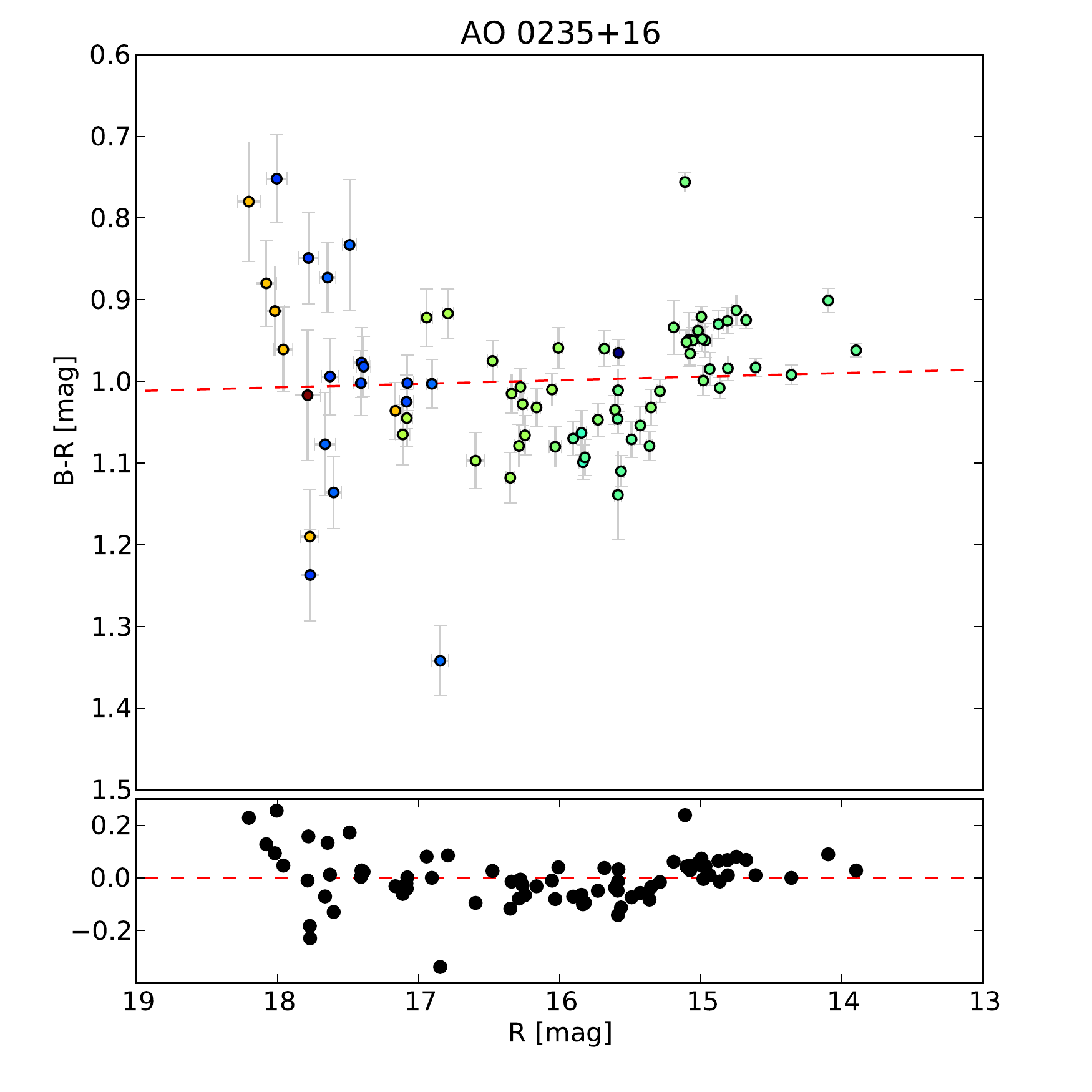}}
\centering{\includegraphics[width=0.43\textwidth]{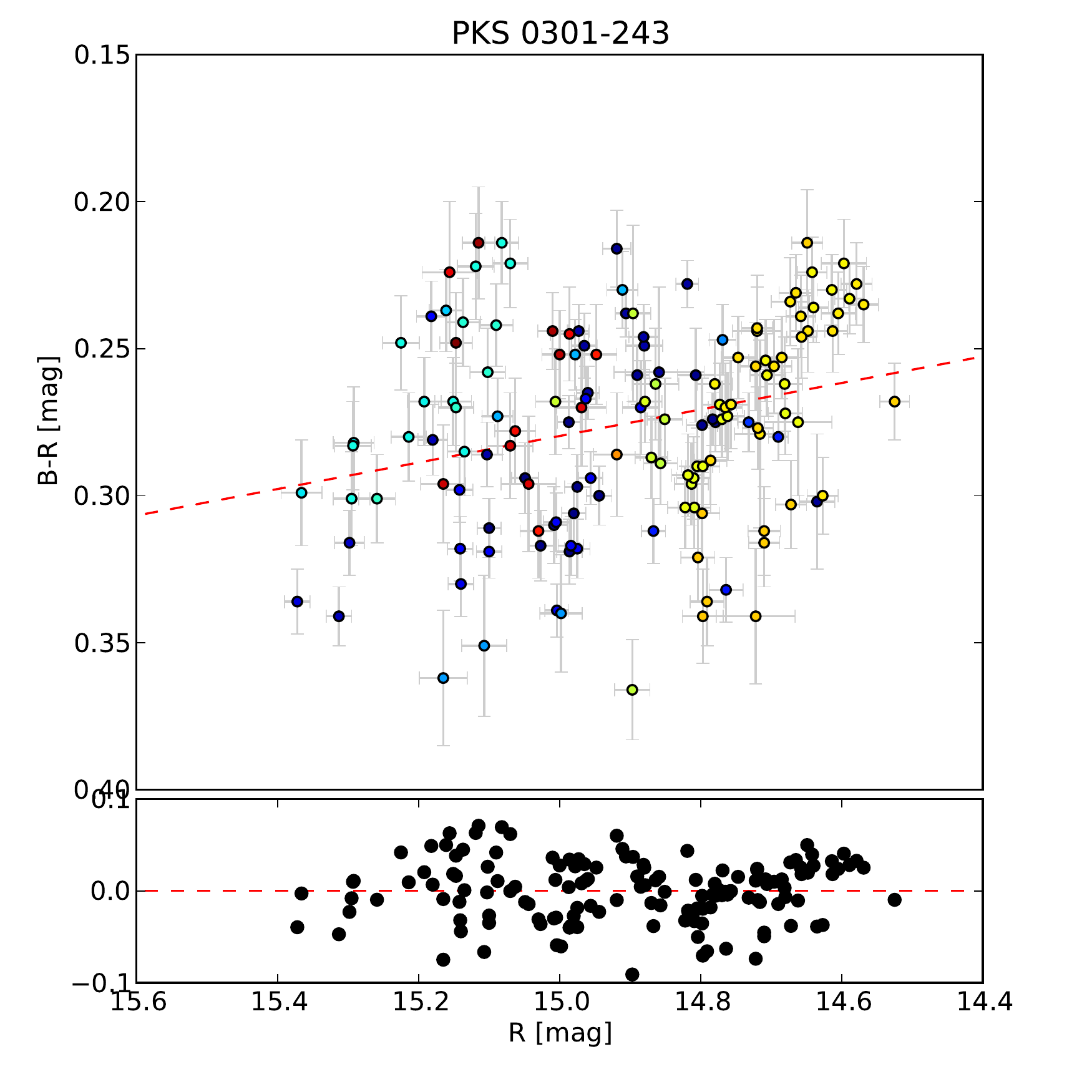}}\\
\caption[]{$B-R$ color vs. $R$-band magnitude diagrams for SHBL~J001355.9--18540, PKS~0048-097, RGB~J0152+017, 1ES~0229+200, AO~0235+16, and PKS~0301--243.  At each panel red dashed lines denote the fitted CM linear correlations (see Table \ref{table_1}). Color of a given data point indicates the time of a given measurement: the earliest pointings are denoted by dark blue symbols and the most recent ones depicted in red, with the rainbow color scale normalized to the entire span of the ATOM observations of a given blazar.}
\label{fig1}
\end{figure*}
 
\begin{figure*}[]
\centering{\includegraphics[width=0.43\textwidth]{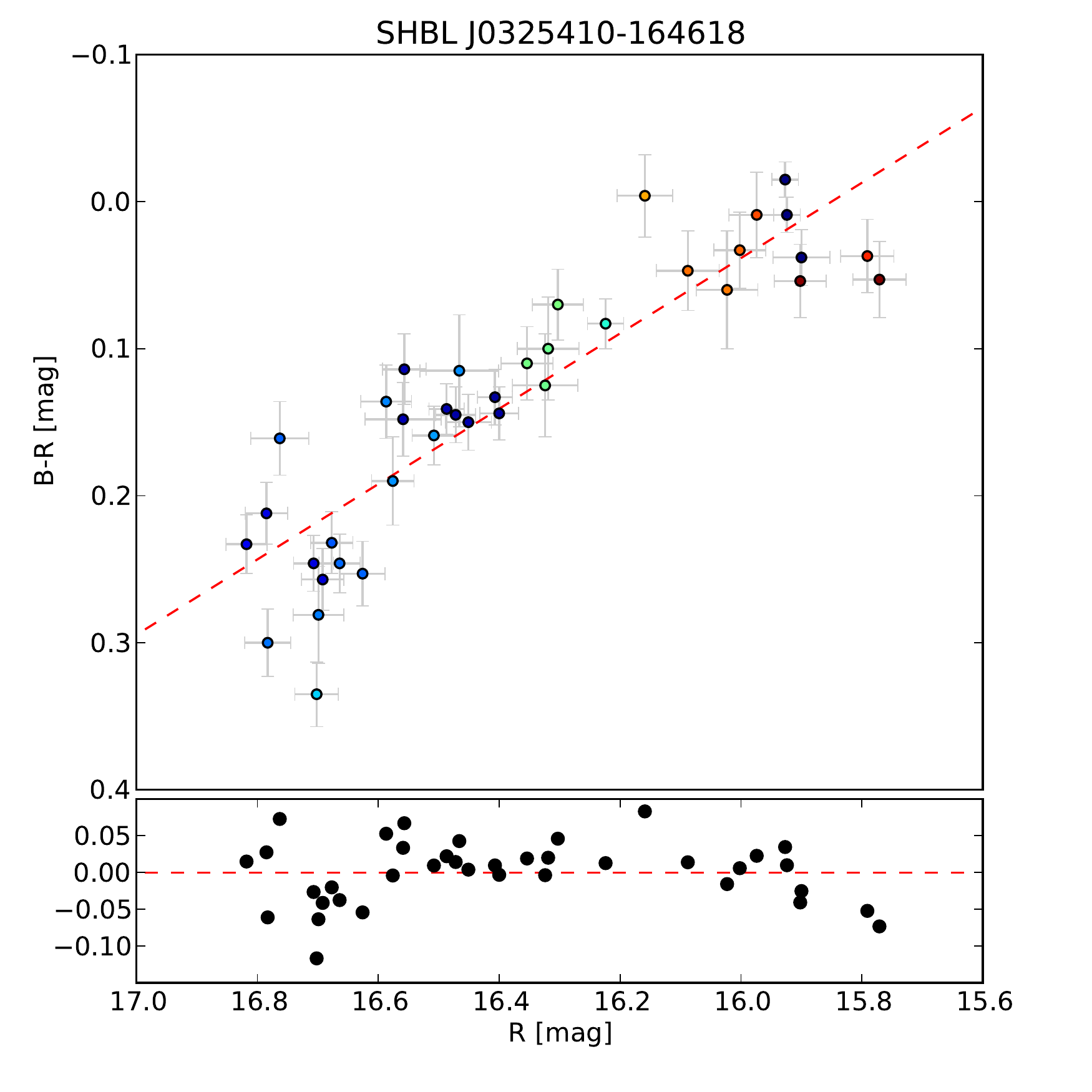}}
\centering{\includegraphics[width=0.43\textwidth]{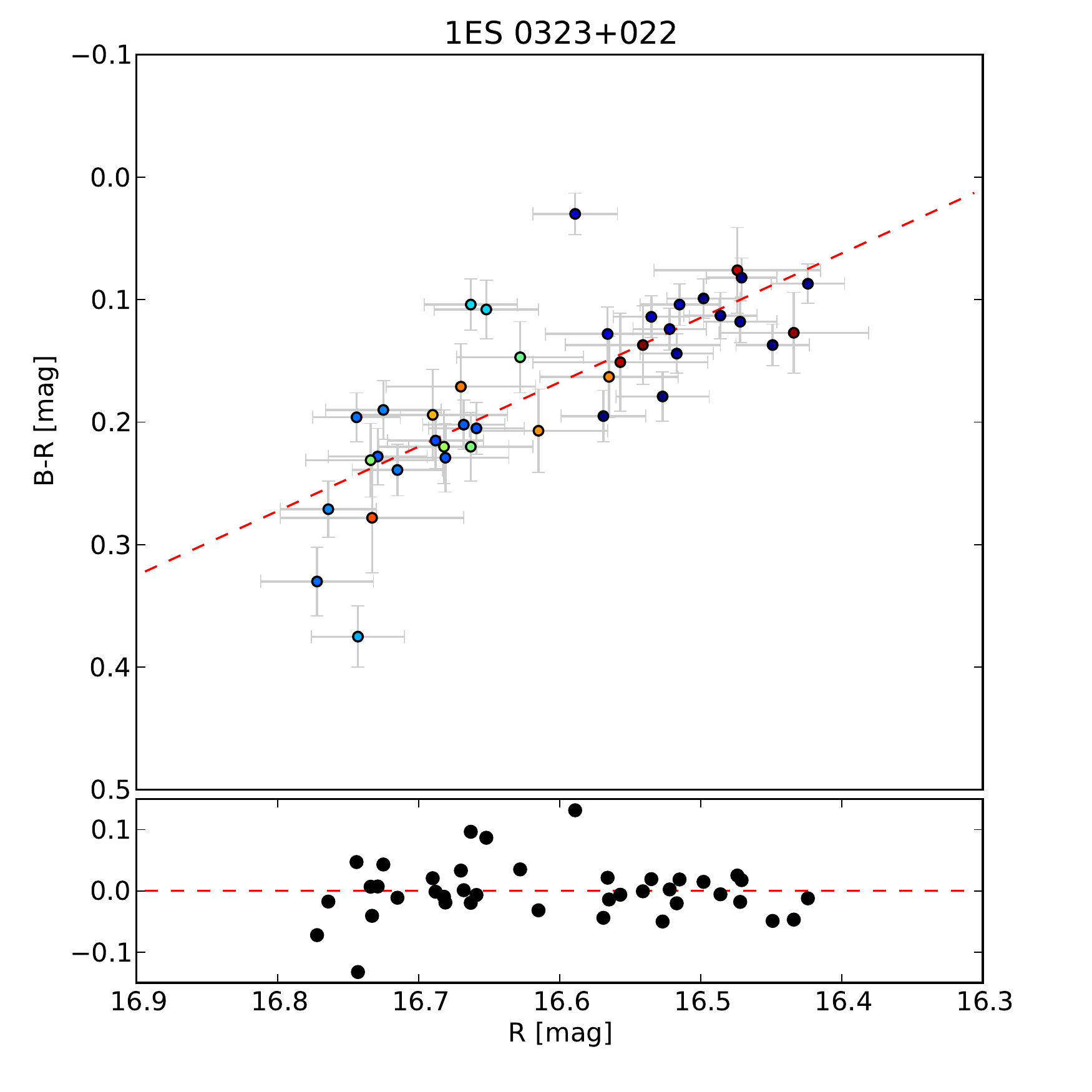}}\\
\centering{\includegraphics[width=0.43\textwidth]{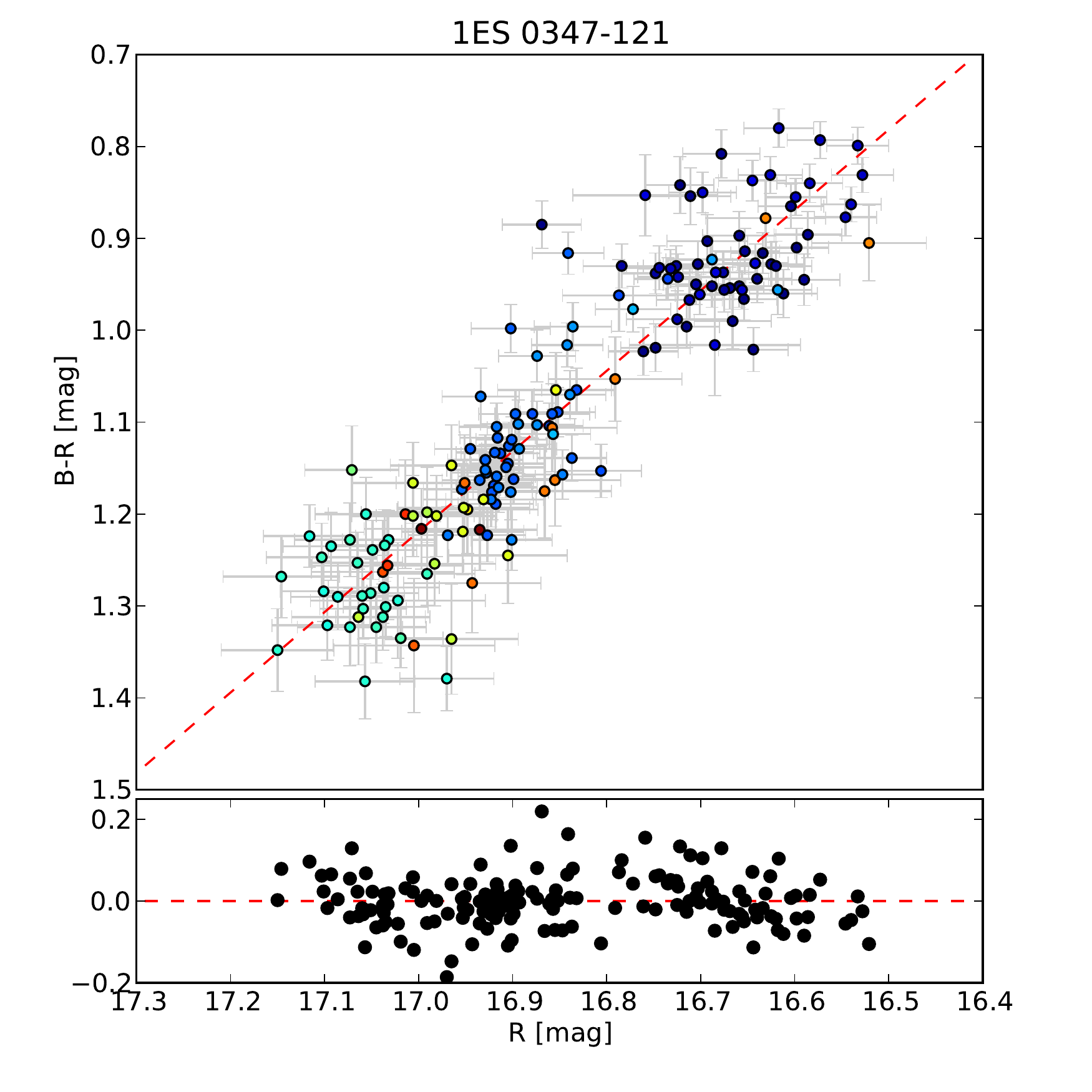}}
\centering{\includegraphics[width=0.43\textwidth]{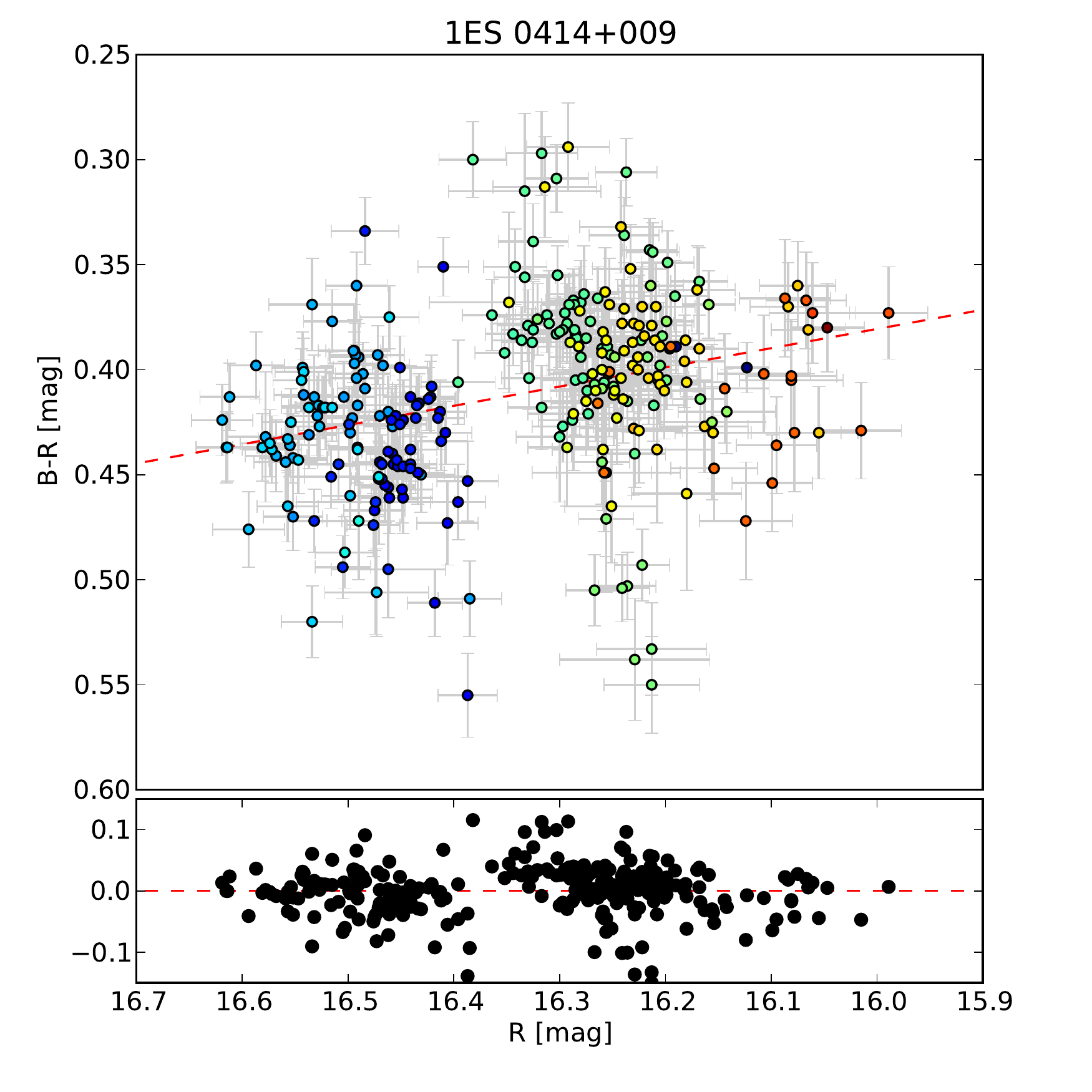}}\\
\centering{\includegraphics[width=0.43\textwidth]{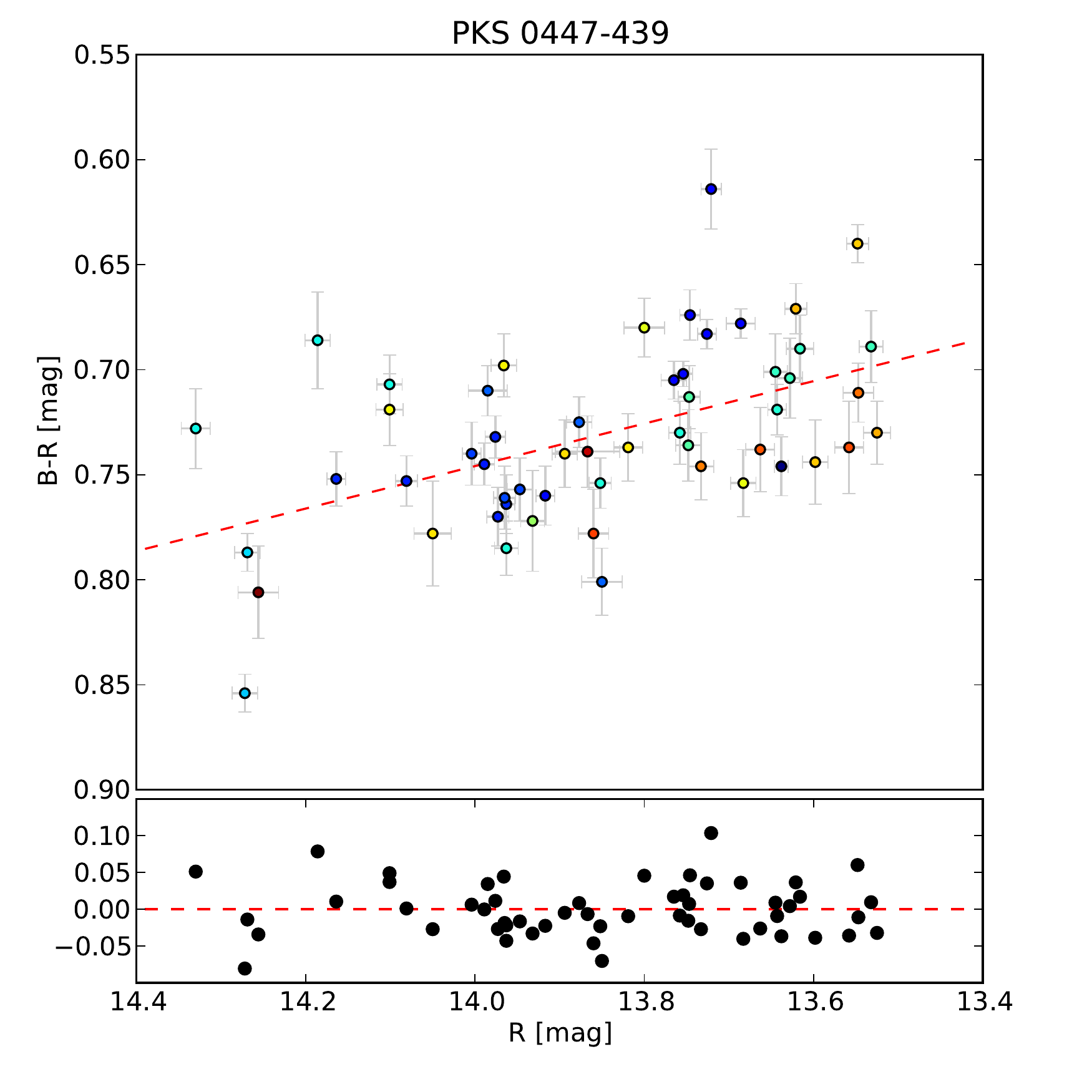}}
\centering{\includegraphics[width=0.43\textwidth]{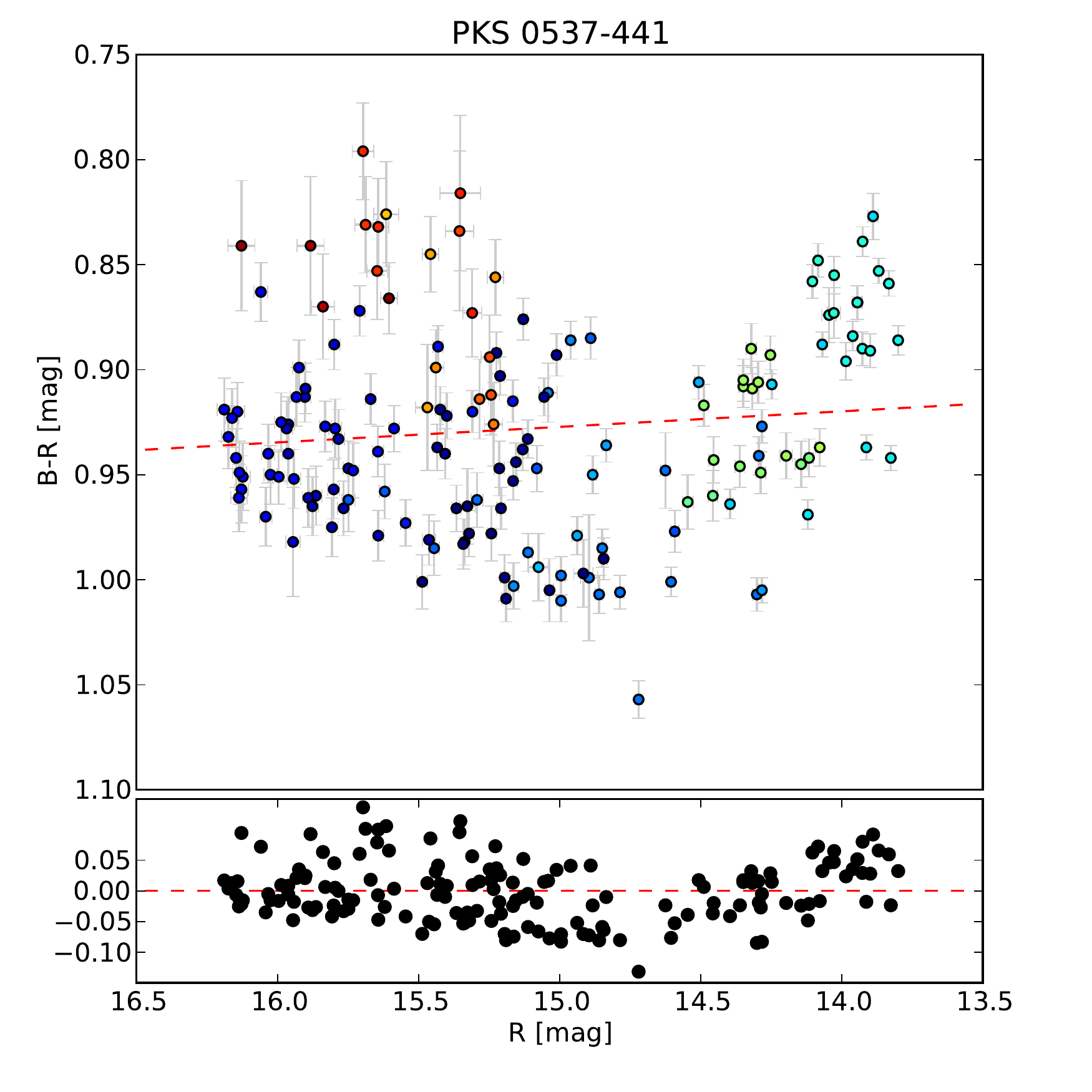}}\\
\caption[]{Same as Figure~\ref{fig1} for SHBL~J032541.0--164618, 1ES~0323+022, 1ES~0347--121,1ES~0414+00.9, PKS~0447--439, and PKS~0537--441.}
\label{fig2}
\end{figure*}

\begin{figure*}[]
\centering{\includegraphics[width=0.43\textwidth]{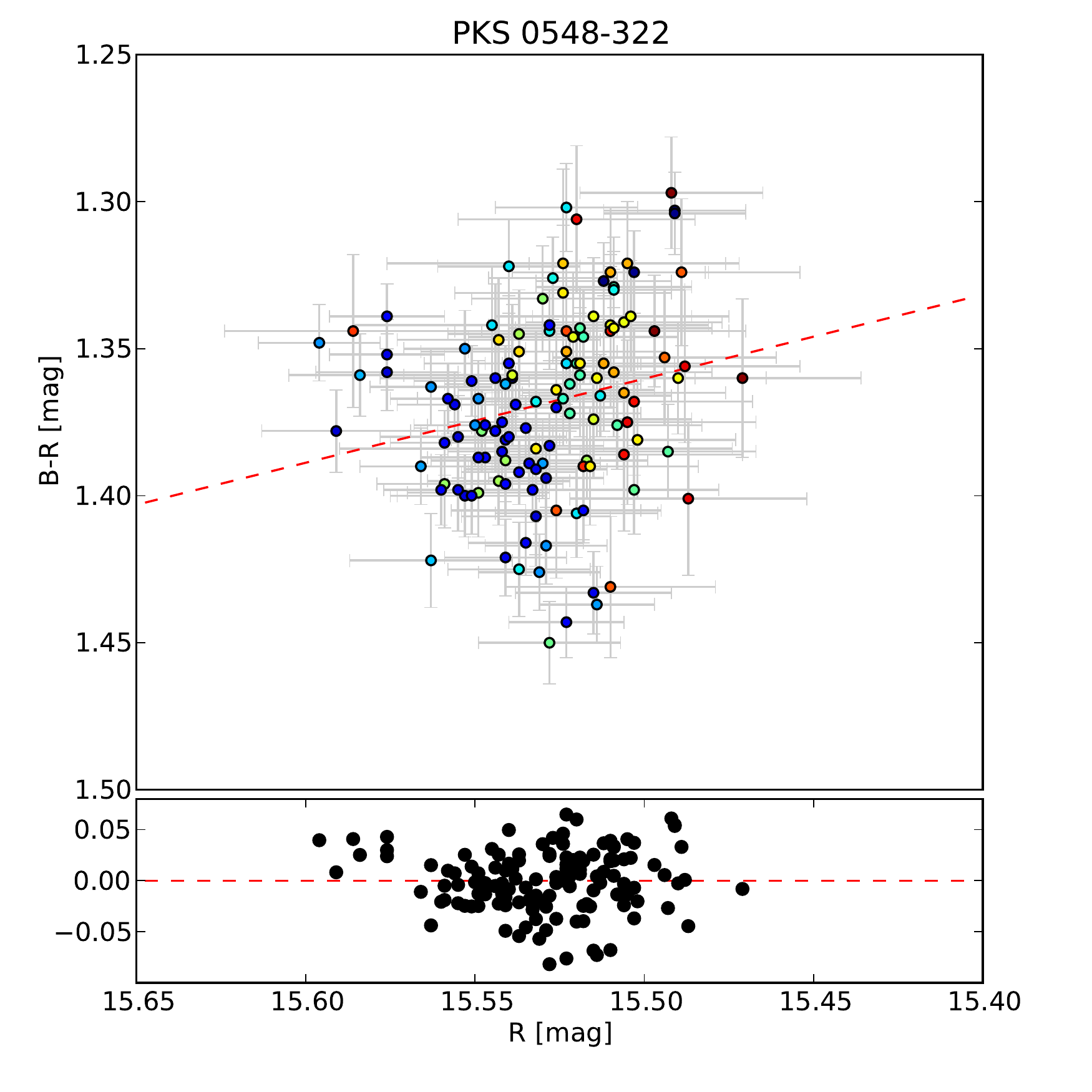}}
\centering{\includegraphics[width=0.43\textwidth]{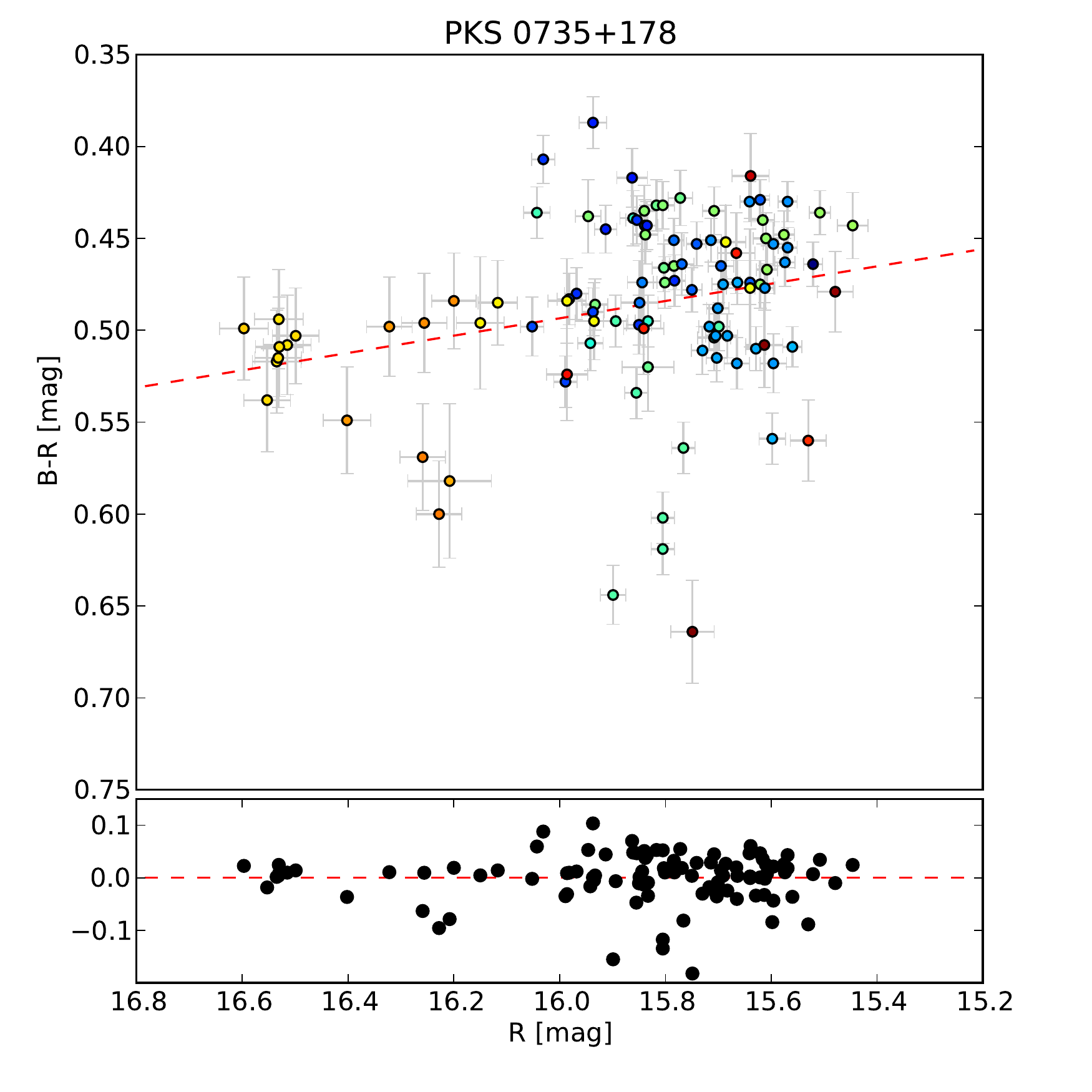}}\\
\centering{\includegraphics[width=0.43\textwidth]{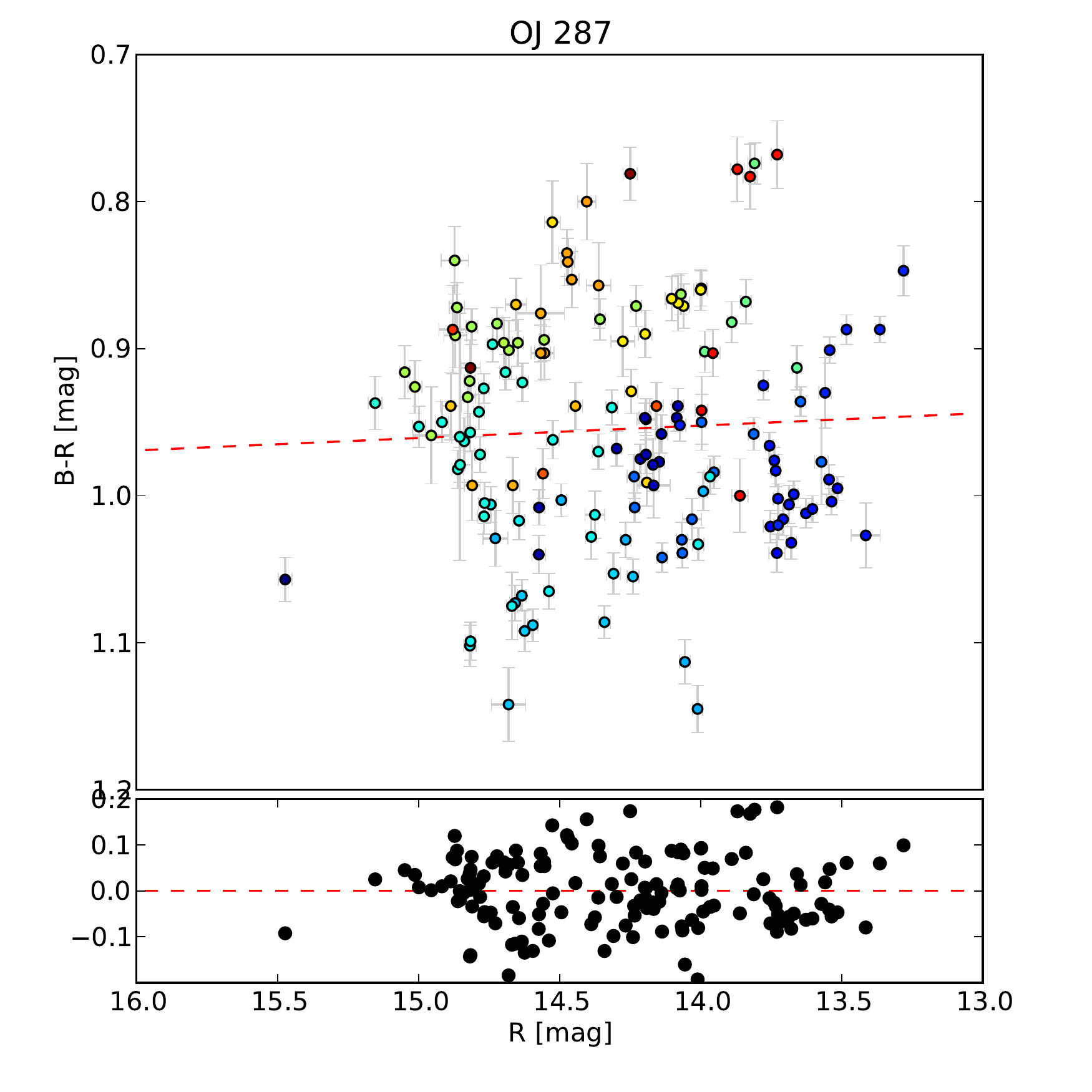}}
\centering{\includegraphics[width=0.43\textwidth]{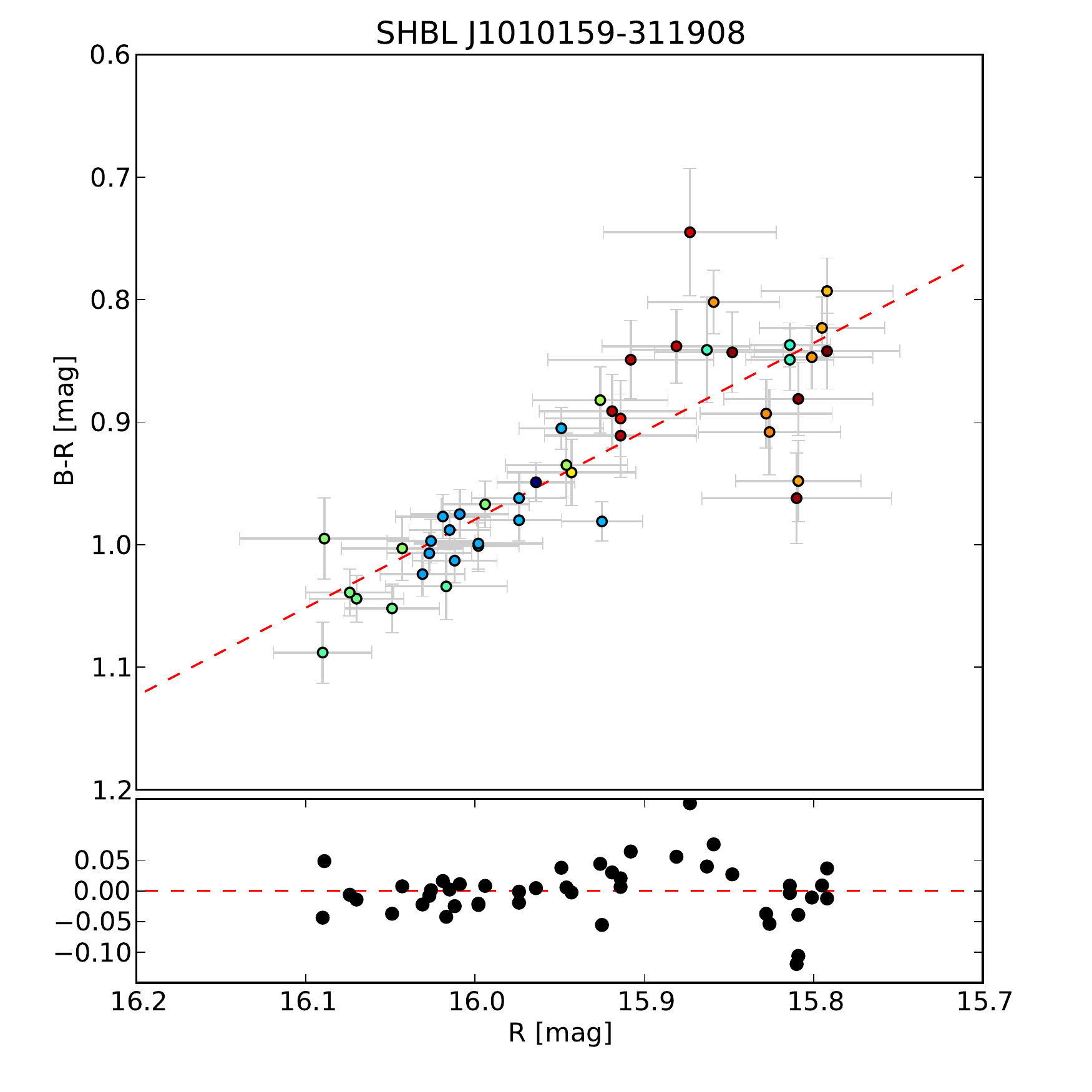}}\\
\centering{\includegraphics[width=0.43\textwidth]{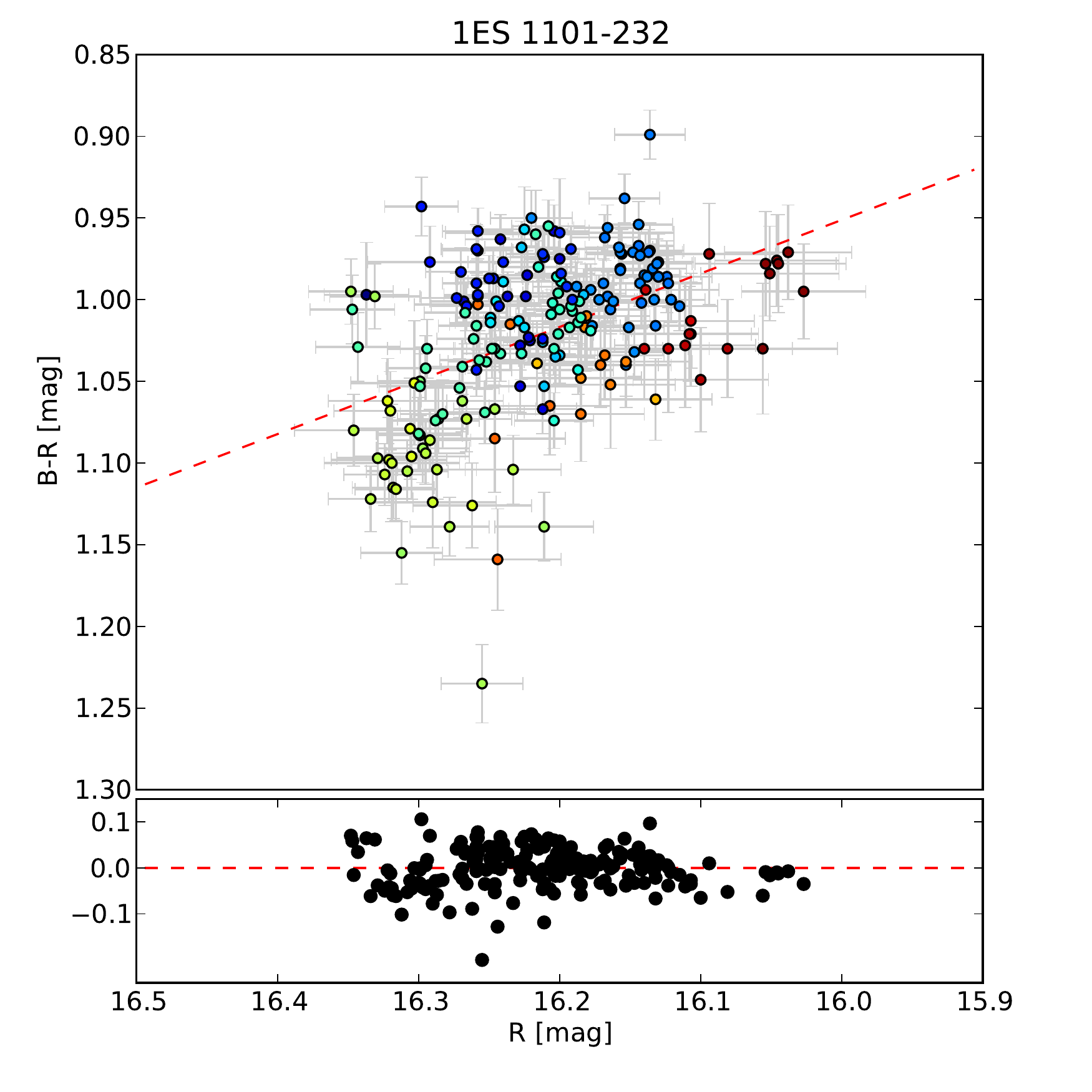}}
\centering{\includegraphics[width=0.43\textwidth]{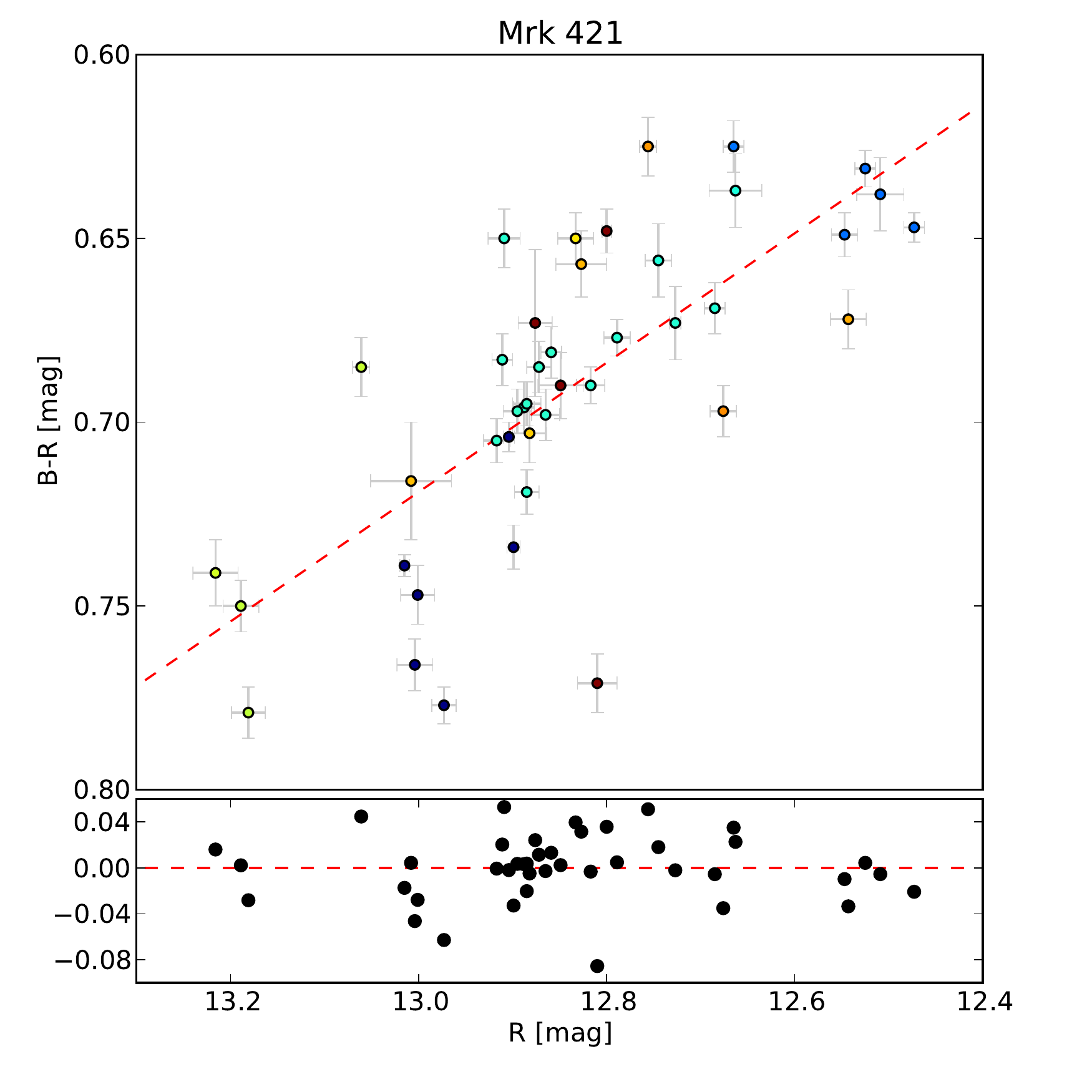}}\\
\caption[]{Same as Figure~\ref{fig1} for PKS~0548--322, PKS~0735+178, OJ~287, SHBL~J101015.9--311908, 1ES~1101--232, and Markarian~421.}
\label{fig3}
\end{figure*}
 
\begin{figure*}[]
\centering{\includegraphics[width=0.43\textwidth]{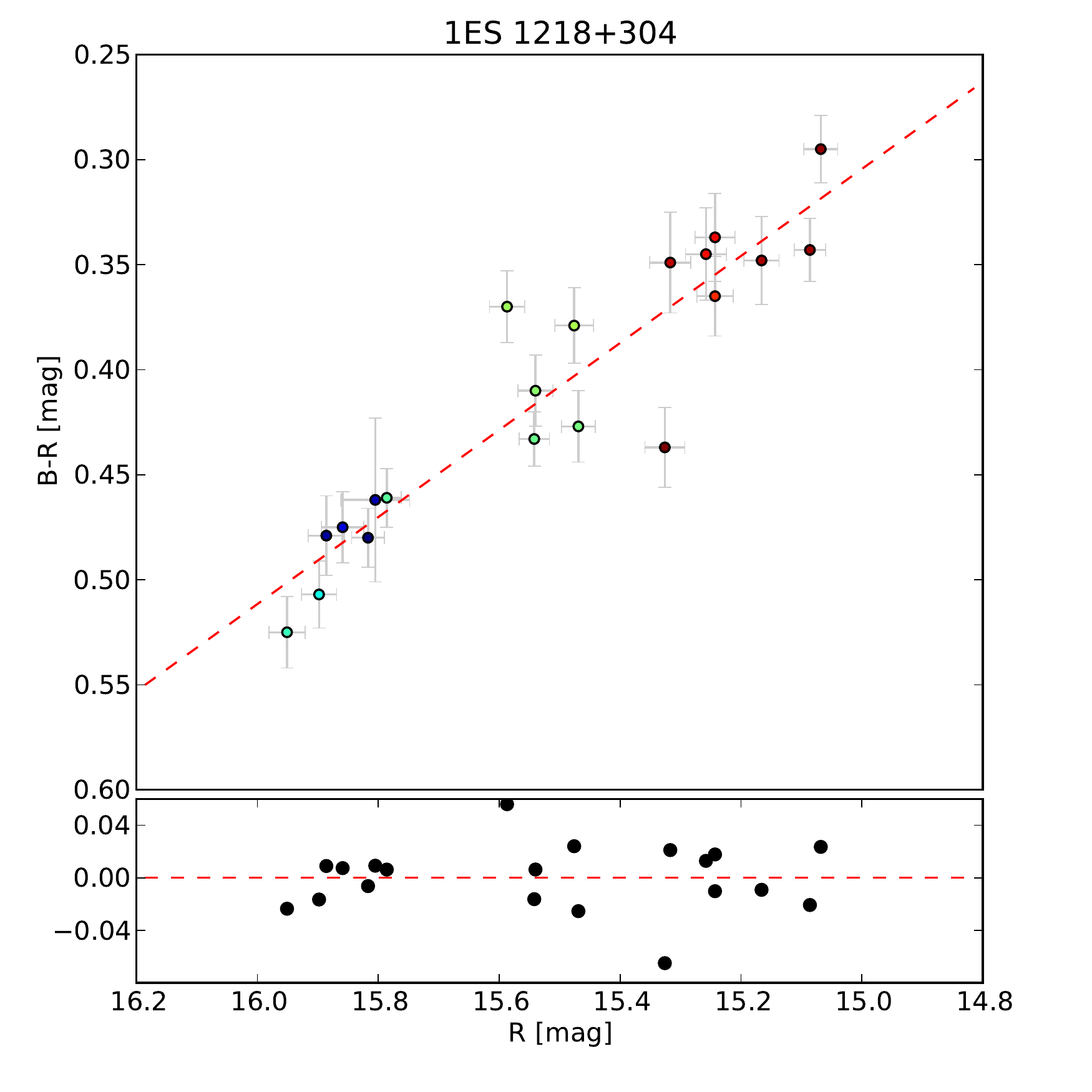}}
\centering{\includegraphics[width=0.43\textwidth]{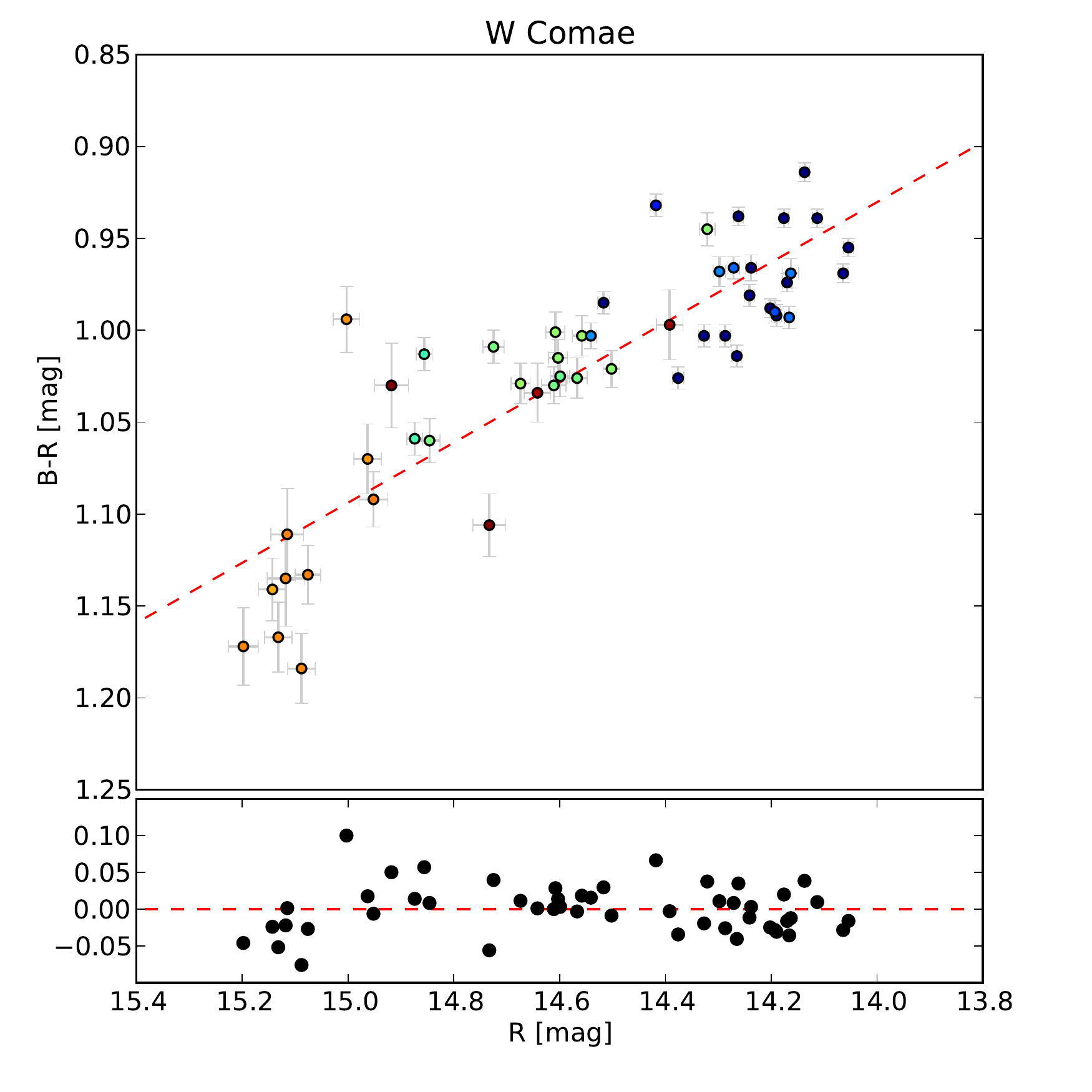}}\\
\centering{\includegraphics[width=0.43\textwidth]{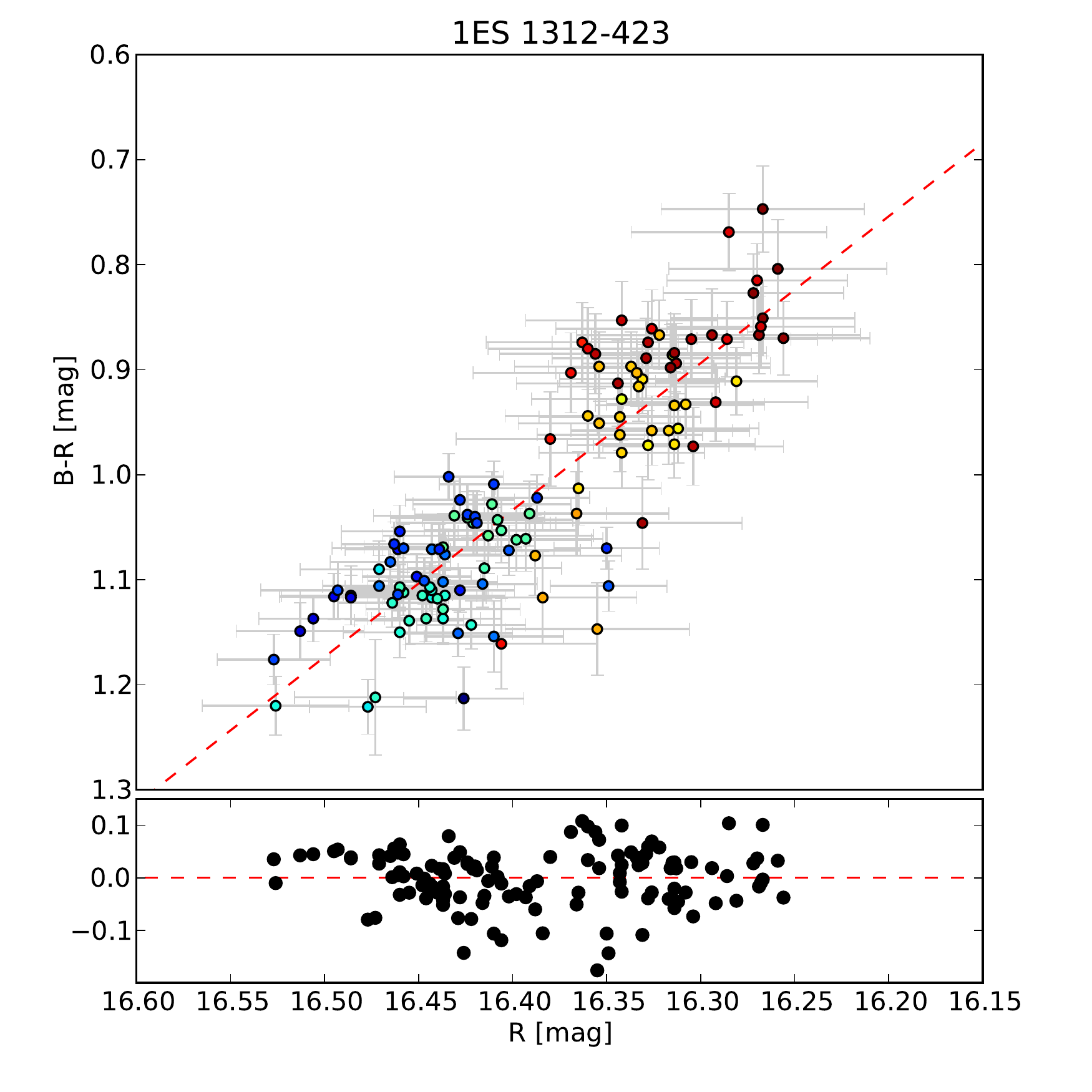}}
\centering{\includegraphics[width=0.43\textwidth]{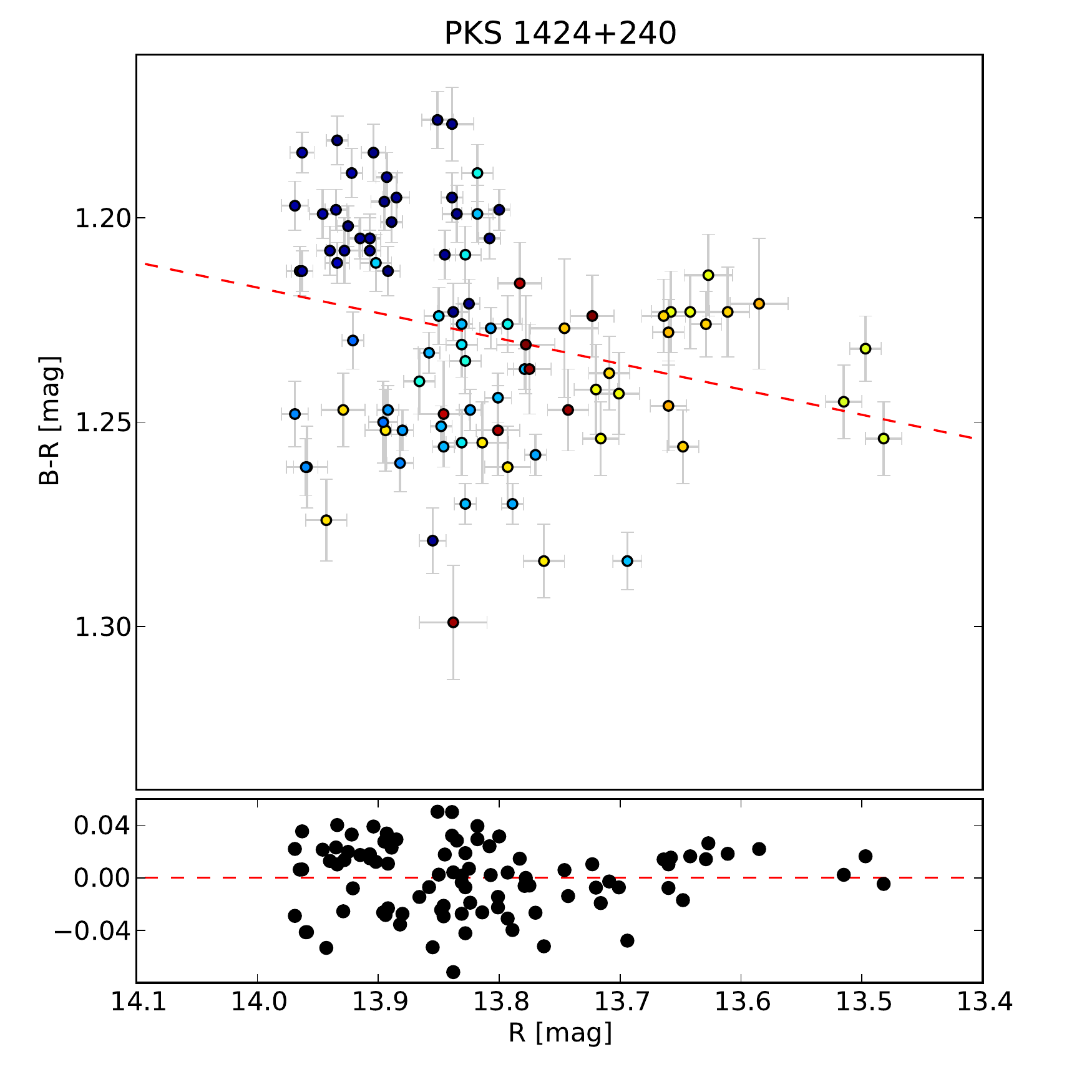}}\\
\centering{\includegraphics[width=0.43\textwidth]{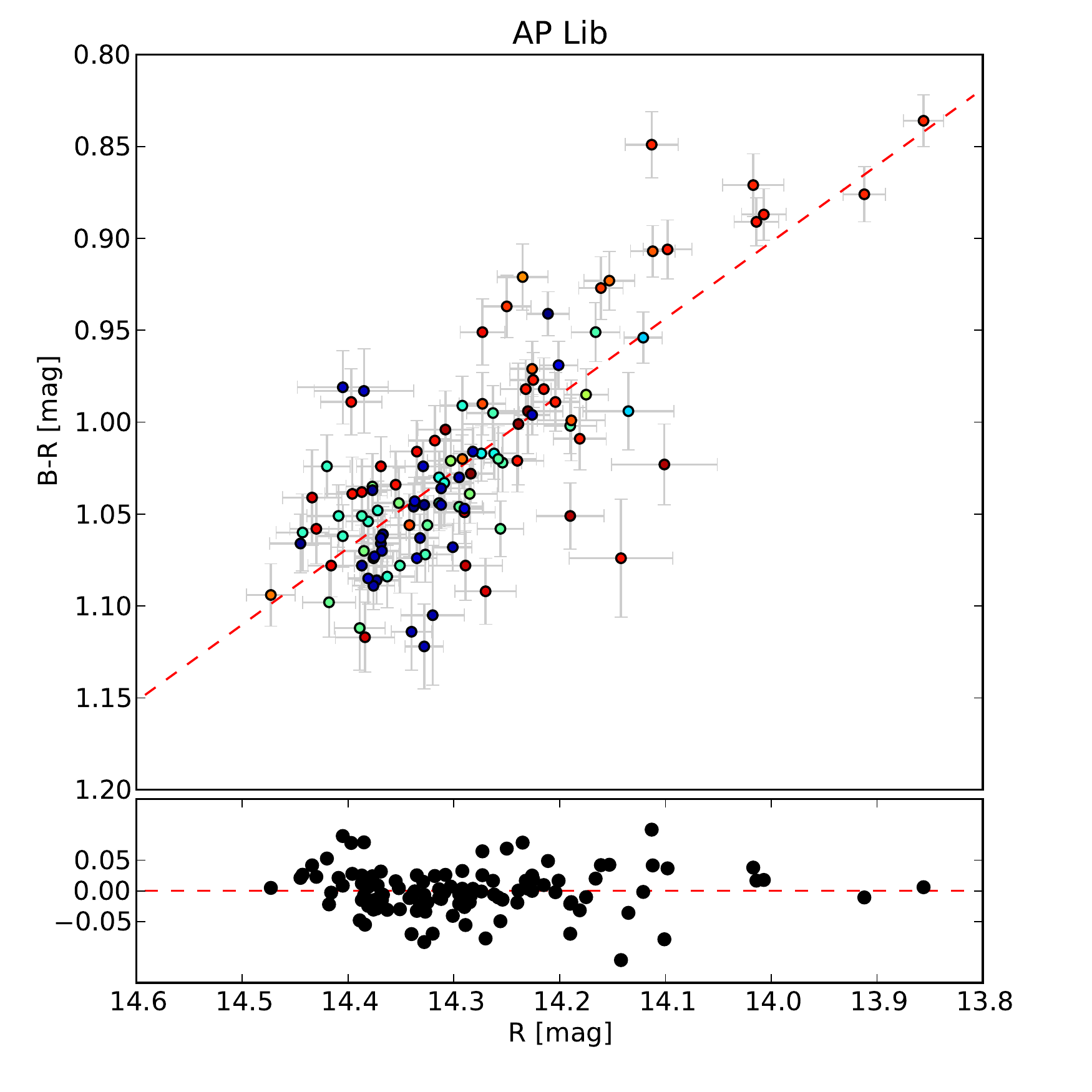}}
\centering{\includegraphics[width=0.43\textwidth]{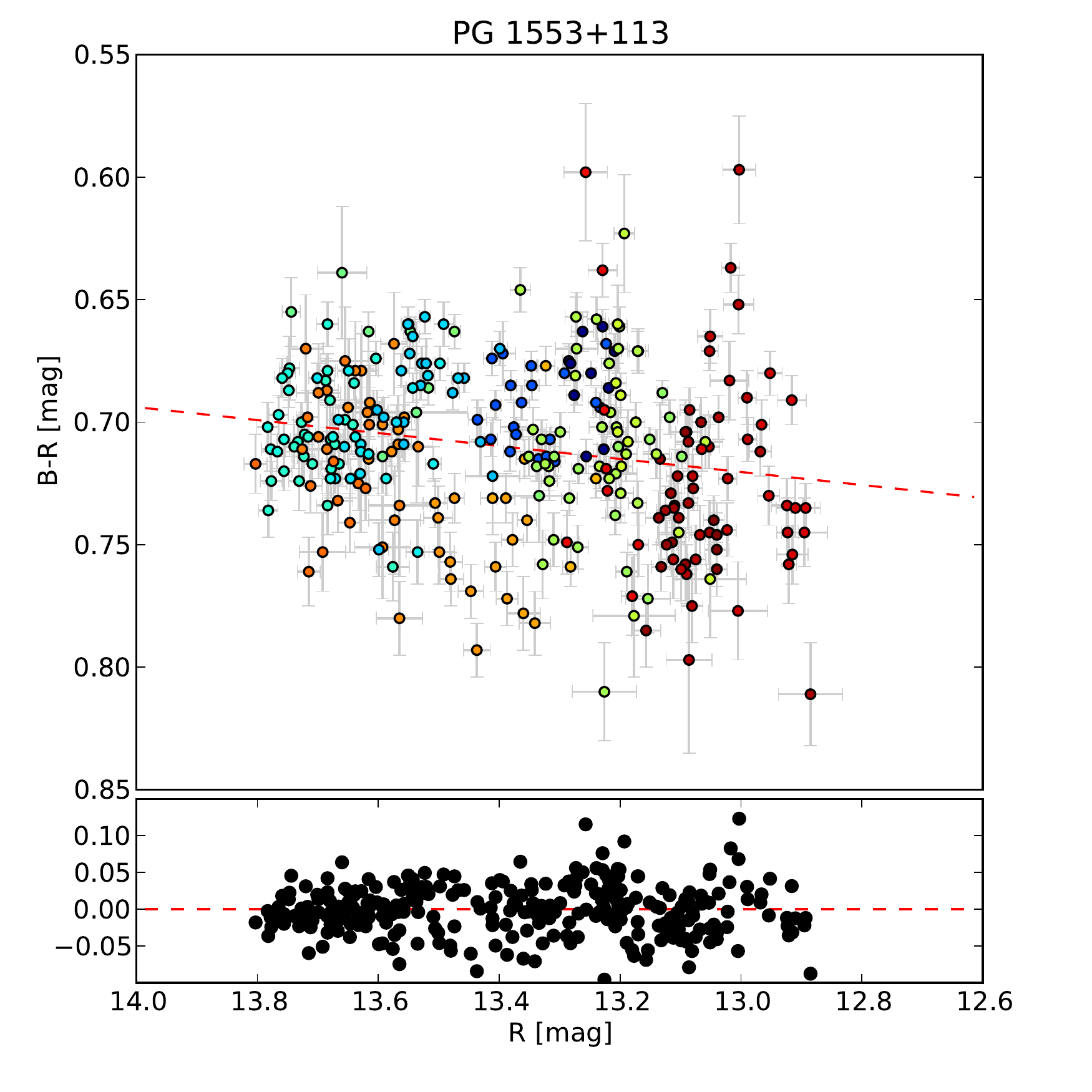}}\\
\caption[]{Same as Figure~\ref{fig1} for 1ES~1218+304, W~Comae, 1ES~1312--423, PKS~1424+240, AP~Librae, and PG~1553+113.}
\label{fig4}
\end{figure*}
  
\begin{figure*}[]
\centering{\includegraphics[width=0.43\textwidth]{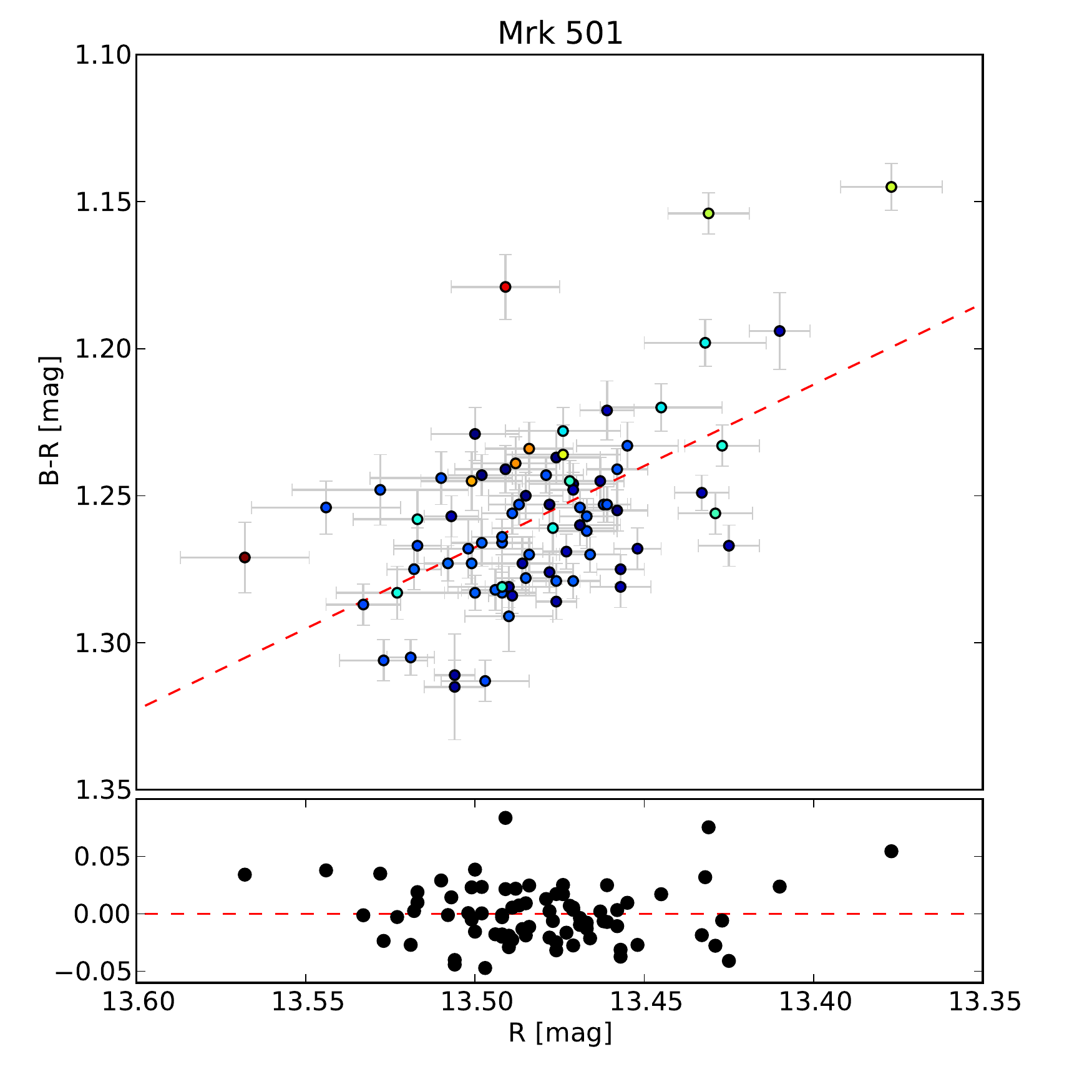}}
\centering{\includegraphics[width=0.43\textwidth]{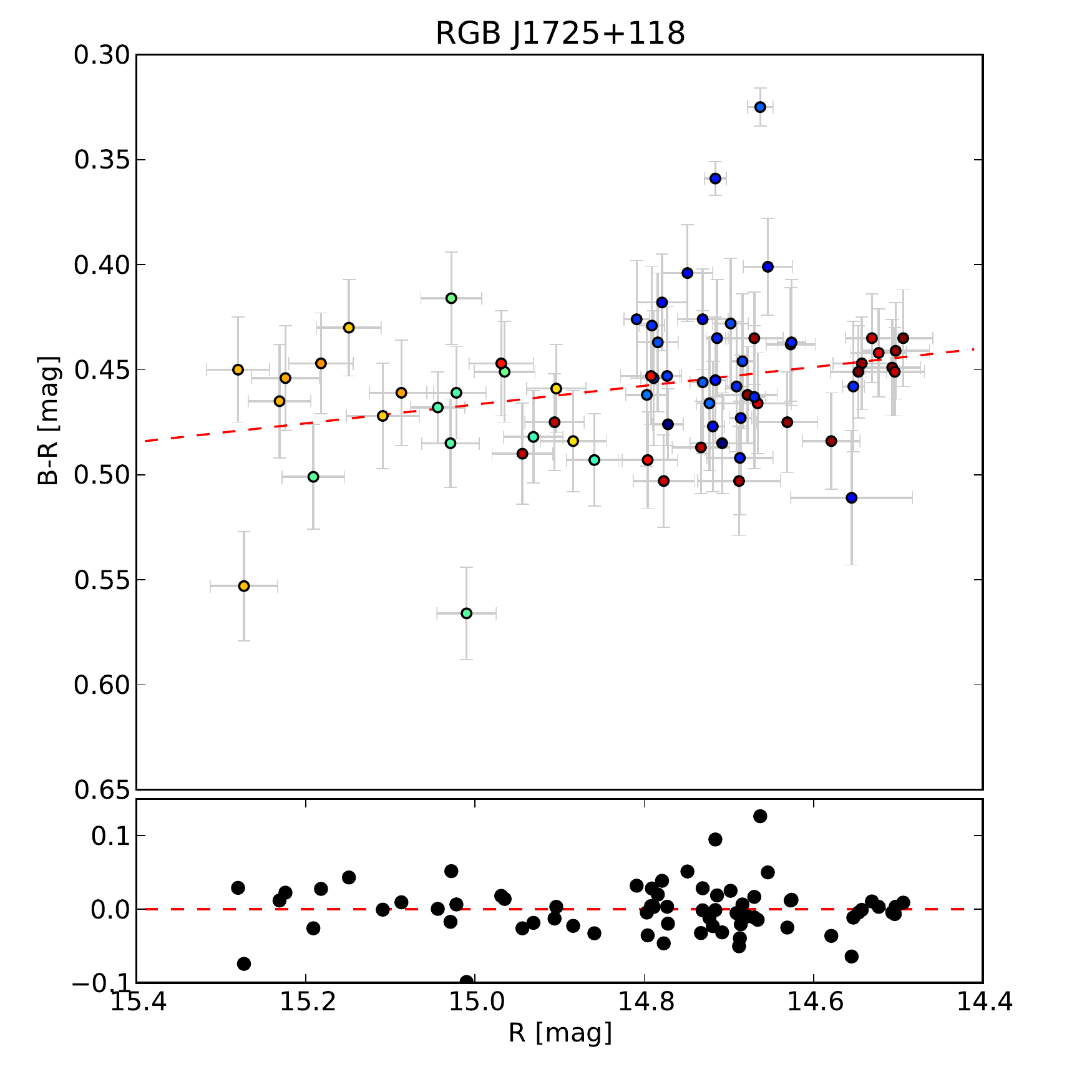}}\\
\centering{\includegraphics[width=0.43\textwidth]{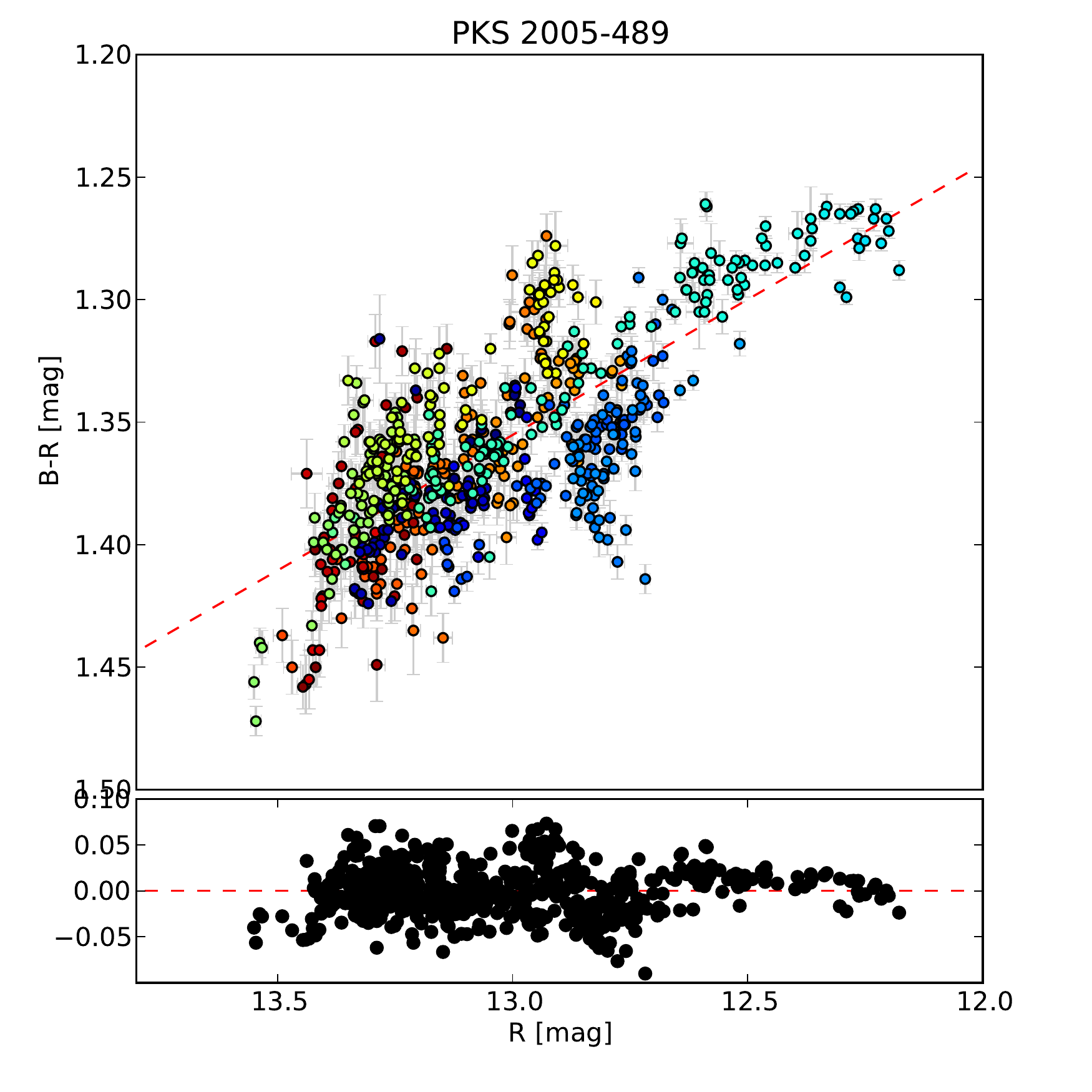}}
\centering{\includegraphics[width=0.43\textwidth]{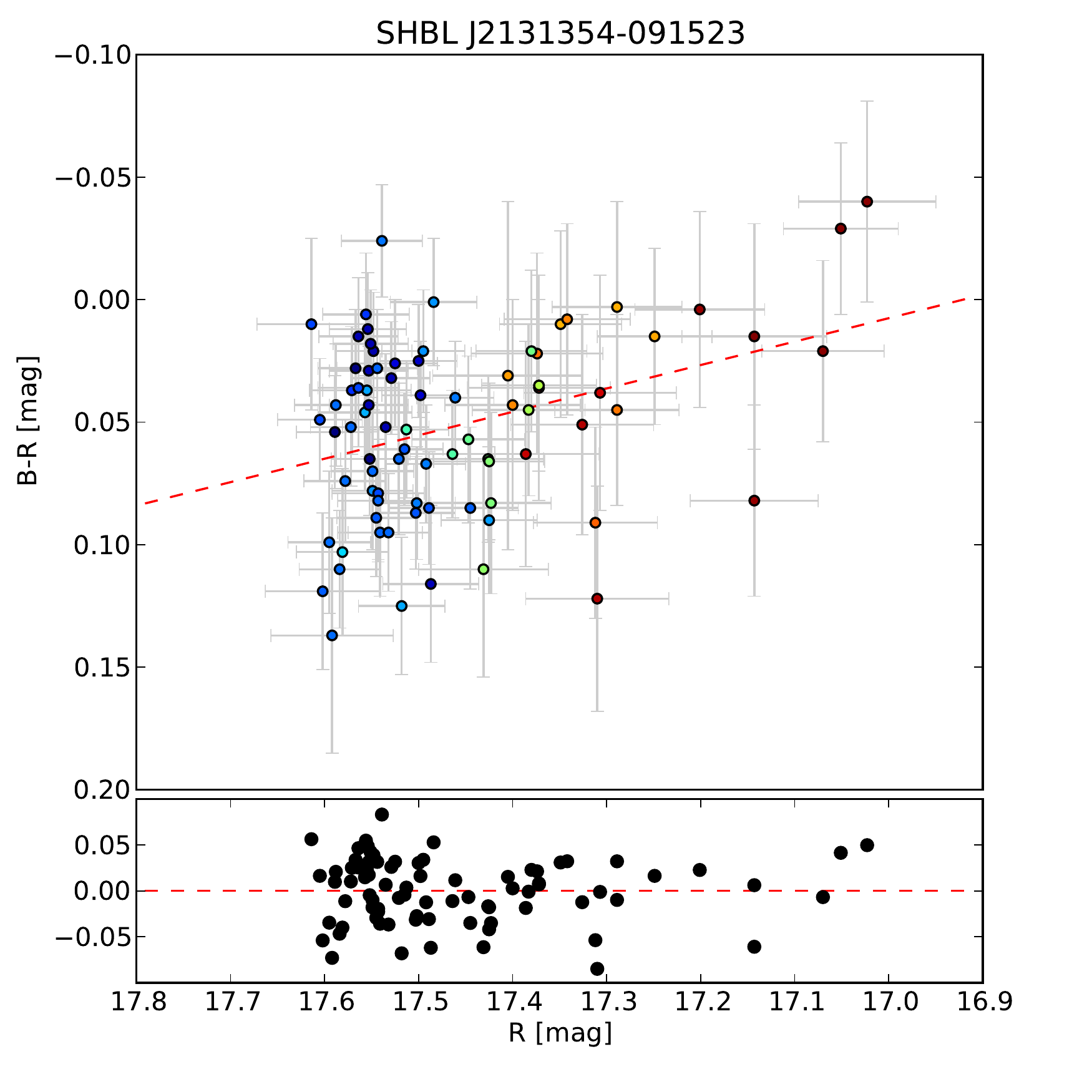}}\\
\centering{\includegraphics[width=0.43\textwidth]{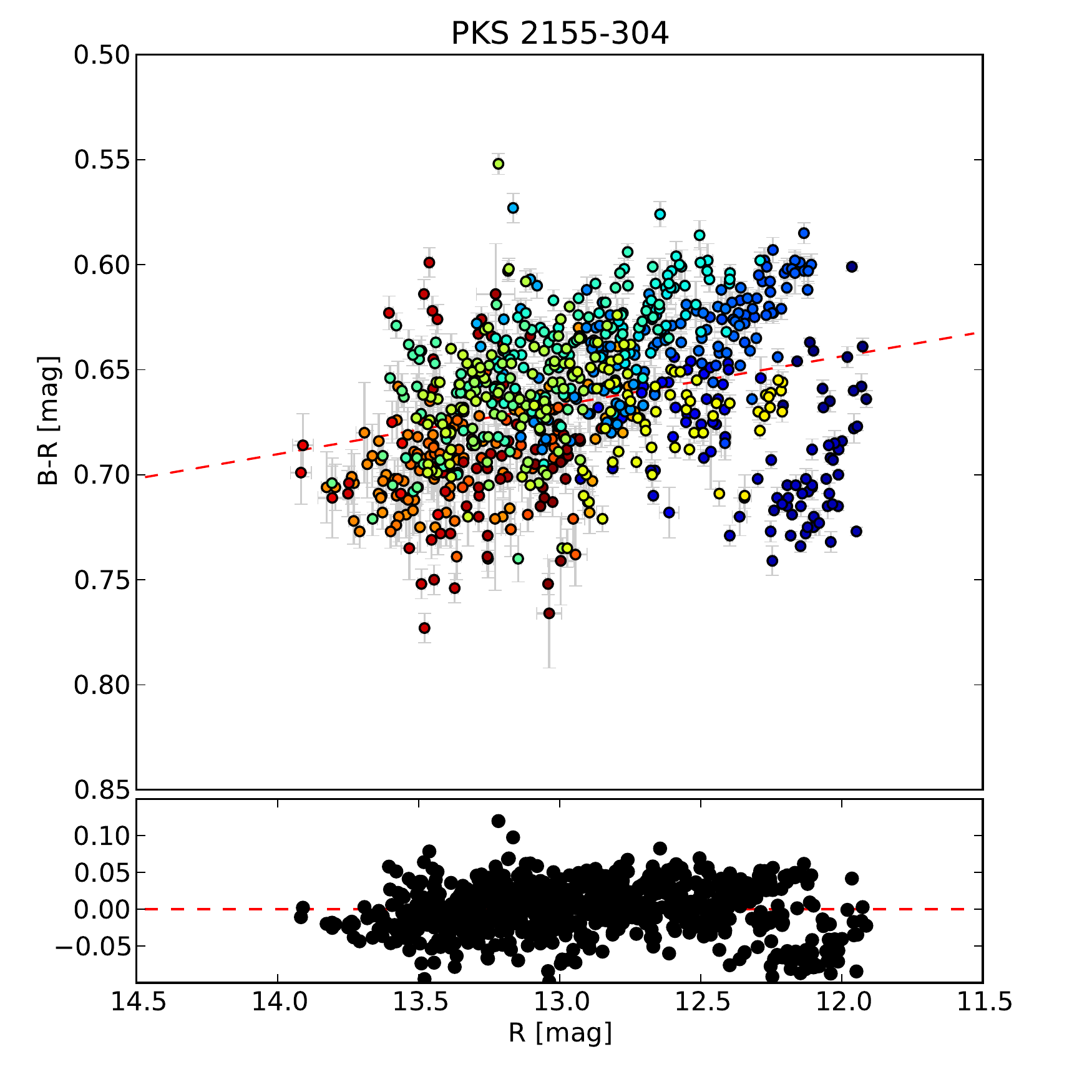}}
\centering{\includegraphics[width=0.43\textwidth]{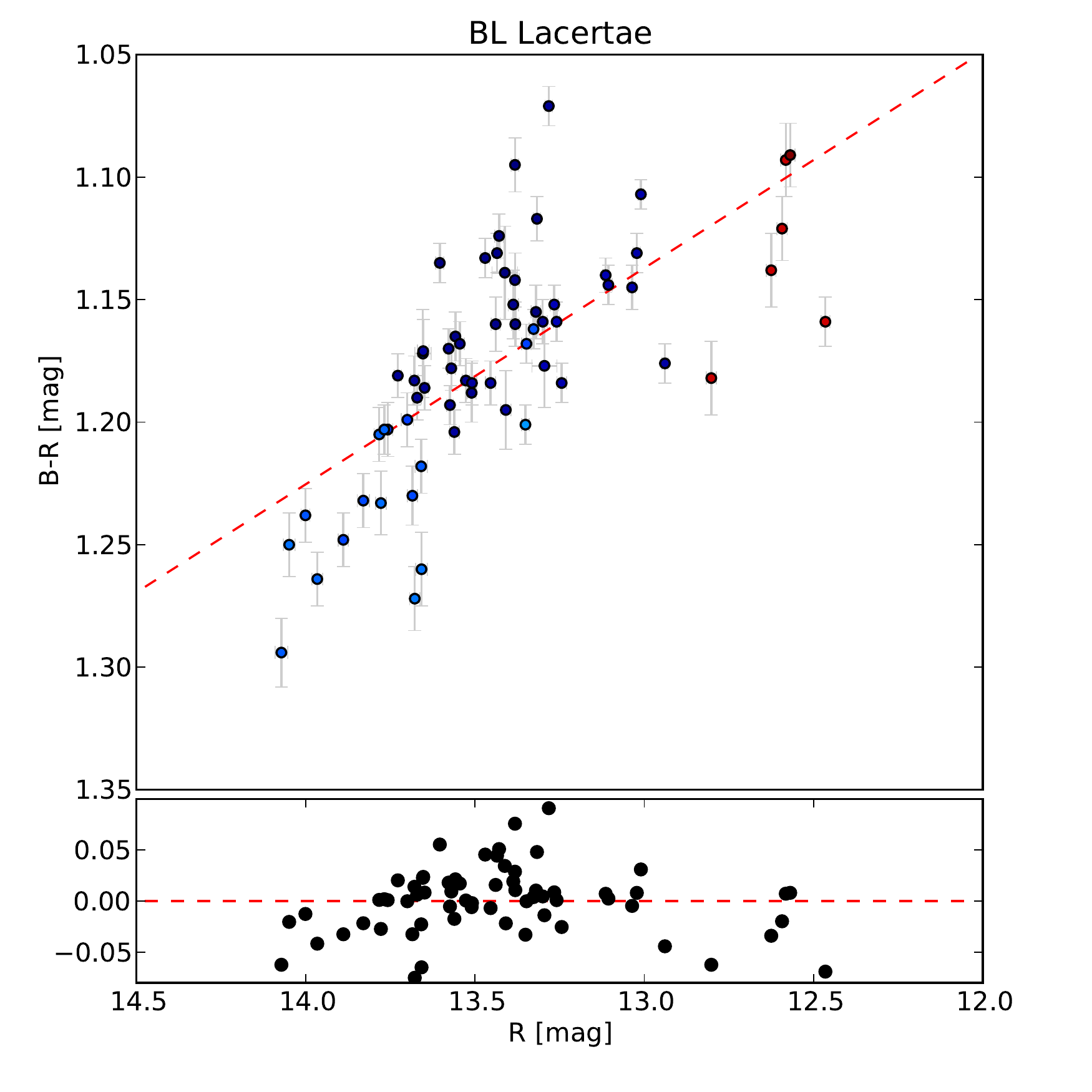}}\\
\caption[]{Same as Figure~\ref{fig1} for Markarian~501, RGB~J1725+118,~PKS 2005--489, SHBL~J213135.4--091523, PKS~2155--304, and BL~Lacertae.}
\label{fig5}
\end{figure*}

\subsection{Color-Magnitude Diagrams}  
\label{CM}

Figures~\ref{fig1}-\ref{fig5} present the $B-R$ colors vs $R$ magnitudes for the analyzed 30 blazars from our sample listed in Table \ref{table_1}. For each object, we evaluate the corresponding CM Pearson's correlation coefficient $C_{(B-R,R)}$, taking into account all the available datapoints, in order to check and to quantify general trends in a spectral appearance at different flux levels. The obtained values of $C_{(B-R,R)}$, together with the emerging slopes of the regression fits, the average colors $\langle B-R \rangle$, and the corresponding spectral indices, are all given in Table \ref{table_1} as well. The average spectral indices are derived here simply as 
\begin{equation}
\langle \alpha_{BR} \rangle = {0.4\, \langle B-R \rangle \over \log(\nu_B / \nu_R)} \, ,
\label{eq_1}
\end{equation}
where $\nu_B$ and $\nu_R$  are effective frequencies of the respective bands \citep{Bessell98}.

In figures~\ref{fig1}-\ref{fig5}, the times of given measurements are represented by the color scale: the earliest pointings are denoted by dark blue symbols and the most recent ones are depicted in red, with the rainbow color scale normalized to the entire span of the ATOM observations of a given blazar. This color-coding demonstrates that the evaluated global correlation coefficients are only rarely reliable proxies of a persistent spectral evolution (in particular in the case of sources displaying a `clustering' of datapoints at different regions of the CM diagrams at distinct epochs; see section~\ref{notes} below). In addition, we note that the time \emph{and} color scales for different objects in the analyzed sample are different, due to the different total time spans and varying spacings of the analyzed ATOM observations (see Table~\ref{table_2}). Hence, the result of the CM regression fits for the analyzed sample should be taken with caution. All in all, however, significant positive CM correlations (
$C_{(B-R,R)} \ll 0.5$), indicating that the sources appear \emph{in general} bluer when brighter, are found for 12 targets, while the negative (and, in fact, statistically not significant) CM correlations are found for only two sources (PG~1553+113 and PKS~1424+240). Below we comment on each analyzed target individually.

\subsection{Notes on Individual Sources}
\label{notes}

\begin{figure}[ht]
\centering{\includegraphics[width=0.49\textwidth]{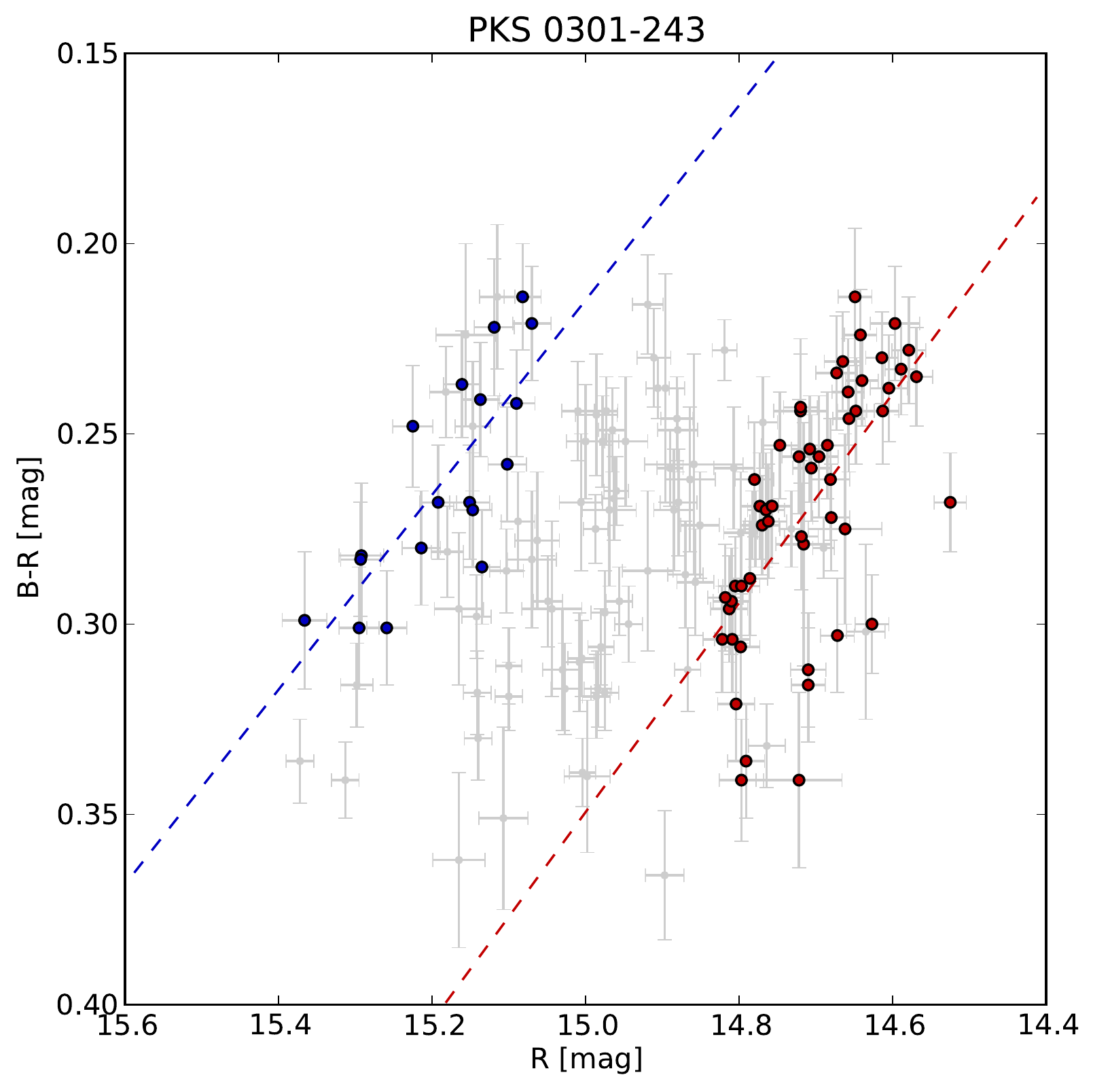}}
\caption[]{$B-R$ vs. $R$ diagram for PKS~0301--243 with the two selected epochs corresponding to the two distinct optical states of the source (red and blue symbols, respectively).}
\label{0301}
\end{figure}

\paragraph{SHBL~J001355.9--18540:}
The source was observed during 138 nights in 2008--2012. The collected data reveal no obvious CM correlations, but the source brightness varied only slightly, within the $\simeq$0.2\,mag range. The source optical spectrum was very steep, corresponding to $\alpha_{BR} > 3.7$.
\paragraph{PKS~0048--097:}
The source was observed during 96 nights in 2009--2012. No clear CM relation found \citep[in agreement with][]{Ikejiri2011}, even though the source brightness varied in a wide, $\sim 3$\,mag range, and the optical spectral index oscillated between $\alpha_{BR} \sim 2$ and 3.
\paragraph{RGB~J0152+017:}
The source was observed during 269 nights in 2007--2012. Clear \bwb behavior is present in the entire dataset, as well as in shorter isolated periods (with the exception of the earliest measurements at the lowest flux level, when a weak \rwb trend could be noted). The observed flux changes were rather minor ($\simeq 0.4$\,mag), and the optical spectrum varied between $\alpha_{BR} \sim 2$ and 3.
\paragraph{1ES~0229+200:}
The source was observed during 184 nights in 2007--2012. No clear CM correlation found, either in the entire dataset or in shorter (e.g., one-year) intervals, but the source brightness varied only slightly within the $\simeq$0.2\,mag range. The optical continuum was very steep, with $\alpha_{BR} > 3.7$.
\paragraph{AO~0235+16:}
The source was observed during 77 nights in 2007--2010. No significant CM correlation can be noted in the entire dataset despite the wide flux range covered ($\sim$4~mag), although during the flaring phase at the highest flux level clear \bwb chromatism is present. \cite{Ikejiri2011} and \cite{Raiteri2001} found significant CM correlations for this blazar, also during its quiescence. The observed spectral variations covered the range from $\alpha_{BR} \sim 2$ up to 3.
\paragraph{PKS~0301--243:}
The source was observed during 145 nights in 2009--2012. The observed \bwb trend in the entire dataset is rather weak, although during the selected shorter time intervals strong CM correlations can be noted, as shown on the additional Figure~\ref{0301}. The flux varied by $\sim 1$\,mag, and the optical spectral index clustered within a relatively narrow range $0.5 \lesssim \alpha_{BR} \lesssim 1$.

\paragraph{SHBL~J032541.0--164618:}
The source was observed during 38 nights in 2007-2011, revealing a flat optical spectrum $0 \lesssim \alpha_{BR} \lesssim 0.7$ and a significant positive CM correlation within the $\sim 1$\,mag flux range covered. 
\paragraph{1ES~0323+022:}
The source was observed during 40 nights in 2008--2012, revealing clear \bwb behavior despite only modest flux changes ($\simeq 0.4$\,mag), in agreement with the results of \cite{Ikejiri2011}. The source spectrum varied between $\alpha_{BR} \sim 0$ and 1.
\paragraph{1ES~0347--121:}
The source was observed during 161 nights in 2007--2012, revealing strong \bwb chromatism within the $\sim1$\,mag flux range covered and optical spectral index changing from $\alpha_{BR} \sim 2$ up to 3.5. 
\paragraph{1ES~0414+00.9:}
The source was observed during 285 nights in 2007--2012. The linear fit to all the collected datapoints does not show any clear CM correlation, but distinct states on the CM diagram can be noted. The flux varied within the $\lesssim 1$\,mag range, and the spectral index oscillated between $\alpha_{BR} \sim 0.7$ and 1.3.
\paragraph{PKS~0447--439:}
The source was observed during 54 nights in 2009--2012, revealing only weak \bwb chromatism in the $\sim 1$\,mag flux variation range covered. The optical slope varied only slightly between $\alpha_{BR} \sim 1.5$ and 2.
\paragraph{PKS~0537--441:}
The source was observed during 160 nights in 2008--2012. The analysis of a longterm monitoring does not reveal any global CM correlation pattern; however, at low flux levels some hints for \rwb behavior can be noted, while at high flux levels \bwb trend seems to be present. Despite large-amplitude flux variations ($\sim 3$\,mag), the optical slope varied only slightly between $\alpha_{BR} \sim 2$ and 2.5.
\paragraph{PKS~0548--322:}
The source was observed during 132 nights in 2007--2012, revealing almost constant flux with $\simeq 0.1$\,mag variations and persistently steep spectrum $\alpha_{BR} \simeq 3.5$.
\paragraph{PKS~0735+178:}
The source was observed during 104 nights in 2008--2012. No obvious global CM correlations found; previously, some hints for \bwb behavior of the blazar were reported by \cite{Gu}. The observed flux variations covered the $\gtrsim 1$\,mag range, and the optical slope varied between $\alpha_{BR} \sim 1$ and 1.5.
\paragraph{OJ~287:}
The source was observed during 148 nights in 2007--2012. No global CM correlations can be noted in the longterm lightcurve, although \bwb chromatism can be seen in shorter isolated intervals, in agreement with the previous results presented in the literature \citep[e.g.,][]{Carini1992,Dai,Ikejiri2011}. The source displayed large, $\sim 2$\,mag flux variations and spectral indices from $\alpha_{BR} \sim 2$ to 3. 
\paragraph{SHBL~J101015.9--311908:} 
The source was observed during 45 nights in 2008--2012, revealing significant positive CM correlation in a relatively narrow flux range covered ($\simeq 0.3$\,mag), with spectral indices $\alpha_{BR} \sim 2-3$.
\paragraph{1ES~1101--232:}
The source was observed during 197 nights in 2007--2012, showing only modest flux variations ($\sim 0.3$\,mag) and weak chromatism with optical slope changing between $\alpha_{BR} \sim 2$ and 3.
\paragraph{Markarian~421:}
The source was observed during 42 nights in 2007--2012. The overall \bwb behavior is present, in the agreement with the results by \cite{Ikejiri2011}. About $\sim 1$\,mag flux variations corresponded to the $\alpha_{BR} \sim 1.5-2.0$ range covered.
\paragraph{1ES~1218+304:}
The source was observed during only 20 nights in 2009, 2010, and 2012. Even though the ATOM observations of the blazar are limited, a strong CM correlation is clearly detected, both in the entire dataset and in shorter time intervals. About $\sim 1$\,mag flux variations corresponded to the $\alpha_{BR} \sim 0.7-1.3$ range covered.
\paragraph{W~Comae:}
The source was observed during 50 nights in 2008--2012, revealing strong \bwb chromatism for $\sim 1$\,mag flux variations and the optical slope changing from $\alpha_{BR} \sim 2$ up to 3.
\paragraph{1ES~1312--423:}
The source was observed during 122 nights in 2008--2011. A strong positive CM correlation can be noted despite very limited ($\sim0.25$\,mag) flux changes. The optical slope varied between $\alpha_{BR} \sim 2$ and 3.
\paragraph{PKS~1424+240:}
The source was observed during 94 nights in 2009--2012. While the global regression fit returns negative CM correlation coefficient, the analysis of shorter intervals reveals some hints for \bwb behavior. The observed flux variations were only models ($\simeq 0.5$\,mag), and the optical spectrum remained basically constant, $\alpha_{BR} \sim 3$.
\paragraph{AP~Librae:}
The source was observed during 112 nights in 2010--2012. Significant positive CM correlation is detected in the entire dataset and also in shorter (one-year-long) intervals. Modest, $\lesssim 1$\,mag flux variations corresponded to the $\alpha_{BR} \sim 2-3$ range covered.
\paragraph{PG~1553+113:}
The source was observed during 294 nights in 2007--2012. The global regression fit indicates negative though statistically not significant CM correlation, in agreement with the results by \cite{Ikejiri2011}; however, in shorter time intervals distinct states following \rwb or \bwb trends can be seen. About $\sim 1$\,mag flux variations corresponded to the relatively narrow, $\alpha_{BR} \sim 1.5-2.0$ range covered.
\paragraph{Markarian~501:}
The source was observed during 79 nights in 2007--2012. \Bwb behavior can be seen despite only weakly varying flux ($\simeq 0.2$\,mag), in agreement with the results by \cite{Ikejiri2011}. The optical spectrum remained basically constant with $\alpha_{BR} \sim 3$.
\paragraph{RGB~J1725+118:}
The source was observed during 69 nights in 2008--2012, revealing no obvious chromatism in the $\sim 1$\,mag flux range covered, due to the persistent optical spectrum with $\alpha_{BR} \sim 1$.
\paragraph{PKS~2005--489:}
The source was observed during 697 nights in 2007--2012. Significant positive CM correlations are seen in the entire dataset as well as in shorter time intervals, which in addition seem to correspond to distinct spectral states of the blazar (each following \bwb trend, though with different regression slopes). About $\sim 1.5$\,mag flux variations corresponded to a relatively narrow, $\alpha_{BR} \sim 3.0-3.5$ range covered.
\paragraph{SHBL~J213135.4--091523:}
The source was observed during 81 nights in 2008--2012, revealing only modest color changes with no obvious relation to the flux changes. The source spectrum remained very flat, $0 \lesssim \alpha_{BR} < 0.5$, within the $\sim 0.6$\,mag flux variation range.
\paragraph{PKS~2155--304:}
The source was observed during 792 nights in 2007--2012. Only weak \bwb trend can be noted in the entire dataset, although at the highest flux level two separate states of the source emerge, each characterized by a positive CM correlation but different optical colors (see Abramowski et al. 2014, in prep, for a further discussion). Large, $\sim 2$\,mag flux variations corresponded to a relatively narrow, $\alpha_{BR} \sim 1.3-1.8$ range covered.
\paragraph{BL~Lacertae:}
The source was observed during 65 nights in 2008, 2009, and 2012. A positive CM correlation can be seen in the entire dataset, albeit with a large scatter \citep[in agreement with][]{Villata2002,Villata2004,Ikejiri2011}. The observed flux changes were significant ($\sim 1.5$\,mag), and the optical spectrum varied from $\alpha_{BR} \sim 2$ up to 3.

\subsection{Multiwavelengh Analysis}
\label {MWL}

In this section we compare the optical properties of ATOM blazars with the high-energy (HE) $\gamma$-ray and high-frequency radio data on the selected targets. The $0.1-100$\,GeV $\gamma$-ray data are taken from the Second Catalog of \emph{Fermi}-LAT Sources \citep[2FGL;][]{2fgl}. The corresponding radio observations were carried out at 15\,GHz by the Owens Valley Radio Observatory (OVRO), which is the 40\,m telescope dedicated to observe \emph{Fermi}-LAT targets \citep{Richards11}. It should be noted that while the ATOM data presented in this paper are from 2007--2012, the 2FGL includes the LAT data collected during the first 24 months of the science phase of the \emph{Fermi} mission which began on 2008 August 4. The analyzed radio data, on the other hand, were collected during the period 2008--2012 in the sources' visibility windows. Hence, the multiwavelength data discussed below can be considered as only `quasi-simultaneous'. 

\begin{figure}[]
\centering{
\includegraphics[width=0.49\textwidth]{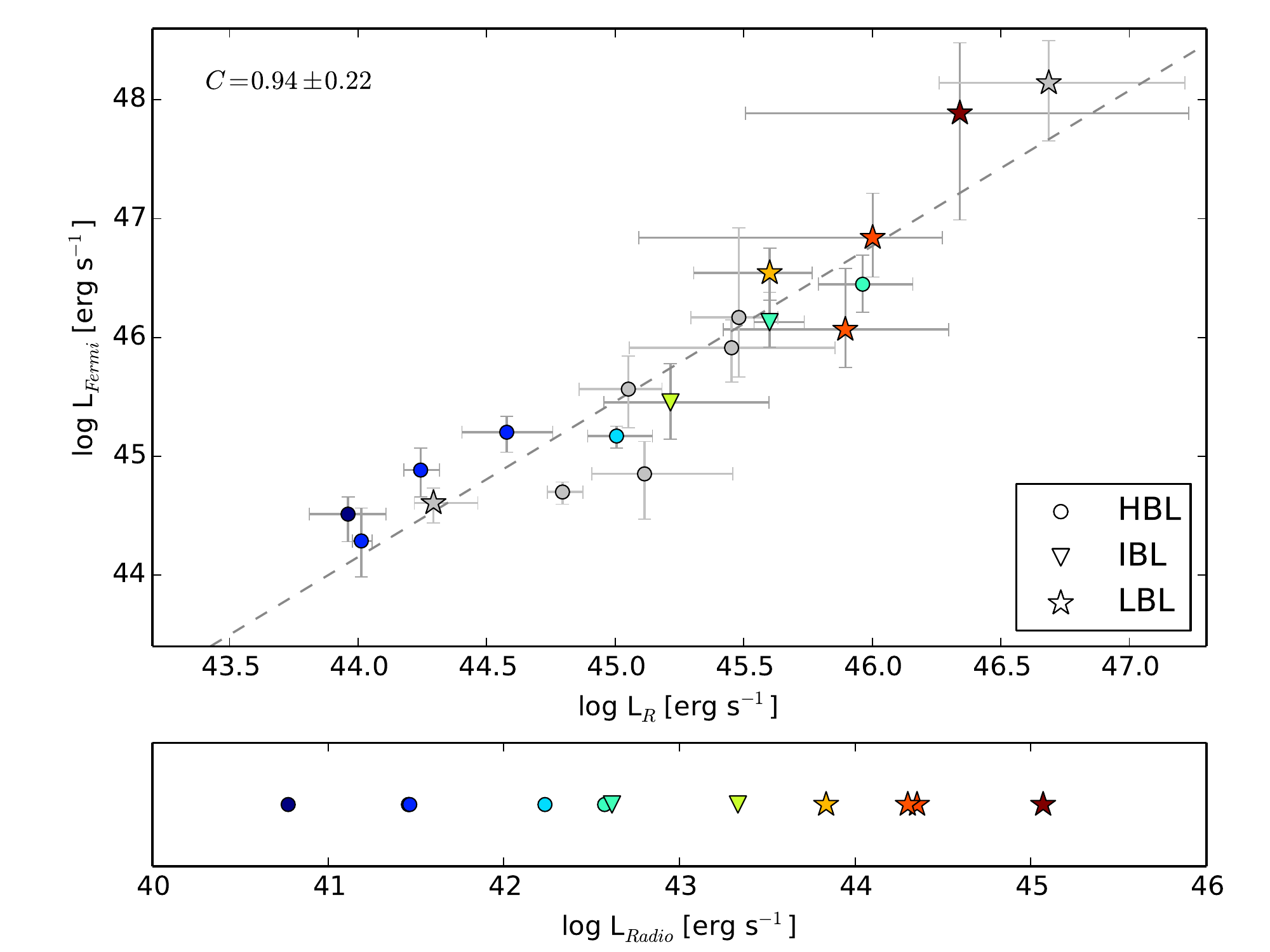}\\[5mm]
\includegraphics[width=0.49\textwidth]{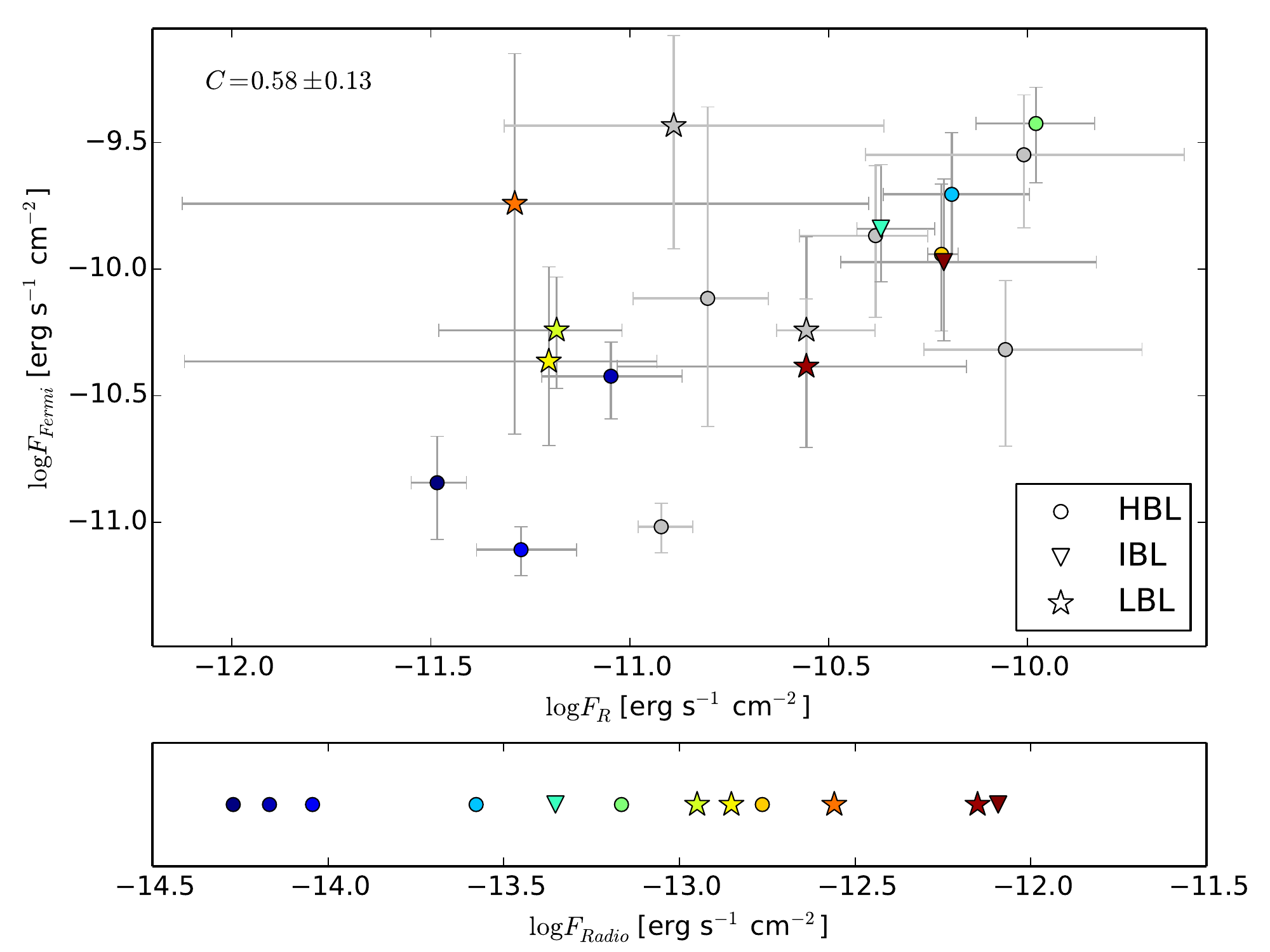}
}
\caption[]{Comparison between the mean HE $\gamma$-ray (LAT) and optical ($R$ filter) luminosities (upper panel) and fluxes (lower panel) for the analyzed BL Lac objects (see Table~\ref{table_mwl}). On each diagram, the color-coding denotes the luminosity or energy flux in the radio band ($15$\,GHz), depicted in the stripes below the panels; if there is no corresponding radio data, gray color is used. Different symbols here are used to denote three different types of BL Lac objects included in the sample, namely HBLs (circles), LBLs (triangles), and LBLs (stars). The formally evaluated correlation coefficients $C$ are provided in the upper-left corners of each panel. The error bars are evaluated as described in \S\,\ref{MWL}.}
\label{LL}
\end{figure}

\begin{figure}[!t]
\centering{
\includegraphics[width=0.49\textwidth]{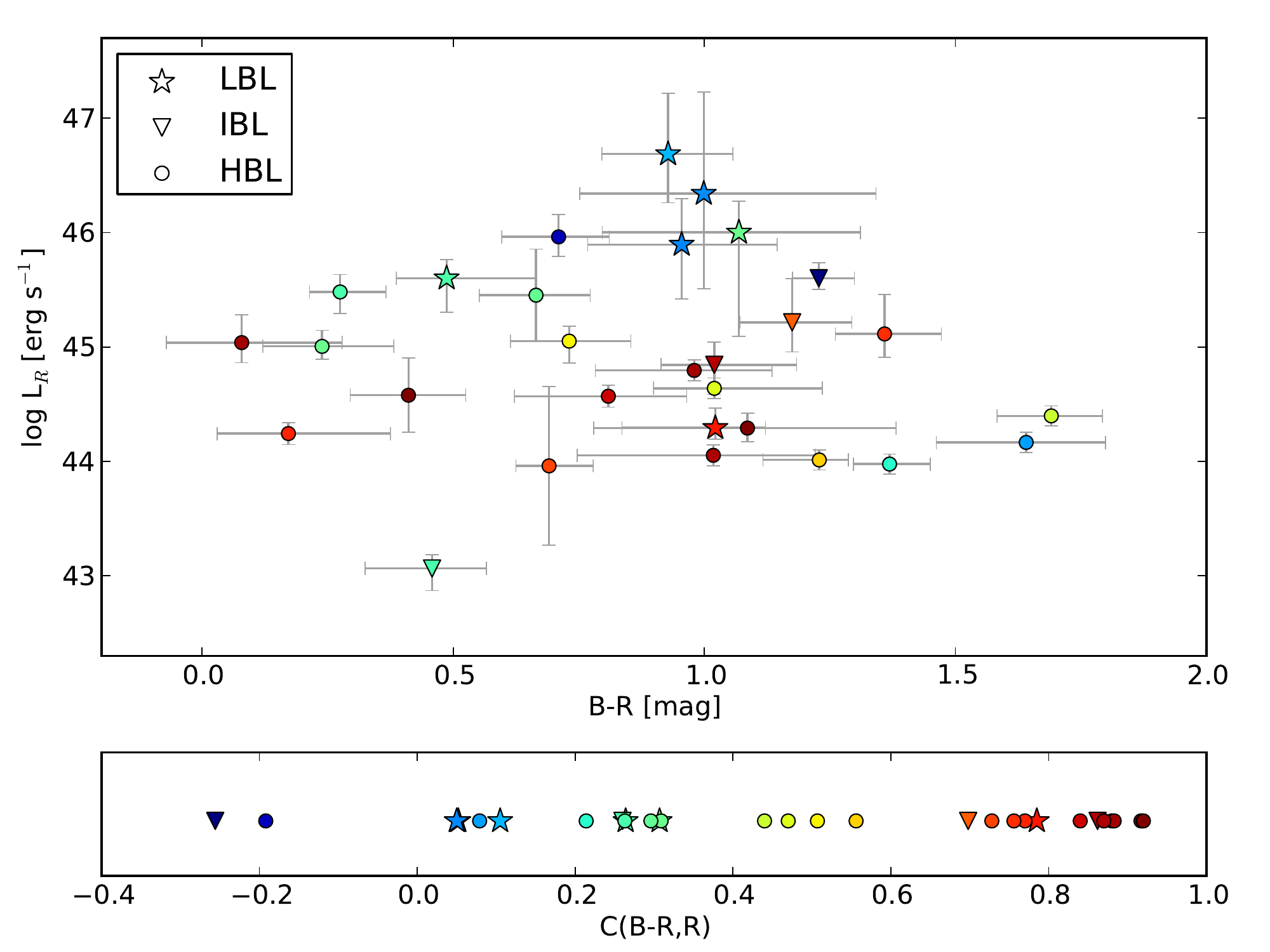}
}
\caption[]{Comparison between mean optical ($R$ filer) luminosities of the analyzed BL Lac objects, and their averaged optical colors $\langle B-R \rangle$ (see Table~\ref{table_mwl}). On the diagram, the color-coding denotes the Pearson's correlation coefficient of the optical CM relation, $C_{(B-R,R)}$, as depicted in the stripe below the panel. Different symbols here are used to denote three different types of BL Lac objects included in the sample, namely HBLs (circles), LBLs (triangles), and LBLs (stars). The error bars are evaluated as described in \S\,\ref{MWL}.}
\label{LC}
\end{figure}

\begin{figure}[!t]
\centering{\includegraphics[width=0.49\textwidth]{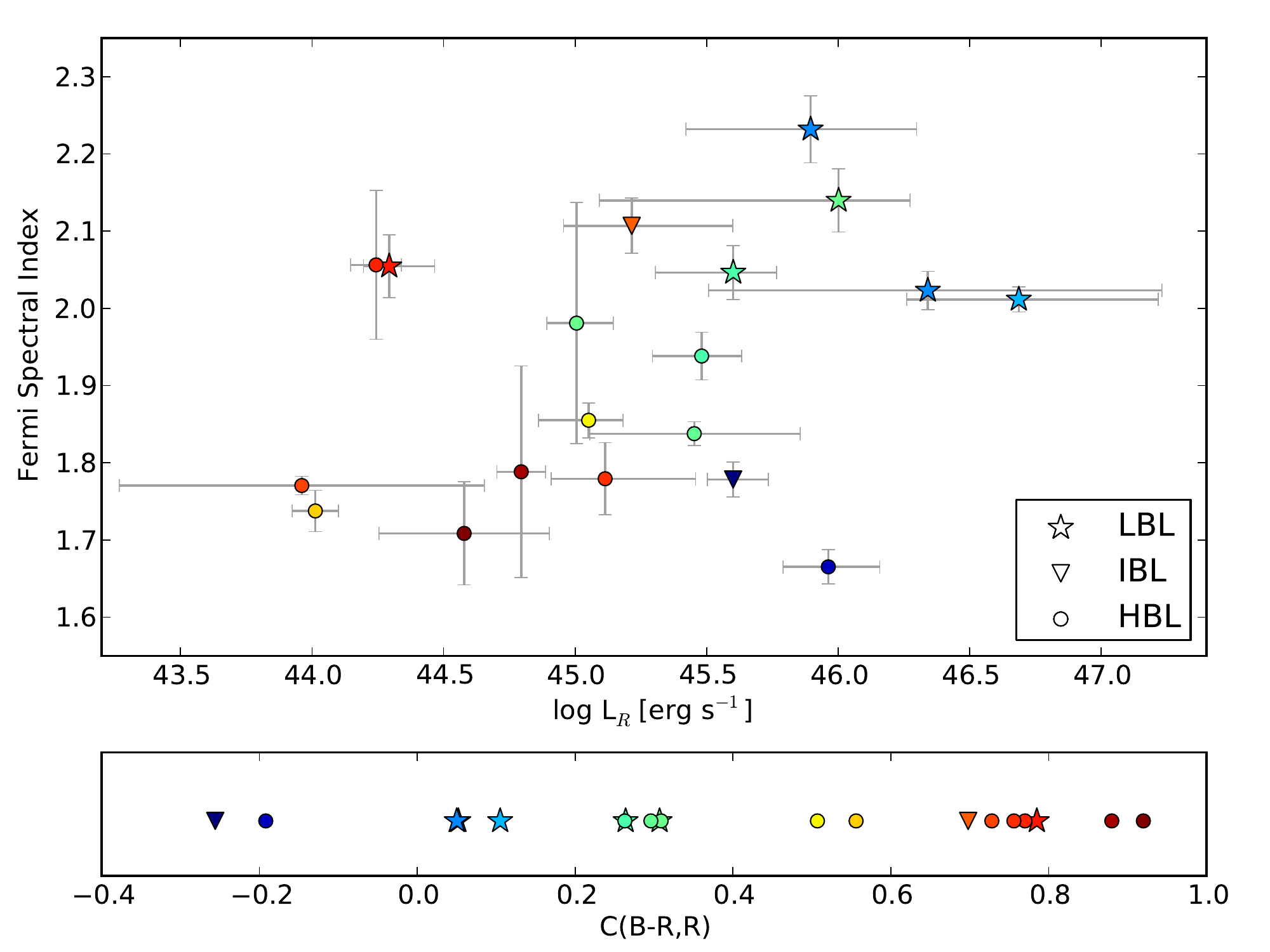}}
\caption[]{Comparison between the HE $\gamma$-ray photon indices for the LAT-detected sources in the sample and their optical luminosities. The color-coding denotes the Pearson's correlation coefficient of the optical CM relation, $C_{(B-R,R)}$, as depicted in the stripe below the diagram. Different symbols here are used to denote three types of BL Lac objects included in the sample: HBLs (circles), LBLs (triangles), and LBLs (stars). The error bars for $L_R$ are evaluated as described in \S\,\ref{MWL}. The error bars for the {\it Fermi}-LAT spectral indices are taken from 2FGL.}
\label{GL}
\end{figure}

To account for different activity states of the studied blazars in the error bars evaluated for fluxes, luminosities, and optical colors (Figures \ref{LL}--\ref{GL}), we consider both the variability range as well as the standard error of a given quantity, and then choose the larger value. In the case of ATOM data, the variability range is estimated as the minimum and maximum values of the  flux in the light curve. For \textit{Fermi}-LAT data minimum and maximum flux based on the \textit{Flux\_History} values are taken from 2FGL, excluding flux upper limits; if the \textit{Flux\_History} includes only the flux upper limits, the variability range is estimated by the standard error of the mean 2FGL flux. The optical and $\gamma$-ray luminosity errors are mainly driven by the distance error, therefore to not underestimate this uncertainty, the larger value of the standard luminosity error and the variability range is used.

First we investigate the mean luminosities of the blazars in the radio ($15$\,GHz), optical ($R$ filter) and $\gamma$-ray (LAT) bands, which are all provided in Table~\ref{table_mwl}. The corresponding luminosity-luminosity relations are presented in the upper panel of Figure~\ref{LL}. In the plot, different symbols are used to denote three different types of BL Lac objects included in the sample, namely HBLs (circles), LBLs (triangles), and LBLs (stars). As shown in the figure, the luminosities obey almost linear correlations, which may be either intrinsic or induced by the redshift dependance \citep[see in this context][]{ghirlanda11,arshakian12,hovatta14}.  Since our blazar sample is scarce and incomplete, and also since we are not considering upper limits for the objects not included in the 2FGL, we cannot apply any well-posted statistical analysis to address this issue in detail. Instead, we simply investigate the corresponding flux-flux correlations, which all appear weaker than the luminosity-
luminosity 
correlations (see the lower panel of Figure~\ref{LL}). Hence we conclude that the observed linear luminosity-luminosity correlations are affected, at least to some extent, by the redshift dependance. In addition, non-simultaneousness of the analyzed multifrequency dataset should be kept in mind in this context as well, as the comparison between mean fluxes in different bands depends on the activity state of a source during the analyzed period of time.

Despite all these caveats, we note however that, as expected, LBLs are in general more luminous than HBLs, with IBL objects falling in between these two types of blazars. Yet interesting outliers can be spotted. In particular, the IBL-classified RGB~J1725+118 is characterized by extremely low radio and optical luminosities; this object does not posses any $\gamma$-ray counterpart in the 2FGL. On the other hand, the HBL-classified PG~1553+113 is characterized by surprisingly high optical \emph{and} $\gamma$-ray luminosities, as shown in the $\gamma$-ray/optical luminosity plot (upper panel in Figure~\ref{LL}). In the same plot one can also see that AP~Librae, an LBL, seems particularly under-luminous for its type. In section~\ref{discussion} we comment in more detail on the two interesting cases of PG~1553+113 and AP~Librae.

The comparison of the source optical luminosities and the average $\langle B-R \rangle$ colors is presented in Figure~\ref{LC}. The plot does not reveal any obvious correlation. One can however notice that in the case of HBL type blazars the $\langle B-R \rangle$ parameter covers a wide range from 0\,mag to 1.7\,mag, which translates to the huge range of average spectral indices from $\langle \alpha _{BR} \rangle \sim 0$ up to $\sim 4$.  In the case of the two other subclasses of BL Lac objects included in the sample, the $\langle B-R \rangle$ ranges occupied are narrower, but this may be solely due to a low number IBLs and LBLs in the gathered sample.

In Figure~\ref{GL} we plot the HE $\gamma$-ray photon index for the LAT-detected sources in the sample versus their optical luminosities. The color-coding here corresponds to the values of the Pearson's correlation coefficient of the optical CM relation.  The diagram, which again may be affected quite substantially by non-simultaneousness of the considered LAT and ATOM datasets, reveals only a weak tendency for the low-power sources (mostly HBLs) to be characterized by more significant CM correlations and flatter HE $\gamma$-ray spectra when compared with the high-power sources (IBLs and LBLs, which typically do not exhibit any obvious \emph{global} CM correlations in the optical band, and tend to have steeper $\gamma$-ray spectra).

\section{Summary and Discussion } 
\label{discussion}

In this paper we analyze the optical data for 30 BL Lac type blazars collected using the ATOM telescope from 2007 till 2012. In particular, we study the evolution of the selected sources on the color-magnitude diagrams in $B$ and $R$ filters. Clear `global' \bwb trends are observed in the gathered datasets for 12 objects in the sample (with the Pearson's correlation coefficients $C_{(B-R,R)}> 0.5$), including 1ES~0323+022, 1ES~0347+121, 1ES~1218+304,  BL~Lacertae, Mrk~421, PKS~2005--489, RGB~J0152+117, SHBL~J032541.0--164618, SHBL~J101015.9--311908, W~Comae, and 1ES~1312--423. 

An important finding of the present analysis is the discovery of separate optical spectral states for several BL Lac objects. Namely, in the case of RGB~J0152+017, AO~0235+16, PKS~0301--243, 1ES~0347--121, 1ES~0414+00.9, OJ~287, 1ES~1101--232, 1ES~1312--423, PKS~1424+240, PG~1553+113, RGB~J1725+118, PKS~2005--489, and PKS~2155--304, the datapoints corresponding to different epochs of the source activity cluster in distinct, often isolated regions of the color--magnitude space. In some cases these distinct spectral states for a given source follow \bwb trends, possibly (but not necessarily) with different correlation slopes or different optical colors, forming in this way separate `branches' on the CM diagrams (e.g., PKS~2155--304, PKS~0301--243). Presence of such distinct states results typically in the overall weakening of the global CM correlations, as observed for example in PKS~0301--243. The analysis of all the available ATOM data for this blazar does not reveal any general \bwb chromatism, even though 
during 
the selected shorter time intervals two distinct branches appear in the CM plot (see Figure~\ref{0301}), separated by the $\simeq 0.4$\,mag gap but following similar \bwb tracks with the correlation coefficients $C_{(B-R,R)}=0.65$ and $0.79$ (for red and blue points in the figure, respectively), and with actually almost identical correlation slopes of $\simeq 0.26 \pm 0.05$.

The significance of \bwb chromatism for the objects analyzed in this paper does not correlate with the optical color, but instead seems to depend (at least partly) on the blazar type, and therefore on the optical luminosity: low-power HBLs are typically characterized by significant positive CM correlations, while high-power LBLs typically display low values of the $C_{(B-R,R)}$ coefficients. Similar conclusions have been presented before by \cite{Ikejiri2011}, even though these authors found positive CM correlations for almost all of the LBL type blazars included in their sample. 

\cite{Ikejiri2011} claimed in addition rather weak (or even absent) correlations between the optical flux and polarization degree changes in the studied sources, and found that the lower-luminosity objects show in general smaller-amplitude variability (in flux, color, and polarization degree) when compared to the higher-luminosity blazars. They concluded that the observed \bwb behavior of blazars arises due to the fact that short-term (days/weeks) flat-spectrum flares are superimposed on slowly-varying (months/years) or even stationary steep-spectrum (redder) components. It is important to note in this context that \citeauthor{Ikejiri2011} did not correct the measured optical fluxes for a contamination from the starlight of host galaxies \citep[which would require the hosts to be well resolved on optical images; e.g.,][]{nilsson07}, or from the continuum emission of accretion disks. Neither did we in the study presented here. 

In principle, a non-negligible contribution from accretion disks in luminous blazars --- which are believed to accrete at high rates, and as such to be characterized by a prominent disk emission peaking at UV frequencies \citep[e.g.,][]{sbarrato12} --- may lead in some cases to the apparent \rwb evolution at optical frequencies, as the flat-spectrum disk emission may become less and less pronounced for the increasing steeper-spectrum synchrotron emission of a blazar jet. This effect, however, is expected to play a role only in the case of FSRQs, or eventually the most luminous LBLs. The starlight contamination due to host galaxies, on the other hand, may be relevant in low-accretion-rate blazars such as HBLs and IBLs.

Host galaxies of BL Lac objects are evolved giant ellipticals, for which the starlight emission peaks at near-infrared ($\sim 1$\,$\mu$m) wavelengths. Spectral features related to the starlight components are often seen directly in the composite SEDs of HBLs, especially during their quiescence phases \citep[e.g., Mrk~501; see][]{LAT-501}. The presence of such may lead to the apparent \bwb evolution at optical frequencies, as the steep-spectrum starlight emission may become less and less pronounced for the increasing flatter-spectrum synchrotron emission of a blazar jet. And this may indeed be the case for at least some of the sources included in our sample. However, we argue that this effect cannot account for all the \bwb behavior seen in the collected ATOM dataset. That is because, as mentioned above, in many cases we observe separate `branches' in the CM diagrams, separated in flux or color but following similar \bwb tracks (see, e.g., Figure~\ref{0301}).

In order to investigate this issue in more detail, we divide the studied sample into the three sub-samples of (i) eleven objects displaying only modest flux variations, $< 1$\,mag, (ii) twelve sources with $R$-band fluxes changing by $\sim 1$\,mag (a factor of a few), and (iii) seven blazars showing large-amplitude variability, $> 1$\,mag. The latter sub-sample is dominated by LBLs (PKS~0048--097, AO~0235+16, PKS~0537--441, and OJ~287), all of which do not display any clear chromatism, and are characterized by similarly steep optical slopes with $2<\alpha _{BR} <3$. The only two `highly-variable' sources showing significant positive CM correlations are the HBL-classified PKS~2005--489 and the IBL-classified BL~Lacertae; in their cases the observed \bwb trends have to be intrinsic to the jets rather than resulting from the host-galaxy contribution. Interestingly, the remaining object in the sub-sample (iii), the HBL-classified PKS~2155--304, displays only weak chromatism but relatively flat spectrum $\alpha _{
BR} < 2$. The `moderately-variable' subsample (ii), on the other hand, is equally split into sources characterized by $C_{(B-R,R)}> 0.5$ and $C_{(B-R,R)}< 0.5$. The optical spectra of the blazars in this group are relatively flat, $0.5 < \alpha _{BR} < 2$, with the exception of steeper-spectrum ($2< \alpha _{BR} < 3$) objects 1ES~0347--121 (HBL), W~Comae (IBL), and AP Lib (LBL); these three `outliers' show particularly clear \bwb trends. Finally, majority of the objects from the `low-variable' sample (i) do not exhibit any significant positive CM correlations. Their ultra-steep ($\alpha _{BR} > 3$ in the case of SHBL~J001355.9--18540, 1ES 0229+200, and PKS~0548--322), or very steep and in addition persistent ($\alpha _{BR} \sim 3$ for PKS~1424+240 and Mrk~501) spectra signal non-negligible, or even dominant host-galaxy contributions to the fluxes measured with ATOM; the prominent exception here is the flat-spectrum SHBL~J213135.4--091523. Only four blazars in the sub-sample (i) reveal positive CM 
correlations (RGB~J0152+017, 1ES~0323+022, SHBL J101015.9--311908, and 1ES~1312--423), and these may be the examples where the observed \bwb trends are due to the underlying host galaxy components as discussed in the previous para.

In this paper we also compare the optical properties of ATOM blazars with the high-energy $\gamma$-ray and high-frequency radio data. We find that the radio, optical, and $\gamma$-ray luminosities of sources included in the sample obey almost linear correlations, which may be either intrinsic or induced by the redshift dependance. We do not find any correlation between the source luminosities and optical colors, or HE spectral indices.  All these multi-wavelength comparisons may be however significantly affected by non-simultaneousness of the analyzed dataset, as well as by a limited number of objects (especially of the IBL and LBL type) included in the studied sample.

In our analysis we noticed that two sources appear exceptional for their types: HBL-classified PG~1553+113 is characterized by a particularly low value of the optical CM correlation coefficient, and at the same time by surprisingly high optical \emph{and} $\gamma$-ray luminosities despite the very flat HE $\gamma$-ray continuum $\Gamma_{\rm HE} \simeq 1.67$ characteristic for an HBL; LBL-classified AP~Librae, on the other hand, seems particularly under-luminous for its type in optical and HE $\gamma$-rays, and displays a significant positive CM correlation. Interestingly, the broad-band SED of AP~Librae is indeed unusual, being characterized by a particularly broad high-energy hump in the SED representation \citep{kaufmann}. To the opposite, one of the main characteristics of the broad-band SED of PG~1553+113 noticed in \cite{MAGIC-1553} is the narrowness of its high-energy hump, which implies a relatively large value of the minimum electron energy within the dominant blazar 
emission zone.

\section*{Acknowledgements}
We are grateful to the Polish National Science Centre for supporting this work  within grants  DEC-2012/04/A/ST9/00083 and DEC-2011/03/N/ST9/01867.
We appreciate the optical observations with ATOM telescope provided by Bagmeet Behera, Gabriele Cologna, Giovanna Pedaletti and Stephanie Schwemmer.
This research has made use of data from the OVRO 40-m monitoring program
\citep{Richards11} which is supported in part by NASA grants NNX08AW31G and NNX11A043G, and
NSF grants AST-0808050 and AST-1109911. A.W. acknowledge support by Polish Ministry of Science and Higher Education in Mobility Plus Program.

\bibliographystyle{aa}
\bibliography{papier_opt}

\begin{thebibliography}{47}
\expandafter\ifx\csname natexlab\endcsname\relax\def\natexlab#1{#1}\fi

\bibitem[{{Abdo} {et~al.}(2010){Abdo}, {Ackermann}, {Agudo}, {Ajello}, {Aller},
  {Aller}, {Angelakis}, {Arkharov}, {Axelsson}, {Bach}, \& et~al.}]{abdo10}
{Abdo}, A.~A., {Ackermann}, M., {Agudo}, I., {et~al.} 2010, \apj, 716, 30

\bibitem[{{Abdo} {et~al.}(2011){Abdo}, {Ackermann}, {Ajello}, {Allafort},
  {Baldini}, {Ballet}, {Barbiellini}, {Baring}, {Bastieri}, {Bechtol}, \&
  et~al.}]{LAT-501}
{Abdo}, A.~A., {Ackermann}, M., {Ajello}, M., {et~al.} 2011, \apj, 727, 129

\bibitem[{{Ackermann} {et~al.}(2011){Ackermann}, {Ajello}, {Allafort},
  {Antolini}, {Atwood}, {Axelsson}, {Baldini}, {Ballet}, {Barbiellini},
  {Bastieri}, {Bechtol}, {Bellazzini}, {Berenji}, {Blandford}, {Bloom},
  {Bonamente}, {Borgland}, {Bottacini}, {Bouvier}, {Bregeon}, {Brigida},
  {Bruel}, {Buehler}, {Burnett}, {Buson}, {Caliandro}, {Cameron}, {Caraveo},
  {Casandjian}, {Cavazzuti}, {Cecchi}, {Charles}, {Cheung}, {Chiang},
  {Ciprini}, {Claus}, {Cohen-Tanugi}, {Conrad}, {Costamante}, {Cutini}, {de
  Angelis}, {de Palma}, {Dermer}, {Digel}, {Silva}, {Drell}, {Dubois},
  {Escande}, {Favuzzi}, {Fegan}, {Ferrara}, {Finke}, {Focke}, {Fortin},
  {Frailis}, {Fukazawa}, {Funk}, {Fusco}, {Gargano}, {Gasparrini}, {Gehrels},
  {Germani}, {Giebels}, {Giglietto}, {Giommi}, {Giordano}, {Giroletti},
  {Glanzman}, {Godfrey}, {Grenier}, {Grove}, {Guiriec}, {Gustafsson},
  {Hadasch}, {Hayashida}, {Hays}, {Healey}, {Horan}, {Hou}, {Hughes},
  {Iafrate}, {J{\'o}hannesson}, {Johnson}, {Johnson}, {Kamae}, {Katagiri},
  {Kataoka}, {Kn{\"o}dlseder}, {Kuss}, {Lande}, {Larsson}, {Latronico},
  {Longo}, {Loparco}, {Lott}, {Lovellette}, {Lubrano}, {Madejski}, {Mazziotta},
  {McConville}, {McEnery}, {Michelson}, {Mitthumsiri}, {Mizuno}, {Moiseev},
  {Monte}, {Monzani}, {Moretti}, {Morselli}, {Moskalenko}, {Murgia},
  {Nakamori}, {Naumann-Godo}, {Nolan}, {Norris}, {Nuss}, {Ohno}, {Ohsugi},
  {Okumura}, {Omodei}, {Orienti}, {Orlando}, {Ormes}, {Ozaki}, {Paneque},
  {Parent}, {Pesce-Rollins}, {Pierbattista}, {Piranomonte}, {Piron}, {Pivato},
  {Porter}, {Rain{\`o}}, {Rando}, {Razzano}, {Razzaque}, {Reimer}, {Reimer},
  {Ritz}, {Rochester}, {Romani}, {Roth}, {Sanchez}, {Sbarra}, {Scargle},
  {Schalk}, {Sgr{\`o}}, {Shaw}, {Siskind}, {Spandre}, {Spinelli}, {Strong},
  {Suson}, {Tajima}, {Takahashi}, {Takahashi}, {Tanaka}, {Thayer}, {Thayer},
  {Thompson}, {Tibaldo}, {Tinivella}, {Torres}, {Tosti}, {Troja}, {Uchiyama},
  {Vandenbroucke}, {Vasileiou}, {Vianello}, {Vitale}, {Waite}, {Wallace},
  {Wang}, {Winer}, {Wood}, {Wood}, \& {Zimmer}}]{2LAC}
{Ackermann}, M., {Ajello}, M., {Allafort}, A., {et~al.} 2011, \apj, 743, 171

\bibitem[{{Aharonian} {et~al.}(2007){Aharonian}, {Akhperjanian}, {Bazer-Bachi},
  {Behera}, {Beilicke}, {Benbow}, {Berge}, {Bernl{\"o}hr}, {Boisson}, {Bolz},
  {Borrel}, {Boutelier}, {Braun}, {Brion}, {Brown}, {B{\"u}hler},
  {B{\"u}sching}, {Bulik}, {Carrigan}, {Chadwick}, {Clapson}, {Chounet},
  {Coignet}, {Cornils}, {Costamante}, {Degrange}, {Dickinson},
  {Djannati-Ata{\"i}}, {Domainko}, {Drury}, {Dubus}, {Dyks}, {Egberts},
  {Emmanoulopoulos}, {Espigat}, {Farnier}, {Feinstein}, {Fiasson},
  {F{\"o}rster}, {Fontaine}, {Funk}, {Funk}, {F{\"u}{\ss}ling}, {Gallant},
  {Giebels}, {Glicenstein}, {Gl{\"u}ck}, {Goret}, {Hadjichristidis}, {Hauser},
  {Hauser}, {Heinzelmann}, {Henri}, {Hermann}, {Hinton}, {Hoffmann}, {Hofmann},
  {Holleran}, {Hoppe}, {Horns}, {Jacholkowska}, {de Jager}, {Kendziorra},
  {Kerschhaggl}, {Kh{\'e}lifi}, {Komin}, {Kosack}, {Lamanna}, {Latham}, {Le
  Gallou}, {Lemi{\`e}re}, {Lemoine-Goumard}, {Lenain}, {Lohse}, {Martin},
  {Martineau-Huynh}, {Marcowith}, {Masterson}, {Maurin}, {McComb}, {Moderski},
  {Moulin}, {de Naurois}, {Nedbal}, {Nolan}, {Olive}, {Orford}, {Osborne},
  {Ostrowski}, {Panter}, {Pedaletti}, {Pelletier}, {Petrucci}, {Pita},
  {P{\"u}hlhofer}, {Punch}, {Ranchon}, {Raubenheimer}, {Raue}, {Rayner},
  {Renaud}, {Ripken}, {Rob}, {Rolland}, {Rosier-Lees}, {Rowell}, {Rudak},
  {Ruppel}, {Sahakian}, {Santangelo}, {Saug{\'e}}, {Schlenker}, {Schlickeiser},
  {Schr{\"o}der}, {Schwanke}, {Schwarzburg}, {Schwemmer}, {Shalchi}, {Sol},
  {Spangler}, {Stawarz}, {Steenkamp}, {Stegmann}, {Superina}, {Tam},
  {Tavernet}, {Terrier}, {van Eldik}, {Vasileiadis}, {Venter}, {Vialle},
  {Vincent}, {Vivier}, {V{\"o}lk}, {Volpe}, {Wagner}, {Ward}, \&
  {Zdziarski}}]{2155flare}
{Aharonian}, F., {Akhperjanian}, A.~G., {Bazer-Bachi}, A.~R., {et~al.} 2007,
  \apjl, 664, L71

\bibitem[{{Aharonian} {et~al.}(2006){Aharonian}, {Akhperjanian}, {Bazer-Bachi},
  {Beilicke}, {Benbow}, {Berge}, {Bernl{\"o}hr}, {Boisson}, {Bolz}, {Borrel},
  {Braun}, {Breitling}, {Brown}, {B{\"u}hler}, {B{\"u}sching}, {Carrigan},
  {Chadwick}, {Chounet}, {Cornils}, {Costamante}, {Degrange}, {Dickinson},
  {Djannati-Ata{\"i}}, {O'C.~Drury}, {Dubus}, {Egberts}, {Emmanoulopoulos},
  {Espigat}, {Feinstein}, {Ferrero}, {Fiasson}, {Fontaine}, {Funk}, {Funk},
  {Gallant}, {Giebels}, {Glicenstein}, {Goret}, {Hadjichristidis}, {Hauser},
  {Hauser}, {Heinzelmann}, {Henri}, {Hermann}, {Hinton}, {Hofmann}, {Holleran},
  {Horns}, {Jacholkowska}, {de Jager}, {Kh{\'e}lifi}, {Komin}, {Konopelko},
  {Kosack}, {Latham}, {Le Gallou}, {Lemi{\`e}re}, {Lemoine-Goumard}, {Lohse},
  {Martin}, {Martineau-Huynh}, {Marcowith}, {Masterson}, {McComb}, {de
  Naurois}, {Nedbal}, {Nolan}, {Noutsos}, {Orford}, {Osborne}, {Ouchrif},
  {Panter}, {Pelletier}, {Pita}, {P{\"u}hlhofer}, {Punch}, {Raubenheimer},
  {Raue}, {Rayner}, {Reimer}, {Reimer}, {Ripken}, {Rob}, {Rolland}, {Rowell},
  {Sahakian}, {Saug{\'e}}, {Schlenker}, {Schlickeiser}, {Schwanke}, {Sol},
  {Spangler}, {Spanier}, {Steenkamp}, {Stegmann}, {Superina}, {Tavernet},
  {Terrier}, {Th{\'e}oret}, {Tluczykont}, {van Eldik}, {Vasileiadis}, {Venter},
  {Vincent}, {V{\"o}lk}, {Wagner}, \& {Ward}}]{crab}
{Aharonian}, F., {Akhperjanian}, A.~G., {Bazer-Bachi}, A.~R., {et~al.} 2006,
  \aap, 457, 899

\bibitem[{{Aleksi{\'c}} {et~al.}(2012){Aleksi{\'c}}, {Alvarez}, {Antonelli},
  {Antoranz}, {Asensio}, {Backes}, {Barrio}, {Bastieri}, {Becerra
  Gonz{\'a}lez}, {Bednarek}, {Berdyugin}, {Berger}, {Bernardini}, {Biland},
  {Blanch}, {Bock}, {Boller}, {Bonnoli}, {Borla Tridon}, {Braun}, {Bretz},
  {Ca{\~n}ellas}, {Carmona}, {Carosi}, {Colin}, {Colombo}, {Contreras},
  {Cortina}, {Cossio}, {Covino}, {Dazzi}, {De Angelis}, {De Caneva}, {De Cea
  del Pozo}, {De Lotto}, {Delgado Mendez}, {Diago Ortega}, {Doert},
  {Dom{\'{\i}}nguez}, {Dominis Prester}, {Dorner}, {Doro}, {Elsaesser},
  {Ferenc}, {Fonseca}, {Font}, {Fruck}, {Garc{\'{\i}}a L{\'o}pez},
  {Garczarczyk}, {Garrido}, {Giavitto}, {Godinovi{\'c}}, {Hadasch},
  {H{\"a}fner}, {Herrero}, {Hildebrand}, {H{\"o}hne-M{\"o}nch}, {Hose},
  {Hrupec}, {Huber}, {Jogler}, {Kellermann}, {Klepser}, {Kr{\"a}henb{\"u}hl},
  {Krause}, {La Barbera}, {Lelas}, {Leonardo}, {Lindfors}, {Lombardi},
  {L{\'o}pez}, {L{\'o}pez-Oramas}, {Lorenz}, {Makariev}, {Maneva},
  {Mankuzhiyil}, {Mannheim}, {Maraschi}, {Mariotti}, {Mart{\'{\i}}nez},
  {Mazin}, {Meucci}, {Miranda}, {Mirzoyan}, {Miyamoto}, {Mold{\'o}n},
  {Moralejo}, {Munar-Adrover}, {Nieto}, {Nilsson}, {Orito}, {Oya}, {Paneque},
  {Paoletti}, {Pardo}, {Paredes}, {Partini}, {Pasanen}, {Pauss},
  {Perez-Torres}, {Persic}, {Peruzzo}, {Pilia}, {Pochon}, {Prada}, {Prada
  Moroni}, {Prandini}, {Puljak}, {Reichardt}, {Reinthal}, {Rhode}, {Rib{\'o}},
  {Rico}, {R{\"u}gamer}, {Saggion}, {Saito}, {Saito}, {Salvati}, {Satalecka},
  {Scalzotto}, {Scapin}, {Schultz}, {Schweizer}, {Shayduk}, {Shore},
  {Sillanp{\"a}{\"a}}, {Sitarek}, {Sobczynska}, {Spanier}, {Spiro},
  {Stamatescu}, {Stamerra}, {Steinke}, {Storz}, {Strah}, {Suri{\'c}}, {Takalo},
  {Takami}, {Tavecchio}, {Temnikov}, {Terzi{\'c}}, {Tescaro}, {Teshima},
  {Tibolla}, {Torres}, {Treves}, {Uellenbeck}, {Vankov}, {Vogler}, {Wagner},
  {Weitzel}, {Zabalza}, {Zandanel}, {Zanin}, {Buson}, {Horan}, {Larsson}, \&
  {D'Ammando}}]{MAGIC-1553}
{Aleksi{\'c}}, J., {Alvarez}, E.~A., {Antonelli}, L.~A., {et~al.} 2012, \apj,
  748, 46

\bibitem[{{Arshakian} {et~al.}(2012){Arshakian}, {Le{\'o}n-Tavares},
  {B{\"o}ttcher}, {Torrealba}, {Chavushyan}, {Lister}, {Ros}, \&
  {Zensus}}]{arshakian12}
{Arshakian}, T.~G., {Le{\'o}n-Tavares}, J., {B{\"o}ttcher}, M., {et~al.} 2012,
  \aap, 537, A32

\bibitem[{{Atwood} {et~al.}(2009){Atwood}, {Abdo}, {Ackermann}, {Althouse},
  {Anderson}, {Axelsson}, {Baldini}, {Ballet}, {Band}, {Barbiellini}, \&
  et~al.}]{LAT}
{Atwood}, W.~B., {Abdo}, A.~A., {Ackermann}, M., {et~al.} 2009, \apj, 697, 1071

\bibitem[{{Begelman} {et~al.}(1984){Begelman}, {Blandford}, \&
  {Rees}}]{begelman84}
{Begelman}, M.~C., {Blandford}, R.~D., \& {Rees}, M.~J. 1984, Reviews of Modern
  Physics, 56, 255

\bibitem[{{Bessell}(1990)}]{Bessell90}
{Bessell}, M.~S. 1990, \pasp, 102, 1181

\bibitem[{{Bessell} {et~al.}(1998){Bessell}, {Castelli}, \& {Plez}}]{Bessell98}
{Bessell}, M.~S., {Castelli}, F., \& {Plez}, B. 1998, \aap, 333, 231

\bibitem[{{Bhatta} {et~al.}(2013){Bhatta}, {Webb}, {Hollingsworth}, {Dhalla},
  {Khanuja}, {Bachev}, {Blinov}, {B{\"o}ttcher}, {Bravo Calle}, {Calcidese},
  {Capezzali}, {Carosati}, {Chigladze}, {Collins}, {Coloma}, {Efimov}, {Gupta},
  {Hu}, {Kurtanidze}, {Lamerato}, {Larionov}, {Lee}, {Lindfors}, {Murphy},
  {Nilsson}, {Ohlert}, {Oksanen}, {P{\"a}{\"a}kk{\"o}nen}, {Pollock}, {Rani},
  {Reinthal}, {Rodriguez}, {Ros}, {Roustazadeh}, {Sagar}, {Sanchez}, {Shastri},
  {Sillanp{\"a}{\"a}}, {Strigachev}, {Takalo}, {Vennes}, {Villata},
  {Villforth}, {Wu}, \& {Zhou}}]{bhatta13}
{Bhatta}, G., {Webb}, J.~R., {Hollingsworth}, H., {et~al.} 2013, \aap, 558, A92

\bibitem[{{Bonning} {et~al.}(2012){Bonning}, {Urry}, {Bailyn}, {Buxton},
  {Chatterjee}, {Coppi}, {Fossati}, {Isler}, \& {Maraschi}}]{bonning12}
{Bonning}, E., {Urry}, C.~M., {Bailyn}, C., {et~al.} 2012, \apj, 756, 13

\bibitem[{{Carini} {et~al.}(1992){Carini}, {Miller}, {Noble}, \&
  {Goodrich}}]{Carini1992}
{Carini}, M.~T., {Miller}, H.~R., {Noble}, J.~C., \& {Goodrich}, B.~D. 1992,
  \aj, 104, 15

\bibitem[{{Chatterjee} {et~al.}(2012){Chatterjee}, {Bailyn}, {Bonning},
  {Buxton}, {Coppi}, {Fossati}, {Isler}, {Maraschi}, \& {Urry}}]{chatterjee12}
{Chatterjee}, R., {Bailyn}, C.~D., {Bonning}, E.~W., {et~al.} 2012, \apj, 749,
  191

\bibitem[{{Clements} \& {Carini}(2001)}]{Clements}
{Clements}, S.~D. \& {Carini}, M.~T. 2001, \aj, 121, 90

\bibitem[{{Dai} {et~al.}(2011){Dai}, {Wu}, {Zhu}, {Zhou}, \& {Ma}}]{Dai}
{Dai}, Y., {Wu}, J., {Zhu}, Z.-H., {Zhou}, X., \& {Ma}, J. 2011, \aj, 141, 65

\bibitem[{{Dai} {et~al.}(2013){Dai}, {Wu}, {Zhu}, {Zhou}, {Ma}, {Yuan}, \&
  {Wang}}]{dai13}
{Dai}, Y., {Wu}, J., {Zhu}, Z.-H., {et~al.} 2013, \apjs, 204, 22

\bibitem[{{Dermer} \& {Schlickeiser}(1993)}]{dermer}
{Dermer}, C.~D. \& {Schlickeiser}, R. 1993, \apj, 416, 458

\bibitem[{{Fossati} {et~al.}(1998){Fossati}, {Maraschi}, {Celotti}, {Comastri},
  \& {Ghisellini}}]{fossati98}
{Fossati}, G., {Maraschi}, L., {Celotti}, A., {Comastri}, A., \& {Ghisellini},
  G. 1998, \mnras, 299, 433

\bibitem[{{Gaur} {et~al.}(2014){Gaur}, {Gupta}, {Wiita}, {Uemura}, {Itoh}, \&
  {Sasada}}]{gaur14}
{Gaur}, H., {Gupta}, A.~C., {Wiita}, P.~J., {et~al.} 2014, \apjl, 781, L4

\bibitem[{{Ghirlanda} {et~al.}(2011){Ghirlanda}, {Ghisellini}, {Tavecchio},
  {Foschini}, \& {Bonnoli}}]{ghirlanda11}
{Ghirlanda}, G., {Ghisellini}, G., {Tavecchio}, F., {Foschini}, L., \&
  {Bonnoli}, G. 2011, \mnras, 413, 852

\bibitem[{{Ghisellini} {et~al.}(1997){Ghisellini}, {Villata}, {Raiteri},
  {Bosio}, {de Francesco}, {Latini}, {Maesano}, {Massaro}, {Montagni}, {Nesci},
  {Tosti}, {Fiorucci}, {Pian}, {Maraschi}, {Treves}, {Comastri}, \&
  {Mignoli}}]{ghisellini97}
{Ghisellini}, G., {Villata}, M., {Raiteri}, C.~M., {et~al.} 1997, \aap, 327, 61

\bibitem[{{Ghosh} {et~al.}(2000){Ghosh}, {Ramsey}, {Sadun}, \&
  {Soundararajaperumal}}]{Ghosh}
{Ghosh}, K.~K., {Ramsey}, B.~D., {Sadun}, A.~C., \& {Soundararajaperumal}, S.
  2000, \apjs, 127, 11

\bibitem[{{Gu} {et~al.}(2006){Gu}, {Lee}, {Pak}, {Yim}, \& {Fletcher}}]{Gu}
{Gu}, M.~F., {Lee}, C.-U., {Pak}, S., {Yim}, H.~S., \& {Fletcher}, A.~B. 2006,
  \aap, 450, 39

\bibitem[{{Hauser} {et~al.}(2004){Hauser}, {M{\"o}llenhoff}, {P{\"u}hlhofer},
  {Wagner}, {Hagen}, \& {Knoll}}]{hauser}
{Hauser}, M., {M{\"o}llenhoff}, C., {P{\"u}hlhofer}, G., {et~al.} 2004,
  Astronomische Nachrichten, 325, 659

\bibitem[{{Hovatta} {et~al.}(2014){Hovatta}, {Pavlidou}, {King}, {Mahabal},
  {Sesar}, {Dancikova}, {Djorgovski}, {Drake}, {Laher}, {Levitan},
  {Max-Moerbeck}, {Ofek}, {Pearson}, {Prince}, {Readhead}, {Richards}, \&
  {Surace}}]{hovatta14}
{Hovatta}, T., {Pavlidou}, V., {King}, O.~G., {et~al.} 2014, ArXiv e-prints

\bibitem[{{Ikejiri} {et~al.}(2011){Ikejiri}, {Uemura}, {Sasada}, {Ito},
  {Yamanaka}, {Sakimoto}, {Arai}, {Fukazawa}, {Ohsugi}, {Kawabata}, {Yoshida},
  {Sato}, \& {Kino}}]{Ikejiri2011}
{Ikejiri}, Y., {Uemura}, M., {Sasada}, M., {et~al.} 2011, \pasj, 63, 639

\bibitem[{{Kaufmann}(2011)}]{kaufmann}
{Kaufmann}, S. 2011, International Cosmic Ray Conference, 8, 199

\bibitem[{{Konigl}(1981)}]{konigl}
{Konigl}, A. 1981, \apj, 243, 700

\bibitem[{{Marscher} \& {Gear}(1985)}]{marscher}
{Marscher}, A.~P. \& {Gear}, W.~K. 1985, \apj, 298, 114

\bibitem[{{Nilsson} {et~al.}(2007){Nilsson}, {Pasanen}, {Takalo}, {Lindfors},
  {Berdyugin}, {Ciprini}, \& {Pforr}}]{nilsson07}
{Nilsson}, K., {Pasanen}, M., {Takalo}, L.~O., {et~al.} 2007, \aap, 475, 199

\bibitem[{{Nolan} {et~al.}(2012){Nolan}, {Abdo}, {Ackermann}, {Ajello},
  {Allafort}, {Antolini}, {Atwood}, {Axelsson}, {Baldini}, {Ballet}, \&
  et~al.}]{2fgl}
{Nolan}, P.~L., {Abdo}, A.~A., {Ackermann}, M., {et~al.} 2012, \apjs, 199, 31

\bibitem[{{Padovani} \& {Giommi}(1995)}]{padovani95}
{Padovani}, P. \& {Giommi}, P. 1995, \apj, 444, 567

\bibitem[{{Raiteri} {et~al.}(2001){Raiteri}, {Villata}, {Aller}, {Aller},
  {Heidt}, {Kurtanidze}, {Lanteri}, {Maesano}, {Massaro}, {Montagni}, {Nesci},
  {Nilsson}, {Nikolashvili}, {Nurmi}, {Ostorero}, {Pursimo}, {Rekola},
  {Sillanp{\"a}{\"a}}, {Takalo}, {Ter{\"a}sranta}, {Tosti}, {Balonek}, {Feldt},
  {Heines}, {Heisler}, {Hu}, {Kidger}, {Mattox}, {McGrath}, {Pati}, {Robb},
  {Sadun}, {Shastri}, {Wagner}, {Wei}, \& {Wu}}]{Raiteri2001}
{Raiteri}, C.~M., {Villata}, M., {Aller}, H.~D., {et~al.} 2001, \aap, 377, 396

\bibitem[{{Raiteri} {et~al.}(2013){Raiteri}, {Villata}, {D'Ammando},
  {Larionov}, {Gurwell}, {Mirzaqulov}, {Smith}, {Acosta-Pulido}, {Agudo},
  {Ar{\'e}valo}, {Bachev}, {Ben{\'{\i}}tez}, {Berdyugin}, {Blinov}, {Borman},
  {B{\"o}ttcher}, {Bozhilov}, {Carnerero}, {Carosati}, {Casadio}, {Chen},
  {Doroshenko}, {Efimov}, {Efimova}, {Ehgamberdiev}, {G{\'o}mez},
  {Gonz{\'a}lez-Morales}, {Hiriart}, {Ibryamov}, {Jadhav}, {Jorstad}, {Joshi},
  {Kadenius}, {Klimanov}, {Kohli}, {Konstantinova}, {Kopatskaya}, {Koptelova},
  {Kimeridze}, {Kurtanidze}, {Larionova}, {Larionova}, {Ligustri}, {Lindfors},
  {Marscher}, {McBreen}, {McHardy}, {Metodieva}, {Molina}, {Morozova},
  {Nazarov}, {Nikolashvili}, {Nilsson}, {Okhmat}, {Ovcharov}, {Panwar},
  {Pasanen}, {Peneva}, {Phipps}, {Pulatova}, {Reinthal}, {Ros}, {Sadun},
  {Schwartz}, {Semkov}, {Sergeev}, {Sigua}, {Sillanp{\"a}{\"a}}, {Smith},
  {Stoyanov}, {Strigachev}, {Takalo}, {Taylor}, {Thum}, {Troitsky}, {Valcheva},
  {Wehrle}, \& {Wiesemeyer}}]{raiteri13}
{Raiteri}, C.~M., {Villata}, M., {D'Ammando}, F., {et~al.} 2013, \mnras, 436,
  1530

\bibitem[{{Richards} {et~al.}(2011){Richards}, {Max-Moerbeck}, {Pavlidou},
  {King}, {Pearson}, {Readhead}, {Reeves}, {Shepherd}, {Stevenson},
  {Weintraub}, {Fuhrmann}, {Angelakis}, {Zensus}, {Healey}, {Romani}, {Shaw},
  {Grainge}, {Birkinshaw}, {Lancaster}, {Worrall}, {Taylor}, {Cotter}, \&
  {Bustos}}]{Richards11}
{Richards}, J.~L., {Max-Moerbeck}, W., {Pavlidou}, V., {et~al.} 2011, \apjs,
  194, 29

\bibitem[{{Saito} {et~al.}(2013){Saito}, {Stawarz}, {Tanaka}, {Takahashi},
  {Madejski}, \& {D'Ammando}}]{saito}
{Saito}, S., {Stawarz}, {\L}., {Tanaka}, Y.~T., {et~al.} 2013, \apjl, 766, L11

\bibitem[{{Sasada} {et~al.}(2008){Sasada}, {Uemura}, {Arai}, {Fukazawa},
  {Kawabata}, {Ohsugi}, {Yamashita}, {Isogai}, {Sato}, \& {Kino}}]{sasada08}
{Sasada}, M., {Uemura}, M., {Arai}, A., {et~al.} 2008, \pasj, 60, L37

\bibitem[{{Sbarrato} {et~al.}(2012){Sbarrato}, {Ghisellini}, {Maraschi}, \&
  {Colpi}}]{sbarrato12}
{Sbarrato}, T., {Ghisellini}, G., {Maraschi}, L., \& {Colpi}, M. 2012, \mnras,
  421, 1764

\bibitem[{{Schlafly} \& {Finkbeiner}(2011)}]{Schlafly}
{Schlafly}, E.~F. \& {Finkbeiner}, D.~P. 2011, \apj, 737, 103

\bibitem[{{Schlegel} {et~al.}(1998){Schlegel}, {Finkbeiner}, \&
  {Davis}}]{Schlegel98}
{Schlegel}, D.~J., {Finkbeiner}, D.~P., \& {Davis}, M. 1998, \apj, 500, 525

\bibitem[{{Sikora} {et~al.}(1994){Sikora}, {Begelman}, \& {Rees}}]{sikora}
{Sikora}, M., {Begelman}, M.~C., \& {Rees}, M.~J. 1994, \apj, 421, 153

\bibitem[{{Urry} \& {Padovani}(1995)}]{urry}
{Urry}, C.~M. \& {Padovani}, P. 1995, \pasp, 107, 803

\bibitem[{{Villata} {et~al.}(2004){Villata}, {Raiteri}, {Kurtanidze},
  {Nikolashvili}, {Ibrahimov}, {Papadakis}, {Tosti}, {Hroch}, {Takalo},
  {Sillanp{\"a}{\"a}}, {Hagen-Thorn}, {Larionov}, {Schwartz}, {Basler},
  {Brown}, {Balonek}, {Ben{\'{\i}}tez}, {Ram{\'{\i}}rez}, {Sadun}, {Boltwood},
  {Carini}, {Barnaby}, {Coloma}, {Ros}, {Dai}, {Xie}, {Mattox}, {Rodriguez},
  {Asfandiyarov}, {Atkerson}, {Beem}, {Bloom}, {Chanturiya}, {Ciprini},
  {Crapanzano}, {de Diego}, {Efimova}, {Gardiol}, {Guerra}, {Kahharov},
  {Kapanadze}, {Karttunen}, {Kato}, {Kimeridze}, {Kudryavtseva}, {Lainela},
  {Lanteri}, {Larionova}, {Maesano}, {Marchili}, {Massone}, {Monroe},
  {Montagni}, {Nesci}, {Nilsson}, {Noble}, {Nucciarelli}, {Ostorero},
  {Papamastorakis}, {Pasanen}, {Peters}, {Pursimo}, {Reig}, {Ryle}, {Sclavi},
  {Sigua}, {Uemura}, \& {Wills}}]{Villata2004}
{Villata}, M., {Raiteri}, C.~M., {Kurtanidze}, O.~M., {et~al.} 2004, \aap, 421,
  103

\bibitem[{{Villata} {et~al.}(2002){Villata}, {Raiteri}, {Kurtanidze},
  {Nikolashvili}, {Ibrahimov}, {Papadakis}, {Tsinganos}, {Sadakane}, {Okada},
  {Takalo}, {Sillanp{\"a}{\"a}}, {Tosti}, {Ciprini}, {Frasca}, {Marilli},
  {Robb}, {Noble}, {Jorstad}, {Hagen-Thorn}, {Larionov}, {Nesci}, {Maesano},
  {Schwartz}, {Basler}, {Gorham}, {Iwamatsu}, {Kato}, {Pullen},
  {Ben{\'{\i}}tez}, {de Diego}, {Moilanen}, {Oksanen}, {Rodriguez}, {Sadun},
  {Kelly}, {Carini}, {Miller}, {Catalano}, {Dultzin-Hacyan}, {Fan}, {Ishioka},
  {Karttunen}, {Kein{\"a}nen}, {Kudryavtseva}, {Lainela}, {Lanteri},
  {Larionova}, {Matsumoto}, {Mattox}, {Montagni}, {Nucciarelli}, {Ostorero},
  {Papamastorakis}, {Pasanen}, {Sobrito}, \& {Uemura}}]{Villata2002}
{Villata}, M., {Raiteri}, C.~M., {Kurtanidze}, O.~M., {et~al.} 2002, \aap, 390,
  407

\bibitem[{{Wagner} \& {Witzel}(1995)}]{wagner}
{Wagner}, S.~J. \& {Witzel}, A. 1995, \araa, 33, 163

\end{thebibliography}

\clearpage

\begin{sidewaystable*}
\caption[]{The analyzed sample of BL Lac objects.}
\centering
\begin{tabular}{c|c|c|c|c|c|c|c|c|c|c}

\hline
\hline
%Object & RA & DEC& Type & $C_{(B-R,R)}$ & A & $<B-R>$ & $\alpha$ & States & $A_B$ & $A_R$ \\

Object & RA & DEC& Type & $C_{(B-R,R)}$ & $a$ & States& $\langle B-R \rangle$  & $A_B$ & $A_R$ & $\langle \alpha_{BR} \rangle$  \\
         (1)       & (2)           & (3)          &  (4) &  (5) & (6) & (7) & (8) & (9) & (10)  & (11) \\   
\hline

SHBL J001355.9--18540& 00:13:55 & $-$18:54:06& HBL & 0.079 $\pm$ 0.069 &  0.12 $\pm$  0.13 & $-$ & 1.641 $\pm$ 0.060 & 0.054 & 0.090 & 4.03 $\pm$   0.15  \\

PKS 0048--097& 00:50:41  & $-$09:29:04& LBL & 0.307 $\pm$ 0.058 & 0.041 $\pm$  0.013  & $-$& 1.069 $\pm$ 0.072 & 0.070 & 0.117 &  2.63 $\pm$  0.18 \\

RGB J0152+017& 	01:52:39  & $+$01:47:17& HBL & 0.880 $\pm$  0.041 & 0.920 $\pm$  0.030 & $+$& 0.980 $\pm$ 0.084 & 0.071 & 0.118 &  2.41 $\pm$  0.21 \\

1ES 0229+200& 02:32:48 & $+$20:17:17& HBL &  0.440 $\pm$ 0.070 & 0.84 $\pm$  0.13 &  $-$ & 1.691$\pm$  0.047 & 0.292 & 0.489 &  4.16 $\pm$  0.12  \\

AO 0235+16& 02:38:38 & $+$16:36:59& LBL & 0.052 $\pm$ 0.048  & 0.004 $\pm$  0.001 &$+$& 0.999 $\pm$ 0.096 & 0.172 & 0.287 &  2.46 $\pm$  0.24 \\

PKS 0301--243& 03:03:26 & $-$24:07:11& HBL & 0.263 $\pm$ 0.037  & 0.045 $\pm$  0.013  &$+$& 0.275 $\pm$ 0.034 & 0.048 & 0.080 &  0.677 $\pm$  0.084  \\

SHBL J032541.0--164618 & 03:25:41& $-$09:15:23& HBL & 0.883 $\pm$ 0.087 & 0.256 $\pm$  0.023  & $-$&  0.079 $\pm$ 0.090 & 0.083 & 0.139 & 0.20 $\pm$  0.22\\

1ES 0323+022 & 03:26:13  & $+$02:25:14& HBL & 0.770 $\pm$ 0.095 &  0.525 $\pm$  0.071  & $-$& 0.172 $\pm$ 0.070 & 0.244 & 0.407&  0.42 $\pm$  0.17\\

1ES 0347--121& 03:49:22 & $-$11:59:26& HBL & 0.917 $\pm$ 0.054 & 0.875 $\pm$  0.030  & $+$& 1.086 $\pm$ 0.154 & 0.102 & 0.170 &  2.67 $\pm$  0.38 \\

1ES 0414+00.9& 	04:16:52  & $+$01:05:23& HBL & 0.309 $\pm$  0.308& 0.092 $\pm$  0.017 & $+$& 0.239 $\pm$ 0.042 & 0.257 & 0.430 & 0.59 $\pm$  0.10 \\

PKS 0447--439& 04:49:24  & $-$43:50:08& HBL &  0.507 $\pm$ 0.061 & 0.101 $\pm$  0.024 & $-$& 0.731 $\pm$ 0.042 & 0.031 & 0.051 & 1.80 $\pm$  0.10 \\

PKS 0537--441& 05:38:50 & $-$44:05:08& LBL &  0.105 $\pm$ 0.021 & 0.0074 $\pm$  0.0056  & $+$&  0.928 $\pm$ 0.049 & 0.082 & 0.137 &  2.28 $\pm$  0.12 \\

PKS 0548--322& 05:50:40  & $-$32:16:17& HBL & 0.214 $\pm$ 0.074  & 0.29 $\pm$  0.11  & $-$& 1.369 $\pm$ 0.031&  0.076 & 0.128 & 3.364  $\pm$  0.076  \\

PKS 0735+178 & 	07:38:07  & $+$17:42:19& LBL & 0.264 $\pm$ 0.048 &  0.047 $\pm$  0.017  & $-$& 0.487 $\pm$ 0.049 & 0.076 & 0.128 & 1.20 $\pm$  0.12 \\

OJ 287& 08:54:49  & $+$20:06:30& LBL & 0.050 $\pm$ 0.020 &  0.009 $\pm$  0.014 & $+$& 0.955 $\pm$ 0.077 & 0.062 & 0.103 & 2.35 $\pm$  0.19 \\

SHBL J101015.9--311908 &10:10:16 & $-$31:19:09& HBL& 0.84 $\pm$ 0.12 &  0.720 $\pm$  0.072 & $-$& 0.809 $\pm$ 0.080 & 0.184 & 0.307 & 1.99 $\pm$  0.20 \\

1ES 1101--232& 11:03:37 &$-$23:29:30& HBL & 0.470 $\pm$ 0.047  & 0.328 $\pm$  0.044 & $+$&  1.020 $\pm$ 0.049 & 0.128 & 0.214& 2.51 $\pm$  0.12\\

Mrk 421& 	11:04:27 &$+$38:12:31& HBL  & 0.728 $\pm$ 0.046 &  0.176 $\pm$  0.026 & $-$& 0.691 $\pm$ 0.042 & 0.033 & 0.055&  1.70 $\pm$  0.10  \\

1ES 1218+304 & 	12:21:21 & $+$30:10:36& HBL  & 0.92 $\pm$  0.13 & 0.207 $\pm$  0.020 & $-$& 0.411 $\pm$ 0.064 & 0.045 & 0.076 & 1.01 $\pm$   0.16  \\

W Comae& 12:21:31 & $+$28:13:58& IBL  &  0.862 $\pm$ 0.055 & 0.163 $\pm$  0.014 & $-$& 1.020 $\pm$ 0.065 & 0.049 & 0.082& 2.51 $\pm$  0.16 \\

1ES 1312--423 & 	13:15:03 & $-$42:36:50& HBL  & 0.870 $\pm$ 0.072 &  1.40 $\pm$  0.07 & $+$& 1.018 $\pm$ 0.109&  0.229 & 0.382 & 2.50 $\pm$   0.27  \\

PKS 1424+240 & 14:27:00 & $+$23:48:00& IBL  &  -0.256 $\pm$ 0.038 & -0.062 $\pm$  0.025 & $+$& 1.228 $\pm$ 0.027 & 0.127 & 0.212 & 3.019  $\pm$  0.066  \\

AP Lib & 15:17:41 & $-$24:22:19& LBL & 0.785 $\pm$ 0.056 & 0.416 $\pm$  0.031  & $-$& 1.022 $\pm$ 0.059 & 0.299 & 0.500&  2.51 $\pm$  0.15 \\

PG 1553+113 & 15:55:43 & $+$11:11:24& HBL & -0.192 $\pm$ 0.022 & -0.027 $\pm$  0.008  & $+$&  0.710 $\pm$ 0.034 & 0.113  &0.189 &  1.746  $\pm$  0.084 \\

Mrk 501&  16:53:52  & $+$39:45:36& HBL & 0.556 $\pm$ 0.062 &  0.553 $\pm$  0.094 & $-$& 1.229 $\pm$ 0.030 & 0.042 & 0.070& 3.021  $\pm$  0.074\\

RGB J1725+118& 17:25:04 & $+$11:52:14& IBL  & 0.260 $\pm$ 0.072 & 0.044 $\pm$  0.020 & $+$&  0.458 $\pm$ 0.035 & 0.372&  0.621&  1.125 $\pm$   0.087 \\

PKS 2005--489& 	20:09:25 &$-$48:49:53& HBL & 0.756 $\pm$ 0.011& 0.1106 $\pm$  0.0036 & $+$& 1.359 $\pm$ 0.039&  0.121 & 0.203 & 3.341  $\pm$  0.096 \\

SHBL J213135.4--091523 & 21:31:35 & $-$09:15:23& HBL &  0.349 $\pm$ 0.091 & 0.096 $\pm$  0.029 & $-$& -0.005 $\pm$ 0.037 & 0.083 & 0.139 & $-$0.011 $\pm$  0.090 \\

PKS 2155--304&  	21:58:52 &$-$30:13:32& HBL & 0.296 $\pm$ 0.007  & 0.023 $\pm$  0.003 & $+$& 0.665$\pm$  0.035 & 0.047 & 0.078 &  1.635  $\pm$  0.086 \\

BL Lacertae & 	22:02:43 &$+$42:16:39& IBL &  0.698 $\pm$ 0.043  & 0.088 $\pm$  0.011 & $-$&  1.175 $\pm$ 0.045&  0.714 & 1.193 &  2.89 $\pm$  0.11 \\

\hline

\end{tabular}
\tablefoot{(1)~Object;
(2)~Right Ascention; 
(3)~Declination;
(4)~Blazar type; 
(5)~Color-magnitude correlation coefficient;
(6)~Color-magnitude correlation slope; 
(7)~Presence of distinct states in the CM diagrams;
(8)~The average value of $B-R$ color [in mag];
(9)--(10)~Galactic extinction cofficients [in mag];
(11)~The average optical spectral index.}
  \label{table_1}
  \end{sidewaystable*}

\clearpage

 \begin{table*}  
\caption[]{ATOM observations.}
\centering
\begin{tabular}{c|c|c|c|c|c|c}

\hline
\hline
Object & 2007 & 2008 & 2009 & 2010 & 2011 & 2012 \\
   
\hline

SHBL J001355.9--18540&  & 11, 12 & 5--12 & 8--12 & 6--12 &  5--12 \\

PKS 0048--097&   &  & 8--12 & 1,  7--12 &6--12  & 5--8, 10  \\

RGB J0152+017&11, 12  & 1,  6--12 & 1,  6--12 & 1,  2,  6--12 & 1,  6--12 & 1,  2,  6--10  \\

1ES 0229+200 &8--12  & 1,  8--12 & 1,  8--12  & 8--11 & 7--11 & 7--10 \\

AO 0235+16& 1,  9--11 & 7,  9--12 & 1,  7,  8 & 8 &  &   \\

PKS 0301--243&  &  &  8--12 & 2,  3,  6--12 & 7--12 &  2,  7--12 \\

SHBL J032541.0--164618 &  & 9--12 & 1,  7--10, 12 & 3,  7,  8 &7,  9, 10, 12  &  1,  7 \\

1ES 0323+022 &  &10--12  & 1,  7--10, 12 & 2,  9-- 12  & 9--11 & 1,  7--10  \\

1ES 0347--121& 8--12 &1,  8--12  &1,  3,  7--12  & 3,  7,  8--11 & 1,  7--11 & 1,  2, 10  \\

1ES 0414+00.9&1,  2,  8--12  &1,  8--12  &1,  3,  7--12  & 2,  8--12 & 1,  3,  7--11 & 10  \\

PKS 0447--439&  &  & 8, 12 & 1--4,  9--12 &  1,  4,  8--11 &1--4,  8, 10   \\

PKS 0537--441&  & 10--12 & 1,  3--9 & 1-- 5,  9--12 &  1,  9--11& 1--4,  8--10  \\

PKS 0548--322& 5,  9--12 & 1, 10--12 & 1,  3--5,  7--12 & 2--5,  9--12 & 1,  2,  4,  5,  9--11 &  1--5,  9, 10 \\

PKS 0735+178 &  &1,  9--12  & 1,  2--5,  9--12 & 1,  2--5, 10--12 &  1,  2,  4,  5,  9--11& 1,  3--5  \\

OJ 287& 2,  5, 11, 12 & 1,  5,  6, 10--12 & 1,  2--6, 10--12 & 3--6, 11, 12 & 1,  3--6, 11 &  2,  4,  5, 11, 12 \\

SHBL J101015.9--311908 &   1 & 5,  6, 11, 12 & 2--7 & 4--7 &  1,  3,  4,  6,  7 \\

1ES 1101--232& 2,  6,  7, 11, 12 & 1,  5--7, 11, 12 & 1,  3--7, 11 & 1,  2-- 7, 12 & 1,  4-- 6 & 3--5,  7  \\

Mrk 421& 5 & 1,  6,  7, 12 & 1,  3--6 & 2 &  &   \\

1ES 1218+304 &  &  & 4--6 & 2,  3--7 & 3--6 & 2-- 8  \\

W Comae&  & 6,  7 & 1,  3--6 & 2,  4--7 &4--6  & 3--5  \\

1ES 1312--423 &  & 9, 12 & 1,  3--8 & 2--7 &1,  2,  4--8  &   \\

PKS 1424+240 &  &  & 6--8 & 2--8 & 4--8 & 3--5  \\

AP Lib &  &  &  &6--8  & 3--9 &  2--9 \\

PG 1553+113 & 8,  9 & 8,  9 & 2--9 & 1,  2--9 & 4--9 & 3,  4--9  \\

Mrk 501&5,  7,  8  & 5--7 & 3--7,  8 & 6,  7,  9 & 4,  5 & 3,  9  \\

RGB J1725+118&  & 10 &3--9  & 6--9 & 5--9 &   3--9 \\

PKS 2005--489& 5--12 & 6--12 & 3--12 & 3--12 & 5--11 &3--10   \\

SHBL J213135.4--091523 &  & 7,  9, 10 & 3--10 & 4-- 8, 10, 11 & 6--10 &  4--7 \\

PKS 2155--304& 5--12  & 1,  6--12 & 1,  3--12 & 4--12 & 5--11 &  4--10 \\

BL Lacertae &  & 8--11 & 5--8, 10 &  &  & 6,  9  \\

\hline

\end{tabular}
\tablefoot{The table provides information about time periods (months and years) of the ATOM observations.}
\label{table_2}
\end{table*}

\clearpage

\begin{sidewaystable*}
\small
\caption[]{Multiwavelength properties of the studied BL Lac objects.}
%\begin{adjustwidth}{-6em}{-7em}
\centering
\begin{tabular}{c|c|c|c|c|c}

\hline
\hline

        Object        &       Distance            &   $L_\textrm{Optical}$    &    $L_\textrm{Radio}$     &     $L_\textrm{Fermi}$    &       $\Gamma_{\rm HE}$    \\
                      &        [Mpc]              &      [erg\,s$^{-1}$]      &       [erg\,s$^{-1}$]     &      [erg\,s$^{-1}$]      &              \\
         (1)          &           (2)             &            (3)            &             (4)           &            (5)            &         (6)       \\           
\hline

SHBL J001355.9--185406 &                   $  413$ & $(1.5 \pm 0.3)\times 10^{44}$ &                              &                              &                 \\
         PKS 0048--097 &                   $ 3659$ & $(1.0 \pm 0.5)\times 10^{46}$ & $(2.3 \pm 0.5)\times 10^{44}$ & $(6.9 \pm 1.4)\times 10^{46}$ & $2.14 \pm 0.04$ \\
        RGB J0152+017 &               $  660^{*}$ & $(6.2 \pm 1.3)\times 10^{44}$ &                              & $(5.0 \pm 1.5)\times 10^{44}$ & $1.79 \pm 0.14$ \\
         1ES 0229+200 &                   $  632$ & $(2.5 \pm 0.5)\times 10^{44}$ & $(2.5 \pm 0.5)\times 10^{40}$ &                              &                 \\
           AO 0235+16 &                   $ 5967$ & $(2.2 \pm 2.0)\times 10^{46}$ & $(1.2 \pm 0.2)\times 10^{45}$ & $(7.7 \pm 1.6)\times 10^{47}$ & $2.02 \pm 0.03$ \\
         PKS 0301--243 &                   $ 1268$ & $(3.0 \pm 0.8)\times 10^{45}$ &                              & $(1.5 \pm 0.3)\times 10^{46}$ & $1.94 \pm 0.03$ \\
HBL J032541.0--164618  &                   $ 1444$ & $(1.1 \pm 0.4)\times 10^{45}$ &                              &                              &                 \\
        1ES 0323+022  &                   $  668$ & $(1.8 \pm 0.4)\times 10^{44}$ & $(2.9 \pm 0.6)\times 10^{41}$ & $(7.7 \pm 1.9)\times 10^{44}$ & $2.06 \pm 0.09$ \\
         1ES 0347--121 &               $  788^{*}$ & $(1.9 \pm 0.5)\times 10^{44}$ & $(1.0 \pm 0.2)\times 10^{41}$ &                              &                 \\
        1ES 0414+00.9 &               $ 1260^{*}$ & $(1.0 \pm 0.2)\times 10^{45}$ & $(1.7 \pm 0.3)\times 10^{42}$ & $(1.5 \pm 0.4)\times 10^{45}$ & $1.99 \pm 0.16$ \\
         PKS 0447--439 &                   $  476$ & $(1.1 \pm 0.3)\times 10^{45}$ &                              & $(3.7 \pm 0.8)\times 10^{45}$ & $1.86 \pm 0.02$ \\
        PKS 0537--441 &                   $ 5613$ & $(4.9 \pm 3.3)\times 10^{46}$ &                              &  $(1.4 \pm 0.3)\times 10^{48}$ & $2.01 \pm 0.02$ \\
         PKS 0548--322 &               $  299^{*}$ & $(9.5 \pm 1.9)\times 10^{43}$ &                              &                              &                 \\
        PKS B0735+178 &                   $ 2256$ & $(4.0 \pm 1.3)\times 10^{45}$ & $(6.8 \pm 1.4)\times 10^{43}$ & $(3.5 \pm 0.7)\times 10^{46}$ & $2.05 \pm 0.04$ \\
               OJ 287 &                   $ 1537$ & $(7.9 \pm 3.6)\times 10^{45}$ & $(2.0 \pm 0.4)\times 10^{44}$ & $(1.2 \pm 0.2)\times 10^{46}$ & $2.23 \pm 0.04$ \\
HBL J101015.9--311908  &                   $  655$ & $(3.7 \pm 0.8)\times 10^{44}$ &                              &                              &                 \\
         1ES 1101--232 &                   $  877$ & $(4.4 \pm 0.9)\times 10^{44}$ &                              &                              &                 \\
              Mrk 421 &  $ 85.15 \pm   67.67^{*}$ & $(0.9 \pm 0.8)\times 10^{44}$ & $(0.8 \pm 0.6)\times 10^{41}$ & $(3.3 \pm 3.2)\times 10^{44}$ & $1.771 \pm 0.012$ \\
        1ES 1218+304  &  $594.00 \pm  206.48^{*}$ & $(3.8 \pm 2.8)\times 10^{44}$ & $(2.9 \pm 2.0)\times 10^{41}$ & $(1.6 \pm 1.1)\times 10^{45}$ & $1.71 \pm 0.07$ \\
              W Comae &               $  517^{*}$ & $(7.0 \pm 2.6)\times 10^{44}$ & $(2.0 \pm 0.4)\times 10^{42}$ &                              &                 \\
        1ES 1312--423  &                   $  485$ & $(1.1 \pm 0.2)\times 10^{44}$ &                              &                              &                 \\
        PKS 1424+240  &               $  882^{*}$ & $(4.0 \pm 0.9)\times 10^{45}$ & $(4.1 \pm 0.8)\times 10^{42}$ & $(1.4 \pm 0.3)\times 10^{46}$ & $1.78 \pm 0.02$ \\
              AP Lib  &               $  243^{*}$ & $(2.0 \pm 0.4)\times 10^{44}$ &                              & $(4.1 \pm 0.9)\times 10^{44}$ & $2.06 \pm 0.04$ \\
         PG 1553+113  & $1090.00 \pm  127.28^{*}$ & $(9.2 \pm 3.0)\times 10^{45}$ & $(3.8 \pm 0.9)\times 10^{42}$ & $(2.8 \pm 0.7)\times 10^{46}$ & $1.67 \pm 0.02$ \\
              Mrk 501 &               $  119^{*}$ & $(1.0 \pm 0.2)\times 10^{44}$ & $(2.9 \pm 0.6)\times 10^{41}$ & $(1.9 \pm 0.4)\times 10^{44}$ & $1.74 \pm 0.03$ \\
        RGB J1725+118 &                   $ 74.6$ & $(1.2 \pm 0.3)\times 10^{43}$ & $(6.0 \pm 1.2)\times 10^{39}$ &                              &                 \\
         PKS 2005--489 &  $ 351.50 \pm  51.62^{*}$ & $(1.3 \pm 0.5)\times 10^{45}$ &                              & $(7.1 \pm 2.2)\times 10^{44}$ & $1.78 \pm 0.05$ \\
HBL J213135.4--091523  &                   $ 2405$ & $(1.1 \pm 0.3)\times 10^{45}$ &                              &                              &                 \\
         PKS 2155--304 &               $  492^{*}$ & $(2.8 \pm 1.3)\times 10^{45}$ &                              & $(8.2 \pm 1.7)\times 10^{45}$ & $1.838 \pm 0.015$ \\
         BL Lacertae  &  $  472.0 \pm   59.4^{*}$ & $(1.6 \pm 0.7)\times 10^{45}$ & $(2.2 \pm 0.5)\times 10^{43}$ & $(2.8 \pm 0.7)\times 10^{45}$ & $2.11 \pm 0.04$ \\

\hline

\end{tabular}\\
\label{table_mwl}
%\end{adjustwidth}
%\justifying \vspace{2mm} \noindent \footnotesize
(1) Source name; (2) Source distance based on NED (if marked with $^{*}$ the metric distance is used, otherwise the luminosity distance provided); (3) Optical luminosity in $R$ band; (4) Radio luminosity at 15\,GHz based on the OVRO monitoring \citep{Richards11}; (5) HE $\gamma$-ray luminosity based on the 2FGL
\citep{2fgl}; (6) HE $\gamma$-ray photon index from the 2FGL.
\end{sidewaystable*} 

\end{document}